\documentclass{article}

\usepackage{caption}
\usepackage{dblfloatfix}
\usepackage{anyfontsize}
\usepackage{tabularx}
\usepackage{enumerate}
\usepackage{booktabs}
\usepackage{makecell}
\usepackage{multirow}
\usepackage{subcaption}
\usepackage{float}
\usepackage{appendix}
\usepackage{amssymb}
\usepackage{graphicx}
\usepackage{threeparttable}
\usepackage{dcolumn}
\usepackage[numbers,sort&compress]{natbib}
\usepackage{bm}
\usepackage[utf8]{inputenc}
\usepackage{amsmath, amsfonts, amsthm}
\usepackage{geometry}
\usepackage{authblk}
\usepackage{hyperref}
\hypersetup{
    colorlinks=true,
    linkcolor=blue,
    urlcolor=blue,
    citecolor=blue,
}

\newcommand{\email}[1]{\href{mailto:#1}{#1}}

\newcommand{\keywords}[1]{\textbf{Keywords:} #1}

\begin{document}

\title{Universality Classes of Interacting Dark Energy from Spontaneous Symmetry Breaking}

\author[1]{Pradosh Keshav MV%
    \thanks{Email: \email{pradosh93keshav@gmail.com}}}

\author[2]{NS Kavya%
    \thanks{Email: \email{kavya.ns@um.edu.my}}}

\author[1]{Kenath Arun%
    \thanks{Corresponding author. Email: \email{kenath.arun@christuniversity.in}}}

\affil[1]{Department of Physics and Electronics, CHRIST (Deemed to be University), Bengaluru, Karnataka 560029, India}
\affil[2]{Institute of Mathematical Sciences, Faculty of Science, Universiti Malaya, 50603 Kuala Lumpur, Malaysia}

\maketitle

\begin{abstract}
Phenomenological models of interacting dark energy (IDE) often treat the late-time activation history of the dark sector coupling as an independent function. We show that in conformally coupled scalar--tensor theories, this freedom is constrained by the local restoring structure of the symmetry-breaking potential. Within the adiabatic tracking regime, the coupling evolution satisfies \(n=3/p\), where \(p\) is the restoring order near the broken minimum thereby organizing distinct symmetry-breaking potentials such as quartic, Coleman--Weinberg, and axion-like forms into a common asymptotic dynamical class (\(p=1\), \(n=3\)). We test this framework using Planck~2018 CMB lensing, redshift-space distortions, and Pantheon+SH0ES supernova data. Current observations provide only limited discrimination between the predicted activation classes and yield no statistically significant evidence for a nonzero interaction with \(|\beta_0|\lesssim0.26\) at \(95\%\) credibility. At the same time, the rigid asymptotic implementation $(n=3)$ is strongly disfavored by the combined geometric and growth constraints indicating that the observable coupling history cannot be identified directly with its asymptotic attractor form. In the heavy-scalar adiabatic regime, the modifications to the growth rate \(f(z)\) and growth factor \(D(z)\) are of opposite sign throughout \(0\le z\le2\), suppressing the net deviation in \(f\sigma_8(z)\) to \(\Delta f\sigma_8/f\sigma_8\lesssim0.3\%\) across the posterior. Standard growth-rate measurements therefore have limited sensitivity to this class of models, shifting the observational focus toward probes that constrain \(f(z)\) and \(D(z)\) independently. Taken together, these results establish a dynamical classification of late-time IDE activation histories and clarify how finite-redshift observables are related to the asymptotic attractor structure and the local restoring properties of the underlying scalar potential.
\end{abstract}

\section{Introduction}
\label{sec:intro}

The ongoing tensions between early and late-time cosmological observations, particularly the $\gtrsim5\sigma$ discrepancy in the Hubble constant $H_0$~\citep{riess2022comprehensive, aghanim2020planck}, suggest that extensions to the standard $\Lambda$CDM model remain well motivated. Recent BAO measurements from DESI DR1 and DR2~\citep{adame2024desi, abdul2025desi}, when combined with CMB and supernova data, further show a statistical preference for dynamical dark energy (DE) over $\Lambda$CDM~\citep{giare2024interacting, you2025dynamical, sabogal2024quantifying}. At the same time, the recent KiDS-Legacy determination, $S_8 = 0.815^{+0.016}_{-0.021}$~\citep{wright2025kids}, substantially reduces the previously reported weak-lensing tension with Planck, shifting the observational focus from whether extensions to $\Lambda$CDM are required to which classes of dark sector models can accommodate the apparent preference for dynamical DE while remaining consistent with increasingly stringent constraints on late-time structure growth.

If DE is dynamical, it is natural to model it at low energies using a scalar degree of freedom\citep{elizalde2004late, nojiri2011unified}. From an effective field theory (EFT) perspective, couplings to matter fields are generically allowed unless forbidden by symmetry~\citep{gleyzes2015effective, gleyzes2016effective, d2016quantum, aoki2026effective}. While couplings to baryons are tightly constrained by equivalence principle tests and fifth-force searches~\citep{touboul2022result}, the dark sector remains comparatively unconstrained, making interacting dark energy (IDE) a natural extension of the standard cosmological model~\citep{clemson2012interacting, farrar2004interacting, sabogal2025sign}. In scalar-tensor models of IDE, the interaction between the scalar field and dark matter (DM) arises at the level of the underlying action \citep{amendola2000coupled, wetterich1988cosmology, elizalde2004late}. A conformal factor $A(\phi)$ induces a field-dependent DM mass, $m_{\rm DM}(\phi)=m_0 \, A(\phi)$, and a dimensionless coupling $\beta(\phi)=M_{\rm Pl}\,d\ln A/d\phi$, both of which play a central role in the dynamics studied here. Such couplings arise naturally in UV-complete settings, including dilaton and Higgs-portal constructions \citep{damour1994string, gasperini1994dilaton, patt2006higgs, katz2014higgs}. 

However, in most phenomenological IDE parameterizations, the coupling kernel is introduced directly at the level of the background equations with an assumed redshift dependence \citep{di2021interacting, yang2018interacting, wang2024further, silva2025new, silva2026one}. While useful for model-independent data exploration, phenomenological coupling kernels with freely prescribed redshift dependence provide limited information about the underlying field dynamics and can develop perturbative instabilities in parts of parameter space~\citep{valiviita2008large, van2024interacting, van2025linear, jackson2009large}. Because the coupling history is not tied to an underlying scalar potential or attractor structure, the functional freedom of these parameterizations can also absorb degeneracies with background cosmological parameters, generating apparent statistical preferences for nonzero interactions that do not necessarily correspond to physically regulated dark sector dynamics. Distinguishing physically regulated interactions from fit improvements driven by excess parametric freedom, therefore, requires constructions in which the coupling history is determined by the underlying potential structure.

A natural realization of such dynamically regulated activation arises when the scalar potential undergoes spontaneous symmetry breaking (SSB)~\citep{bamba2013spontaneous, hinterbichler2011symmetron, tan2019dark}. In screening mechanisms such as the chameleon~\citep{khoury2004chameleon, burrage2018tests} and symmetron~\citep{hinterbichler2010screening, upadhye2013symmetron}, the environmental density controls the effective scalar mass, suppressing fifth forces in high-density regions while allowing them in low-density environments. Here, we focus on the complementary cosmological regime in which the dilution of the CDM density drives the field toward a symmetry-broken minimum. In this regime, the evolution follows a density-controlled attractor trajectory, and the scalar--DM coupling remains strongly suppressed through most of cosmic history before activating smoothly at late times as the system approaches the broken phase. We show that this mechanism leads to a universality classification for late-time IDE in which the coupling evolution is governed by the attractor dynamics of the underlying scalar theory. This classification parallels the organization of screening mechanisms~\citep{joyce2015beyond, brax2021testing, vardanyan2023modeling}, where the dynamics are likewise determined by the local structure of the scalar potential, but applies here in the opposite regime of the low-density symmetry-broken phase. A companion paper~\citep{mv2025spontaneous} derives the full perturbation equations for a representative symmetry-breaking realization and confronts that specific realization with observational data. In this work, rather than constraining a single potential, we derive the relation $n=3/p$, linking the late-time activation history to the local restoring structure of the scalar potential and thereby organizing a broad class of symmetry-breaking models into common asymptotic dynamical classes.

We implement the corresponding coupling correction in a modified Boltzmann solver~\citep{blas2011cosmic, class_ide} and investigate whether current cosmological observations can distinguish the resulting activation classes by treating the activation index $n$ as a free parameter. In the adiabatic tracking limit, the late-time evolution near the broken minimum is described by a universal normal form in which the activation profile approaches a logistic form with steepness parameter $n=3/p$, where $p$ denotes the local restoring order of the scalar potential. Potentials as diverse as the quartic Mexican-hat, Coleman--Weinberg, and axion-like forms all belong to the same $p=1$ class and therefore exhibit identical asymptotic activation behavior. Moreover, the $p=1$ class is radiatively preferred, since one-loop corrections generically drive higher-order ($p>1$) minima toward the non-degenerate $p=1$ form unless protected by additional symmetries. In this regime, deviations from $\Lambda$CDM arise primarily at the perturbative level~\citep{barros2019coupled, barros2020spherical}, where opposite-sign modifications to the growth rate and growth factor strongly suppress the observable signal in $f\sigma_8$. We compare a six-parameter realization with free $n$ to a five-parameter realization with fixed $n=3$ using Planck~2018 CMB lensing~\citep{aghanim2020planck}, redshift-space distortion (RSD) measurements~\citep{alam2021completed}, and the Pantheon+SH0ES Type~Ia supernova (SN) sample~\citep{brout2022pantheon+}. The present model does not produce the background-level $w(z)\neq-1$ signature directly probed by DESI BAO measurements (see, e.g., \citep{giare2024interacting, pan2026interacting}); connecting the adiabatic IDE regime studied here to the DESI preference for dynamical DE likely requires extensions beyond the perturbative attractor limit considered in this work.

The remainder of this paper is organized as follows. Sections~\ref{sec:covariant_framework}--\ref{sec:logistic_activation} derive the logistic normal form and the associated universality structure, while Section~\ref{sec:methodology} implements the resulting IDE framework in a modified Boltzmann solver. Observational constraints, microphysical interpretation, and the implications of this classification for future observational strategy are presented in Sections~\ref{sec:observ_constraints}--\ref{sec:conclusion}.

\section{General Formalism}
\label{sec:covariant_framework}
We work in the Einstein frame, where the conformal coupling between the scalar field \(\phi\) and DM is introduced through a field-dependent mass rescaling; this choice makes the energy-exchange structure transparent and simplifies the adiabatic tracking analysis that follows.

Consider a canonical scalar field $\phi$ with a standard two-derivative kinetic term conformally coupled to the DM sector. The minimal covariant action can be written as
\begin{align}
S = &\int d^4x \sqrt{-g}\left[\frac{M_{\rm Pl}^2}{2} R- \frac{1}{2} g^{\mu\nu} \partial_\mu \phi \partial_\nu \phi- V(\phi) \right] \\
&+ S_{\rm DM}\!\left[A^2(\phi) g_{\mu\nu}, \psi_{\rm DM}\right] + S_{\rm b}\!\left[g_{\mu\nu}, \psi_{\rm b}\right]
+ S_{\rm r}\!\left[g_{\mu\nu}, \psi_{\rm r}\right],
\label{eq:action}
\end{align}
where $M_{\rm Pl}=(8\pi G)^{-1/2}$ is the reduced Planck mass, $A(\phi)$ is the conformal coupling function acting only on the DM sector, $S_{\rm DM}$ denotes the dark matter action, and $S_{\rm b}$ and $S_{\rm r}$ are the minimally coupled baryon and radiation actions, respectively. Here $V(\phi)$ is the scalar potential whose local structure will be shown to determine the universality class. The DM sector $\psi_{\rm DM}$ is taken to be a non-relativistic species with a field-dependent mass $m_{\rm DM}(\phi)=A(\phi)m_0$; its microscopic realization may represent a Dirac fermion $\psi_{\rm DM}$ with Yukawa-type mass term $A(\phi)m_0\bar{\psi}_{\rm DM}\psi_{\rm DM}$ or a heavy scalar field. In either case, only the non-relativistic limit enters the cosmological dynamics.

At the background level, the DM behaves as pressureless dust, so that $T_{\rm DM}=-\rho_{\rm DM}$~\citep{damour1992tensor,faraoni2004scalar,lerner2009gauge,lebedev2023dark}. Baryons $\psi_{\rm b}$ and radiation $\psi_{\rm r}$ couple directly to the Einstein-frame metric $g_{\mu\nu}$, consistent with fifth-force and equivalence-principle constraints on visible matter~\citep{touboul2022result}. Moreover, since the trace of the radiation energy-momentum tensor (EMT) vanishes ($T_{\rm r}=0$), the conformal interaction induces energy exchange only between the scalar and DM sectors.

The dimensionless coupling function is defined as
\begin{equation}
\beta(\phi) \equiv M_{\rm Pl}\frac{d\ln A(\phi)}{d\phi}.
\label{eq:beta_def}
\end{equation}
When the scalar becomes heavy, it tracks the instantaneous minimum of the effective potential rather than rolling freely. As it evolves along a density-controlled adiabatic trajectory $\phi_{\rm ad}(\rho_{\rm DM})$, the coupling becomes time-dependent~\citep{zaregonbadi2023cosmological}. In this work, we focus on late-time interactions of the scalar with DM, and hence we prefer the tracking solution
\begin{equation}
\beta_{\rm eff}(a)\equiv \beta\!\left(\phi_{\rm ad}(a)\right),
\end{equation}
with \(\phi_{\rm ad}(a)\) determined by the evolving DM density.

Varying the action with respect to $\phi$ yields the scalar equation of motion,
\begin{equation}
\square \phi - V'(\phi)
= \frac{\beta(\phi)}{M_{\rm Pl}}\, T_{\rm DM},
\label{eq:scalar_eom}
\end{equation}
where $T_{\rm DM}=g_{\mu\nu}T^{\mu\nu}_{\rm DM}$ is the trace of the DM EMT and prime denotes differentiation with respect to $\phi$. For non-relativistic matter, $T_{\rm DM}= g_{\mu \nu}T^{\mu \nu}_{\rm DM} =-\rho_{\rm DM}$ (with negligible pressure), and the right-hand side becomes a source term proportional to the local DM energy density.

The total EMT is conserved while allowing energy exchange between the scalar and DM sectors. On a spatially flat Friedmann-Lemaître-Robertson-Walker (FLRW) background \[ds^2 = -dt^2 + a^2(t)d\vec{x}^{\,2},\] the homogeneous equations take the form
\begin{align}
&3M_{\rm Pl}^2 H^2 = \rho_{\rm DM} + \rho_\phi + \rho_{\rm b} + \rho_{\gamma} + \rho_{\nu},
\label{eq:friedmann} \\
&\dot{\rho}_{\rm DM} + 3H\rho_{\rm DM} = -\frac{\beta(\phi)}{M_{\rm Pl}}\dot{\phi}\rho_{\rm DM},
\label{eq:dm_continuity} \\
&\ddot{\phi} + 3H\dot{\phi} + V'(\phi) = \frac{\beta(\phi)}{M_{\rm Pl}}\rho_{\rm DM},
\label{eq:phi_background}
\end{align}
where $H\equiv \dot a/a$, $\rho_\phi$ is the scalar field density, \( \rho_{\rm b}\) is the baryon density, \(\rho_{\gamma}\) and \(\rho_{\nu}\) are radiation components. These equations describe a coupled system in which \(\beta>0\) and \(\dot{\phi}>0\) correspond to energy transfer from DM to the scalar. The interaction terms proportional to $\beta(\phi)$ vanish in the symmetric phase \(\beta\to0\) and activate smoothly as the scalar approaches the broken minimum.

We now focus on a density-tracking regime in which the scalar field evolves slowly around a minimum of the effective potential \(V_{\rm eff}\) (see \citep{khoury2004chameleon,brax2012unified}). Defining \(m_{\rm eff}^2(\phi)\equiv \left.{\partial^2 V_{\rm eff}}/{\partial \phi^2}\right|_{\phi=\phi_{\rm ad}}\), the heavy field hierarchy \(m_{\rm eff}^2 \gg H^2\) leads to a rapid relaxation to the instantaneous minimum. In this tracking regime, the adiabatic conditions
\begin{equation}
\left|\frac{\ddot{\phi}}{3H\dot{\phi}}\right|\ll1,
\qquad
\left|\frac{\dot{\phi}}{H\phi}\right|\ll1,
\qquad
\dot{\phi}^2 \ll V(\phi),
\label{eq:adiabatic_conditions}
\end{equation}
reduce the scalar equation of motion. The first two conditions ensure that the field evolves slowly relative to the Hubble rate, while the third condition keeps the kinetic energy subdominant to the potential energy. In the heavy field limit, the dominant balance in~\eqref{eq:scalar_eom} becomes $|3H\dot{\phi}| \ll |V'(\phi)|$, giving the algebraic balance condition
\begin{equation}
V'(\phi_{\rm ad})
= \frac{\beta(\phi_{\rm ad})}{M_{\rm Pl}}\rho_{\rm DM}.
\label{eq:adiabatic_balance}
\end{equation}
The balance condition defines the density-controlled trajectory $\phi_{\rm ad}(\rho_{\rm DM})$ corresponding to the regime where the \(\phi\) closely tracks the density-dependent minimum of the \(V_{\rm eff}\). In this sense, one could obtain the balance condition by minimizing the effective potential
\begin{equation}
V_{\rm eff}(\phi;\rho_{\rm DM}) = V(\phi) - \frac{\rho_{\rm DM}}{M_{\rm Pl}} \int^\phi \beta(\tilde\phi)\, d\tilde\phi,
\label{eq:Veff_general}
\end{equation}
and, for constant coupling $\beta(\phi)=\beta_c$, we have
\begin{equation}
V_{\rm eff}(\phi;\rho_{\rm DM}) = V(\phi) - \frac{\beta_c}{M_{\rm Pl}}\rho_{\rm DM}\phi.
\label{eq:simplified_eff_potential}
\end{equation}

\medskip

The specialization to a $\mathbb{Z}_2$-symmetric polynomial potential and the derivation of the autonomous flow equation are provided in Appendix~\ref{app:polynomial}. The resulting attractor dynamics, which constitute the main result of the paper, are analyzed in the next section.

\section{Attractor Structure and Logistic Normal Form}
\label{sec:attractor_structure}

\subsection{Local asymptotics and attractor eigenvalue}
\label{subsec:local_stability}

Consider a scalar potential $V(\phi)$ with an analytic symmetry-breaking minimum at $\phi = v$. In the adiabatic regime, the late-time evolution is controlled by the local behavior of $V(\phi)$ near the minimum. The balance condition is therefore determined by the leading nonvanishing term in the expansion of $V'(\phi)$ about $\phi = v$.  

For a smooth potential with analytic minimum at $\phi = v$, the first derivative expands as
\begin{equation}
V'(\phi)=\kappa_p (\phi - v)^p+\mathcal{O}\!\left((\phi-v)^{p+1}\right),
\label{eq:local_expansion}
\end{equation}
where $p\ge1$ is the order of the first non-vanishing term in the expansion of $V'(\phi)$ about $\phi=v$, and $\kappa_p\neq0$. For $p>1$, one has $V''(v)=0$, so the leading restoring force is nonlinear; stability requires that the first non-vanishing even derivative of $V$ at $\phi=v$ be positive \citep{strogatz2024nonlinear} (specifically, $\partial^{2p}V/\partial\phi^{2p}|_{\phi=v} > 0$ for the minimum to be dynamically stable under the flow).

Recall that in the adiabatic tracking regime, the field is constrained by the balance condition~\eqref{eq:adiabatic_balance}, which defines the density-controlled trajectory. For simplicity, we momentarily specialize to a constant coupling $\beta(\phi)=\beta_c$, so that \(V'(\phi_{\rm ad}) =({\beta_c}/{M_{\rm Pl}})\rho_{\rm DM}\). This simplification does not affect the asymptotic scaling, which we verify in Appendix~\ref{app:adiabatic_asymptotics}. The scaling result therefore carries over to the general field-dependent case  $\beta_{\rm eff}(a)=\beta(\phi_{\rm ad}(a))$, which we adopt throughout this work.

Defining the displacement from the minimum by
\[
\delta(a)\equiv \phi_{\rm ad}(a)-v, \qquad |\delta|\ll v,
\]
and substituting Eq.~\eqref{eq:local_expansion} into~\eqref{eq:adiabatic_balance} gives, to leading order,
\begin{equation}
\kappa_p \, \delta^p=\frac{\beta_c}{M_{\rm Pl}}\,\rho_{\rm DM}(a)+ \mathcal{O}(\delta^{p+1}).
\label{eq:delta_balance_general}
\end{equation}
Solving Eq.~\eqref{eq:delta_balance_general} for $\delta$ then gives
\begin{equation}
\delta(a)\propto\rho_{\rm DM}(a)^{1/p} \propto a^{-\,\frac{3}{p}},
\label{eq:delta_scaling_general}
\end{equation}
so the displacement from the minimum decays as a power law determined by the local restoring order $p$. The scaling holds for $|\delta| \ll v$, i.e., near the broken minimum. At earlier times, when $\rho_{\rm DM}$ is large and $\delta \sim v$, the balance condition~\eqref{eq:delta_balance_general} must be solved numerically; the analytic power-law behavior emerges only asymptotically. A detailed analysis of the early-time evolution is provided in Appendix~\ref{app:adiabatic_asymptotics}.

Differentiating Eq.~\eqref{eq:delta_scaling_general} with respect to $\ln a$ gives $d\ln\delta/d\ln a = -3/p$ at leading order. Expressing this as a flow equation for $\delta$ itself yields the linearized dynamics about the fixed point,
\begin{equation}
\frac{d\delta}{d\ln a} = -\frac{3}{p}\,\delta + \mathcal{O}(\delta^{\,2}),
\label{eq:linear_flow_general}
\end{equation}
and hence $\delta=0$ is a hyperbolic fixed point of the reduced one-dimensional autonomous system with eigenvalue
\begin{equation}
\gamma_p = -\frac{3}{p}.
\label{eq:eigenvalue_general}
\end{equation}
The magnitude of this eigenvalue decreases with increasing $p$, noting that flatter minima approach the attractor more slowly. This behavior depends only on the scaling $\rho_{\rm DM}\propto a^{-3}$ and the local analytic structure of the potential near $\phi=v$, and it is insensitive to the global nonlinear shape of $V(\phi)$. This conclusion holds provided the tracking hierarchy remains valid, and no additional time-dependent sources modify the dilution law.

\subsection{Logistic normal form near the attractor}
\label{subsec:logistic_normal_form}

The previous section showed that the late-time approach to the symmetry-breaking minimum is governed by a one-dimensional normal form. To make this structure explicit, introduce the dimensionless order parameter
\[
x(a)\equiv \frac{\phi_{\rm ad}(a)}{v}, \qquad x\to 1 \quad \text{as} \quad a\to\infty,
\]
where $v$ is the vacuum expectation value (VEV) of the scalar field. Recalling the displacement $\delta(a)$ defined above, we have $x = 1 + \delta/v$.

Using Eq.~\eqref{eq:eigenvalue_general}, the linearized late-time evolution can be written as Eq.~\eqref{eq:linear_flow_general}, valid for $|\delta|\ll v$. Dividing by $v$ then gives
\begin{equation}
\frac{dx}{d\ln a} = -\frac{3}{p}\,(x-1) + \mathcal{O}\!\left((x-1)^2\right).
\label{eq:x_linear_flow}
\end{equation}
The logistic form is obtained by reparametrizing the physical branch using the inverse variable \(y(a)\equiv {1}/{x(a)}.\) This is useful since it maps the relevant domain $x\in[1,\infty)$ to $y\in(0,1]$, with the late-time attractor at $y=1$. Physically, $y(a) \equiv v/\phi_{\rm ad}(a)$ measures how close the field is to its broken-phase vacuum: $y \to 0$ when the field is far from the minimum (early times, symmetric phase), and $y \to 1$ as it relaxes to $v$ (late times, broken phase). As we will see, $y(a)$ maps directly onto the normalized activation profile of the coupling.

Differentiating $y=1/x$ with respect to $\ln a$ gives
\[
\frac{dy}{d\ln a} = -\frac{1}{x^2}\frac{dx}{d\ln a}.
\]
Substituting Eq.~\eqref{eq:x_linear_flow} and expanding near the attractor, where $x=1+\mathcal{O}(\delta)$ and $x^{-2}=1+\mathcal{O}(x-1)$, yields
\[
\frac{dy}{d\ln a} = \frac{3}{p}\,\frac{x-1}{x^2} + \mathcal{O}\!\left((x-1)^2\right) = \frac{3}{p}\,y(1-y) + \mathcal{O}\!\left((1-y)^2\right).
\]
Note that the identity $y(1-y) = (x-1)/x^2$ holds exactly as an algebraic relation between $y = 1/x$ and $x$. Substituting and retaining only the leading-order contribution near the attractor $y \to 1$ (equivalently $x \to 1$, where higher-order terms in $(1-y)$ are suppressed) yields the local normal form
\begin{equation}
\frac{dy}{d\ln a} = \frac{3}{p}\,y(1-y) + \mathcal{O}\!\left((1-y)^2\right),
\label{eq:logistic_normal_form}
\end{equation}
which is a logistic differential equation, with $\ln a$ acting as the time coordinate and $3/p$ setting the intrinsic growth rate. The fixed point $y=0$ is unstable, and $y=1$ is stable, with the stability eigenvalue given by Eq.~\eqref{eq:eigenvalue_general}. The corresponding logistic index is therefore \(n={3}/{p}\): the activation index determined by the local restoring order of the scalar potential. This defines universality classes, where potentials with the same local restoring order $p$ share the same late-time normal form and asymptotic scaling exponent, although their finite-redshift evolution may differ.

Equation~\eqref{eq:logistic_normal_form} is a local normal form of the late-time attractor and should not be interpreted as an exact global solution. Away from the symmetry-breaking minimum, higher-order nonlinearities and departures from strict adiabatic tracking modify the evolution. Nevertheless, the logistic behavior emerges generically from the local analytic structure of $V(\phi)$, independent of its global form. The activation index $n$ is therefore fixed by the local restoring order $p$, establishing a direct correspondence between the local analytic structure of the scalar potential and the phenomenological coupling histories commonly employed in the IDE literature (see e.g. \citep{linder2004reconstructing,linder2005many,di2017constraining,di2017can,keshav2024interacting,V:2025oex,di2020nonminimal}).

\subsection{Polynomial realization of the universality classes}
\label{subsec:polynomial_realization}

The results of Secs.~\ref{subsec:local_stability} and \ref{subsec:logistic_normal_form} show that the late-time dynamics of the scalar field \(\phi\) along the controlled trajectory are governed by the local restoring order $p$ of the scalar potential. We now show that the $\mathbb{Z}_2$-symmetric polynomial family (commonly used in chameleon and symmetron models) provides an explicit realization of these universality classes.

Consider the symmetry-breaking potential
\begin{equation}
V(\phi)=\frac{\lambda}{2q}\bigl(\phi^2-v^2\bigr)^q, \qquad q\ge2,
\label{eq:poly_potential}
\end{equation}
which corresponds to the standard Mexican-hat potential for $q=2$ (with the normalization $\lambda/2q$ chosen so that the vacuum energy vanishes at $\phi=\pm v$ for all $q$) and to higher-order EFT deformations for $q>2$ \citep{ghilencea2019spontaneous, katz2014higgs}. Differentiating \eqref{eq:poly_potential} gives
\[
V'(\phi) = \lambda \phi (\phi^2-v^2)^{q-1}.
\]
Writing $\phi = v + \delta$ with $|\delta| \ll v$ and expanding near the minimum, we have $\phi \approx v$ and $\phi^2 - v^2 \approx 2v\delta$. Substituting these approximations into $V'(\phi)$ yields, to leading order,
\[
V'(\phi) \propto \delta^{\,q-1},
\]
where the first non-vanishing term in the local expansion of $V'(\phi)$ is of order $p = q-1$. Carrying the expansion through to extract the coefficient gives $\kappa_{q-1} = \lambda\,2^{q-1}v^{q}$, which sets the magnitude of the restoring force and enters into the effective mass $m_{\rm eff}^2 \sim V''(\phi_{\rm ad})$.

By substituting $p=q-1$ into Eqs.~\eqref{eq:delta_scaling_general}, \eqref{eq:eigenvalue_general}, and using \(n={3}/{p}\), the displacement from the minimum scales as \(\delta(a) \propto a^{-\,{3}/{q-1}}.\) The corresponding attractor eigenvalue and logistic index are
\begin{equation}
\gamma_p = -\frac{3}{q-1}, \qquad n = \frac{3}{q-1}.
\label{eq:n_poly}
\end{equation}

It follows that larger values of $q$ correspond to flatter minima, leading to a slower approach to the attractor and broader activation profiles. A broader activation profile means the coupling turns on more gradually over a wider redshift range, which is associated with a weaker growth modification.

\begin{figure}
\centering
\begin{subfigure}[t]{0.48\textwidth}
\centering
\includegraphics[width=\linewidth]{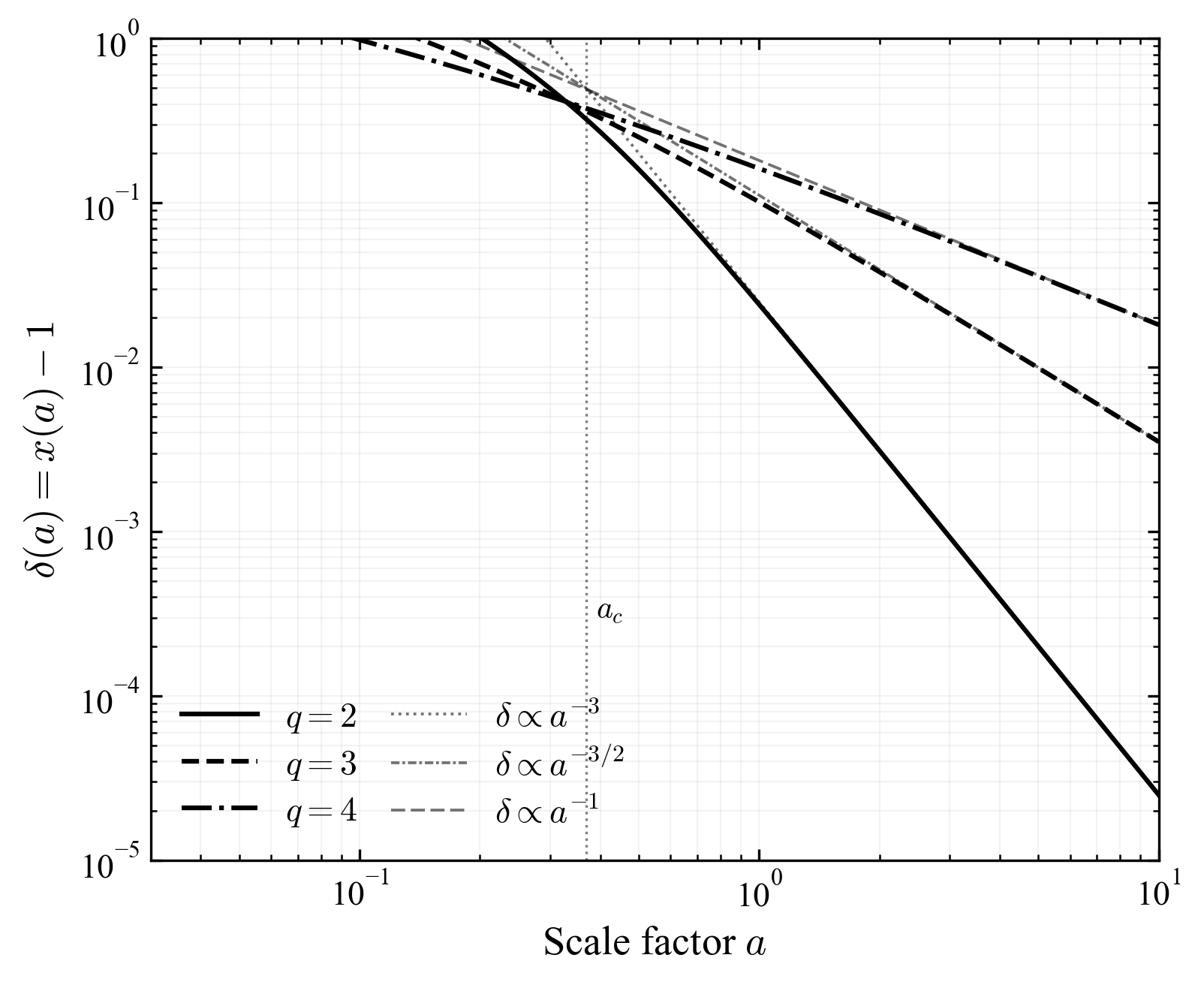}
\subcaption{}
\label{fig:universal_attractor_a}
\end{subfigure}
\hfill
\begin{subfigure}[t]{0.48\textwidth}
\centering
\includegraphics[width=\linewidth]{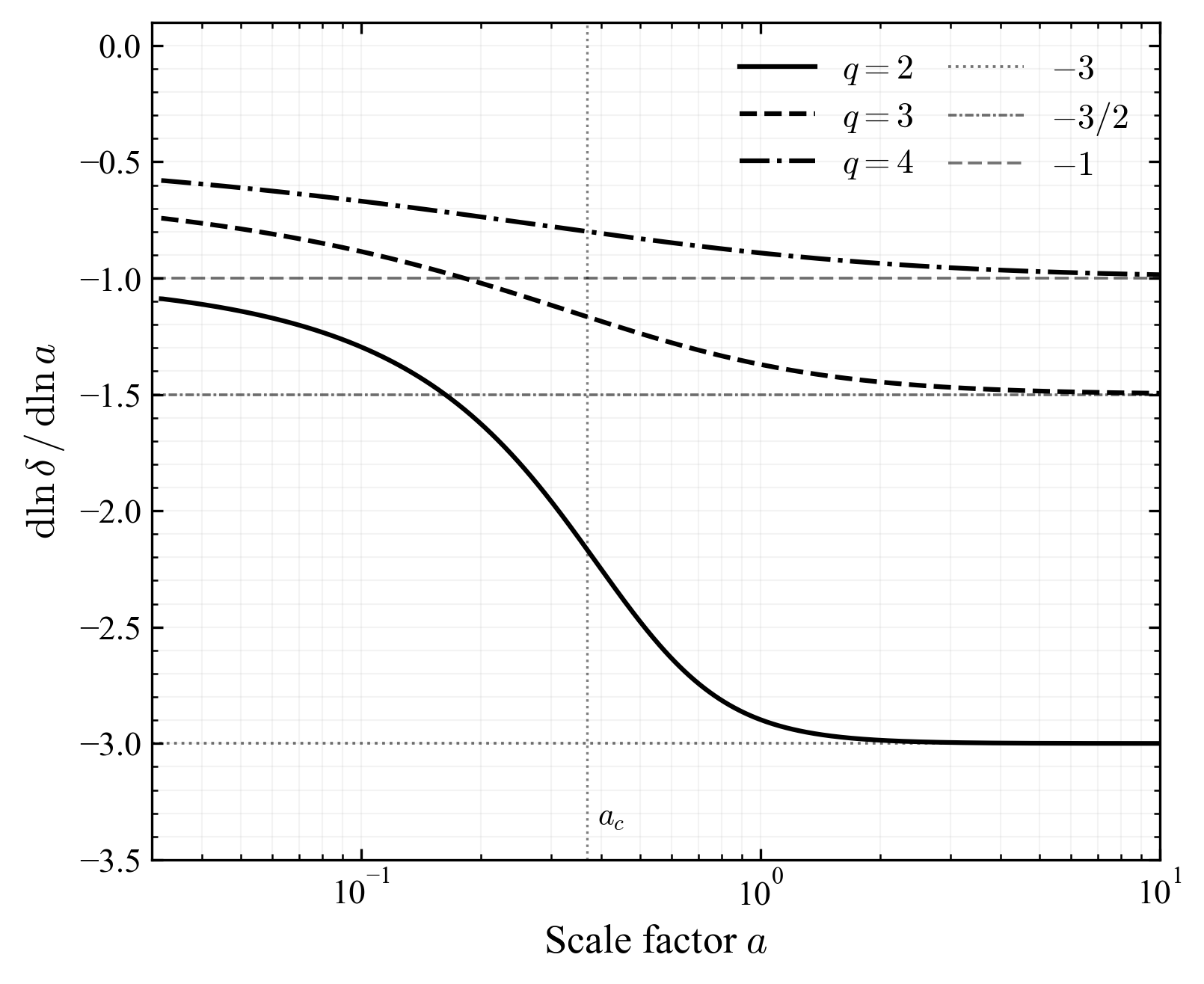}
\subcaption{}
\label{fig:universal_attractor_b}
\end{subfigure}
\caption{\small Late-time attractor behavior for the polynomial potentials~\eqref{eq:poly_potential}. (a)~Evolution of the displacement from the broken minimum, $\delta(a)$, for $q=2,3,4$. The curves approach distinct power-law trajectories, with faster decay for lower $q$. (b)~Instantaneous scaling exponent $d\ln\delta/d\ln a$ for the same models. The evolution approaches constant asymptotic values, with each curve settling to a different attractor eigenvalue determined by the corresponding universality class shown in Eq.~\eqref{eq:n_poly}.}
\label{fig:universal_attractor}
\end{figure}

Furthermore, the polynomial family does not form a continuous deformation of a single model, but instead defines a discrete set of dynamical classes labeled by the restoring order $p=q-1$. The quartic case $q=2$ corresponds to $p=1$ and yields the canonical eigenvalue $\gamma_p=-3$ and index $n=3$, representing the fastest approach to the attractor within this polynomial family. For $q>2$, the minimum becomes increasingly flat at tree level, and the approach to the attractor slows accordingly, with smaller $|\gamma_p|$ and broader activation profiles. This behavior is illustrated in Fig.~\ref{fig:universal_attractor}, which shows the displacement $\delta(a)=x(a)-1$ and the instantaneous scaling exponent $d\ln\delta/d\ln a$ for $q=2,3,4$. At late times, the solutions converge to the scalings of Eq.~\eqref{eq:delta_scaling_general}, and the exponents asymptote to the constant values $-3$, $-3/2$, and $-1$. The broken minimum is therefore a hyperbolic fixed point with eigenvalue $\gamma_p = -3/(q-1)$ for each universality class.

The local curvature of the potential also reflects this structure: flatter minima correspond to a suppressed curvature near the vacuum, leading to a time-dependent mass scale. For instance, we could expand the second derivative of \eqref{eq:poly_potential} near the minimum
\begin{equation}
V''(\phi_{\rm ad}) \propto \lambda v^{2q-2} \left(\frac{\delta}{v}\right)^{q-2},
\label{eq:poly_curvature_clean}
\end{equation}
so that, substituting the scaling $\delta(a)\propto a^{-3/(q-1)}$ and reducing $v^{2q-2}/v^{q-2}=v^q$, the curvature along the tracking trajectory becomes
\begin{equation}
m_{\rm eff}^2(a) \sim \lambda v^{q}\, a^{-\,\frac{3(q-2)}{q-1}}.
\label{eq:poly_mass_scaling}
\end{equation}
For $q=2$, the curvature is constant, and the minimum is non-degenerate. However, for $q>2$ the vacuum curvature vanishes at tree level, and the restoring force is generated dynamically along the trajectory. The mass scale, therefore, decreases toward late times, making the hierarchy condition $m_{\rm eff}^2 \gg H^2$ most restrictive near the transition epoch.

The validity of the controlled perturbative regime requires both the adiabatic hierarchy and a sufficiently weak interaction strength. For a large microscopic scale $\lambda v^{2q-2}$ or a small coupling, these conditions hold throughout the epoch of our interest. Within this regime, the polynomial class provides a natural realization of the general attractor structure, in which the late-time exponent is fixed entirely by the local flatness of the symmetry-breaking minimum. In this sense, all analytic potentials with the same local order $p$ belong to the same asymptotic dynamical class within the adiabatic tracking regime, irrespective of their global form. Its domain of validity and relation to wider symmetry-breaking potential are discussed in the following section.

\subsection{Universality and domain of validity}
\label{subsec:universality}

The asymptotic universality classification and the associated relation $n=3/p$ are valid under the following assumptions:
\begin{enumerate}
    \item The potential admits an isolated and analytic minimum at $\phi=v$, so that the local expansion of Eq.~\eqref{eq:local_expansion} is valid and the local restoring order $p$ is well defined. Non-analytic or cuspy potentials invalidate this expansion and fall outside the scope of the present framework.
    
    \item The scalar field evolves in the adiabatic tracking regime $m_{\rm eff}\gg H$, where the field rapidly relaxes to the instantaneous minimum and the algebraic balance condition governs the dynamics. Violation of this hierarchy introduces kinetic corrections that modify the attractor behavior and the resulting activation profile.

    \item The system remains effectively single-field, with no additional degrees of freedom introducing independent dynamical time scales. Multi-field dynamics can generate coupled attractors and departures from the single-parameter universality structure derived here.
\end{enumerate}
\medskip

We also note that the classification is defined at the level of the local restoring structure of the scalar potential. The radiative stability of a given class is a separate question and depends on the symmetries protecting the corresponding minimum. In particular, generic one-loop corrections tend to lift higher-order degeneracies and drive the effective potential toward the $p=1$ class unless protected by additional symmetry. The $p=1$ class, therefore, occupies a distinguished role as the most radiatively robust realization of the late-time attractor, independent of the observational constraints discussed in the subsequent analysis.

In this context, varying the activation index $n$ probes different local restoring behaviors near the minimum of the scalar potential. Representative scalar potentials belonging to these universality classes are shown in Table~\ref{tab:universality_mapping}. The class $(p=1)$ (or $(n=3)$) includes potentials with non-degenerate analytic minima, such as quartic Mexican-hat~\citep{curtin2021twin,alvarez2019einstein} and Coleman--Weinberg~\citep{coleman1973radiative,pradosh2025loop} potential. Although the axion potential~\citep{turner1990windows,dine1981simple} is periodic and its minima occur at $\phi=2\pi n f$, corresponding to a discrete shift symmetry rather than spontaneous breaking of a continuous symmetry with a nonzero VEV, its local expansion about each minimum is quadratic with $V''\neq0$, placing it in the same $p=1$ universality class. These potentials exhibit the canonical attractor exponent $(\gamma_p=-3)$ and share identical late-time scaling behavior. Higher-order potentials with $(q>2)$ form distinct classes with $(p=q-1)$ and $(n=3/(q-1))$, corresponding to progressively flatter minima and slower relaxation toward the attractor. The radiative stability of these higher-order classes, however, depends on the symmetry structure of the underlying theory (see Section~\ref{sec:discussion}).

\begin{table*}[t]
\centering
\small
\caption{\small Representative scalar potentials and their associated universality classes in the adiabatic regime. Potentials with identical local restoring order $p$ exhibit identical asymptotic attractor behavior within the adiabatic regime, irrespective of their global potential structure.}
\label{tab:universality_mapping}
\setlength{\tabcolsep}{6pt}
\renewcommand{\arraystretch}{2}
\begin{tabular}{l l c c l}
\toprule
Potential Form & $V(\phi)$ & $p$ & $\gamma_p$ &
\hspace{2cm} Physical Interpretation \\
\midrule

Quartic &
$\dfrac{\lambda}{4}(\phi^2 - v^2)^2$ &
$1$ & $-3$ &
\makecell[l]{Non-degenerate minimum; $V''(v) = 2\lambda v^2 \neq 0$ \\
with a linear restoring force.}
\\[6pt]

Coleman--Weinberg &
$\dfrac{\lambda}{4}\phi^4
\!\left(\ln\dfrac{\phi^2}{v^2} - \dfrac{1}{2}\right)$ &
$1$ & $-3$ &
\makecell[l]{Radiatively generated symmetry breaking; \\
locally quadratic minimum with finite $V''(v)$.}
\\[6pt]

Axion-like &
$\Lambda^4\!\left[1 - \cos(\phi/f)\right]$ &
$1$ & $-3$ &
\makecell[l]{Periodic potential; $V''(v) = \Lambda^4/f^2 \neq 0$ \\
near the minimum.}
\\[6pt]

Higher-order ($q > 2$) &
$\dfrac{\lambda}{2q}(\phi^2 - v^2)^q$ &
$q-1$ & $-\dfrac{3}{q-1}$ &
\makecell[l]{Degenerate tree-level minimum; $V''(v) = 0$ and \\ 
the restoring force arises from higher-order terms.}
\\[6pt]

\bottomrule
\end{tabular}
\end{table*}

\section{Late-Time Effective Coupling and Logistic Activation}
\label{sec:logistic_activation}

The preceding section showed that the late-time evolution of the scalar field along the density-controlled trajectory is governed by the logistic normal form with index $n = 3/p$. We now adopt the logistic function as a late-time effective parametrization of the activation history over the observational window $0\leq z\leq 2$. This is not an exact solution of the full early-universe dynamics, but a controlled extension of the asymptotic normal form, and its validity is supported by the numerical verification of the adiabatic hierarchy in Appendix~\ref{app:adiabatic_asymptotics}.

Near the broken minimum, the scalar evolution can be written as \( y(a) \equiv {v}/{\phi_{\rm ad}(a)} = 1/x(a),\) which evolves from $y \to 0$ at early times to $y \to 1$ at late times. As shown in Section~\ref{subsec:logistic_normal_form}, the local dynamics are governed by the logistic normal form \({dy}/{d\ln a} = n\,y(1-y),\) whose logistic solution, obtained by separation of variables with the boundary condition $y(a_c) = 1/2$, is
\begin{equation}
y(a) = \frac{(a/a_c)^n}{1 + (a/a_c)^n},
\label{eq:logistic_solution}
\end{equation}
where $a_c$ denotes the characteristic transition scale defined by $y(a_c)=1/2$. Physically, $a_c$ marks the epoch at which the DM-sourced tilt of effective potential, $(\beta_c/M_{\rm Pl})\rho_{\rm DM}\phi$, becomes comparable in magnitude to the intrinsic restoring force of $V(\phi)$ at its minimum. At this scale, the field transitions from a displacement-dominated early-time trajectory to a broken minimum attractor. Note that $a_c$ denotes the dynamical activation epoch, and not the onset of SSB itself - the latter is a static feature of the potential present from the outset.

In the scalar-tensor framework, the effective coupling is evaluated along the tracking solution $\phi_{\rm ad}(a)$. Since the interaction strength is controlled by the displacement from the symmetry-breaking minimum, the effective coupling follows the same activation profile as the order parameter. We therefore relate the order-parameter dynamics to the coupling strength, and expand $\beta(\phi)$ about the broken minimum $\phi=v$:
\begin{equation}
\beta(\phi_{\rm ad}) = \beta(v) + \beta'(v)\,\delta + \mathcal{O}(\delta^2),
\end{equation}
where $\delta \equiv \phi_{\rm ad} - v$. Because $y = v/\phi_{\rm ad} = 1/(1+\delta/v)$, we have $\delta/v = (1-y)/y$ at leading order near $y\to 1$. The correction $\beta'(v)\delta$ is therefore of order $\mathcal{O}(1-y)$ relative to $\beta_0\equiv\beta(v)$. Retaining only the zeroth-order term and using $y(a)\to 1$ at late times, the minimal leading-order approximation is
\begin{equation}
\beta_{\rm eff}(a) = \beta_0\, y(a),
\label{eq:beta_mapping}
\end{equation}
with $\beta_0 \equiv \beta(v)$ setting the asymptotic coupling strength at late times. Substituting Eq.~\eqref{eq:logistic_solution} into Eq.~\eqref{eq:beta_mapping} yields the logistic coupling profile
\begin{equation}
\beta_{\rm eff}(a) = \beta_0 \frac{a^n}{a^n + a_c^n}.
\label{eq:beta_logistic}
\end{equation}

At early times $a\to0$, the logistic profile gives $\beta_{\rm eff}\to0$. The logistic form is a late-time normal form derived asymptotically near the symmetry-broken minimum and employed here as a controlled extension over the observationally relevant regime $z\lesssim 2$, rather than as an exact solution of the early-universe dynamics. The early-time behavior of the coupling, and its consequences for BBN and recombination constraints, is addressed quantitatively in Appendix~\ref{app:adiabatic_asymptotics}. In the heavy-field adiabatic regime, the effective coupling saturates to the constant value $\beta_0$ for $a\gg a_c$, recovering the functional form of a constant-coupling model familiar from coupled quintessence  \citep{amendola2008quintessence,amendola2014multifield,potting2022coupled,pourtsidou2013models,bean2008_prd}. The underlying scalar dynamics nevertheless remain in adiabatic tracking rather than slow roll, so this correspondence applies at the level of the coupling history and not the full field evolution.

\section{Methodology and Numerical Implementation}
\label{sec:methodology}

In this section, we present the numerical implementation, the hierarchy of models considered, outline the perturbative consistency conditions, and describe the implementation of the coupling in the Boltzmann solver \texttt{CLASS}~\citep{blas2011cosmic}. We first define the hierarchy of cosmological models and their parameterizations in Section~\ref{subsec:model_hierarchy}. The full derivation of the perturbation equations is presented in the companion paper~\citep{mv2025spontaneous}. For completeness, the perturbation equations implemented in the modified \texttt{CLASS} solver are summarized in Section~\ref{subsec:perturbation_summary}, while their self-consistency is established through the perturbative filtering procedure of Section~\ref{subsec:perturbative_filtering}, ensuring that all subsequent numerical results remain within the domain of validity of the underlying effective description.

\subsection{Model hierarchy and parameterization}
\label{subsec:model_hierarchy}

We consider two nested models compared against the standard $\Lambda$CDM model. In the first case, $n$ is treated as a free parameter, and in the second, it is fixed to the canonical value $n=3$. The prior choices and likelihood are defined as follows. 

{\bf (i) Six-parameter IDE model:} The full six-parameter (6p) model is defined by
\begin{equation*}
\Theta_{\rm 6p} = (\Omega_m,\; h,\; \sigma_8,\; \beta_0,\; a_c,\; n).
\end{equation*}
We adopt the following prior ranges:
\[
\Omega_m \in (0.2,0.4),\quad
h \in (0.55,0.85),\quad
\sigma_8 \in (0.6,1.0),
\]
\[
\beta_0 \in (-0.3,0.3),\quad
a_c \in (0.1,1.0),\quad
n \in (0.5,5.0).
\]
The prior $n \in (0.5, 5.0)$ corresponds to restoring orders $p = 3/n \in [0.6, 6]$.

We note that the posterior accumulation at the upper boundary $n = 5$ observed is therefore partly prior-driven, which is discussed in detail in Section~\ref{subsec:flexible_6p_results}. Extending the upper bound beyond $n=5$ produces no qualitative change in the inferred coupling constraints, since the posterior boundary accumulation is weakly sensitive to steep activation histories and not due to any physical preference for arbitrarily large $n$ (see Appendix~\ref{sec:diagnostics}). Although the physically motivated integer classes are $p = 1, 2, 3, \dots$ (giving the discrete values $n = 3, 3/2, 1, \dots$), we deliberately choose a wider range than these benchmarks. This allows the data to constrain $n$ freely without being forced onto a discrete grid. Values $n > 3$ correspond to $p < 1$, which do not appear in the polynomial family of Eq.~\eqref{eq:poly_potential}. We retain them nonetheless to avoid truncating the prior at the boundary of the physically motivated region, which could bias the posterior inference for $n$.

{\bf (ii) Fixed-index IDE model:} As a rigid benchmark for assessing the role of activation-history flexibility, we additionally consider a realization in which the index is fixed to its asymptotic value $n=3$. This model is not intended as the preferred physical realization of the framework, but serves as a diagnostic reference for quantifying the observational consequences of imposing the asymptotic attractor form throughout the observational window. The sampled parameter vector becomes
\begin{equation*}
\Theta_{\rm 5p} = (\Omega_m,\; h,\; \sigma_8,\; \beta_0,\; a_c).
\end{equation*}

{\bf (iii) $\Lambda$CDM limit:}
In the limit $\beta_0 \rightarrow 0$, the effective coupling vanishes and the model reduces to standard $\Lambda$CDM,
\begin{equation*}
\Theta_{\Lambda{\rm CDM}}= (\Omega_m,\; h,\; \sigma_8).
\end{equation*}

The baryon density $\omega_b = 0.02237$, scalar spectral index $n_s=0.965$, and optical depth $\tau=0.054$ are fixed to their \emph{Planck} 2018 best-fit values \citep{aghanim2020planck}. Since the datasets employed here offer little leverage on the primary CMB acoustic physics, we assume that the uncertainties of these fixed parameters are sufficiently small that they do not propagate significantly into the IDE constraints. The internal mapping $\sigma_8 \to A_s$ within \texttt{CLASS} is performed at fixed $(n_s, \tau, \Omega_b)$, so that $\sigma_8$ serves as the effective amplitude parameter throughout.

The modified \texttt{CLASS} solver \citep{class_ide} evolves the full system to compute linear observables, including the matter power spectrum \(P(k, z)\), the growth rate \(f(z)\), and the CMB lensing potential \(C_L^{\phi\phi}\). The analysis is restricted to linear scales $k \lesssim 0.1\,h\,\mathrm{Mpc}^{-1}$ at $z=0$; this cut becomes increasingly conservative at higher redshifts, where the nonlinear scale $k_{\mathrm{NL}}(z)$ shifts to smaller values. For the RSD data \citep{alam2021completed}, the linear treatment remains adequate because conservative \(k\)-cuts are applied (typically \(k_{\mathrm{max}} \sim 0.1 \, h\,\mathrm{Mpc}^{-1}\) at the effective redshift of each measurement), restricting the analysis to scales where linear perturbation theory is reliable. Nonlinear corrections to the matter power spectrum are not expected to significantly affect our conclusions, as the interaction primarily modifies the growth history on linear scales.

\subsection{Summary of the perturbation equations}
\label{subsec:perturbation_summary}

For self-containedness, we summarize the linear perturbation equations implemented in the modified solver~\cite{blas2011cosmic}, in the conformal Newtonian gauge,
\begin{equation}
ds^2 = a^2(\tau)\left[-(1+2\Psi)d\tau^2 + (1-2\Phi)\delta_{ij}dx^idx^j\right].
\end{equation}

The scalar field perturbation $\delta\phi$ evolves according to the standard perturbed Klein--Gordon equation for a canonical field,
\begin{equation}
\delta\phi'' + 2\mathcal{H}\delta\phi' + \left(k^2 + a^2 V''(\phi_{\rm ad})\right)\delta\phi = 0,
\label{eq:deltaphi_pert}
\end{equation}
where $\mathcal H\equiv a'/a$ and primes denote conformal-time derivatives. In the present implementation, this equation carries no explicit source term proportional to the coupled CDM density perturbation.

The coupled CDM sector receives a direct correction to its continuity and Euler equations, proportional to the background exchange rate
\begin{equation}
\Gamma(a) \equiv \frac{\beta_{\rm eff}(a)}{M_{\rm Pl}}\,\dot{\bar\phi}(a),
\label{eq:gamma_rate}
\end{equation}
the same rate entering the background continuity equation~\eqref{eq:dm_continuity}. In cosmic time,
\begin{align}
\dot\delta_{\rm cdm} + \frac{\theta_{\rm cdm}}{a} - 3\dot\Psi &= a\,\Gamma(a)\,\delta_{\rm cdm},
\label{eq:delta_cdm_pert}\\
\dot\theta_{\rm cdm} + H\theta_{\rm cdm} - \frac{k^2}{a}\Phi &= a\,\Gamma(a)\,\theta_{\rm cdm},
\label{eq:theta_cdm_pert}
\end{align}
where $\Phi$ sources the Poisson equation $k^2\Phi = -4\pi G a^2\bar\rho_m\delta_m$ with the standard $G$; baryons are minimally coupled and obey the corresponding equations with the right-hand side set to zero.

Both correction terms are proportional to the perturbation variable itself, with no additional $k$-dependence beyond that already present in the uncoupled equations; the coupling therefore acts as a scale-independent, density-proportional source, rather than through a modification of the gravitational sourcing term. Equations~\eqref{eq:delta_cdm_pert}--\eqref{eq:theta_cdm_pert} are obtained by converting the solver's native conformal-time equations to cosmic time; a dedicated derivation of the full perturbed system, including a self-consistently sourced $\delta\phi$, is left for future work.
\subsection{Controlled perturbative regime}
\label{subsec:perturbative_filtering}

We restrict to a regime in which the background evolution remains close to $\Lambda$CDM. In this regime, the adiabatic tracking solution is valid, and the dynamics are reduced to a one-dimensional autonomous system from which the logistic normal form emerges.

From ~\eqref{eq:dm_continuity}, the fractional deviation of the DM dilution from the standard $a^{-3}$ law per Hubble time is $\dot{\rho}_{\rm DM}/(H\rho_{\rm DM}) + 3 = -(\beta\dot{\phi})/(M_{\rm Pl}H)$, which motivates defining
\begin{equation}
\epsilon(a) \equiv \frac{\beta(\phi)\dot{\phi}}{M_{\rm Pl} H},
\label{eq:epsilon_def_method}
\end{equation}
In the perturbative regime, one requires \(|\epsilon(a)| \ll 1,\) so that the cumulative deformation of the background remains small over the observationally relevant epoch. This condition is satisfied when the effective mass hierarchy $m_{\rm eff}^2 \gg H^2$ holds and the coupling $\beta(\phi)$ remains sufficiently small, i.e., $\epsilon(a)\sim 3\beta^2 H^2/m_{\rm eff}^2$. Consistency of the posterior solutions with the adiabatic hierarchy is verified numerically in Appendix~\ref{app:verification_adiabatic}.

We further impose a conservative post-chain filtering as a conservative bound with sub-percent deviations in both $\rho_{\rm DM}(a)$ and $H(a)$ over $0 \leq z \leq 2$,
\begin{equation}
\max_a |\epsilon(a)| < \epsilon_\star,
\qquad
\epsilon_\star = 10^{-2}.
\label{eq:epsilon_bound_method}
\end{equation}
The qualitative conclusions presented in Section~\ref{sec:observ_constraints} are not expected to change significantly for $\epsilon_\star$ in the range $[5\times 10^{-3}, 2\times 10^{-2}]$, since the posterior geometry is driven primarily by the likelihood rather than the precise value of the filtering threshold. Within this regime, the background evolution is
\begin{align*}
&\rho_{\rm DM}(a)  = \rho_{{\rm DM},0} a^{-3}\left[1+\mathcal{O}(\epsilon_\star)\right], \\
H^2(a) &= H_0^2\left[\Omega_m a^{-3} + (1-\Omega_m)\right] + \mathcal{O}(\epsilon_\star),
\end{align*}
where \(\Omega_{\rm m}= \Omega_{\rm DM}+\Omega_{\rm b}\). 

\medskip
We note that Eq.~\eqref{eq:epsilon_bound_method} is not imposed as a prior in the parameter estimation. Instead, it is imposed as a consistency condition for the sampled cosmologies to remain within the regime where the perturbative treatment and linear background approximation are valid. For each posterior sample, $\epsilon(a)$ is evaluated along the tracking solution $\phi_{\rm ad}(a)$ over the redshift interval $0 \le z \le 2$. The maximum value of $|\epsilon(a)|$ is then used to determine whether the sample lies within the regime. 

The fraction of posterior samples satisfying this condition for each dataset combination is reported in Table~\ref{tab:epsilon_acceptance}. The independence of the filtering criterion from the data likelihood is verified by comparing the marginalized posteriors of $(\Omega_m, h, \sigma_8)$ before and after filtering for the $\Lambda$CDM model, where the acceptance fraction is $1.0$ by construction. For the IDE models, the filtered posteriors are interpreted as the posterior \emph{conditional on} the adiabatic tracking regime, rather than as a modification of the unconditional posterior.  The perturbative filter, therefore, acts as a perturbative consistency condition rather than as a driver of the inferred background cosmology. As demonstrated in Appendix~\ref{sec:diagnostics}, varying the filtering threshold changes the acceptance fraction but leaves the marginalized constraints on $(\Omega_m,h,\sigma_8)$ statistically unchanged.

\begin{table}[ht]
\centering
\caption{\small Fraction of posterior samples satisfying $\max_a |\epsilon(a)| \leq 10^{-2}$ for each dataset combination, using the corrected curvature $m_{\rm eff}^2$ of Eq.~\eqref{eq:meff_dimensionless_app} and, for the fixed-index model, excluding one non-converged chain replicate identified in Appendix~\ref{sec:mcmc_diagnostics} (Planck+RSD+SN only). The acceptance fraction for $\Lambda$CDM is unity by construction since no perturbative filter applies to the uncoupled limit.}
\label{tab:epsilon_acceptance}
\setlength{\tabcolsep}{10pt}
\renewcommand{\arraystretch}{1.4}
\begin{tabular}{llc}
\toprule
Dataset & Model & Acceptance Fraction \\
\midrule
\multicolumn{3}{c}{\textit{RSD only}} \\
\midrule
& $\Lambda$CDM & 1.000 \\
& IDE~(6p)     & 0.444 \\
& IDE~($n=3$)  & 0.332 \\
\midrule
\multicolumn{3}{c}{\textit{Planck+RSD}} \\
\midrule
& $\Lambda$CDM & 1.000 \\
& IDE~(6p)     & 0.444 \\
& IDE~($n=3$)  & 0.336 \\
\midrule
\multicolumn{3}{c}{\textit{Planck+RSD+SN}} \\
\midrule
& $\Lambda$CDM & 1.000 \\
& IDE~(6p)     & 0.437 \\
& IDE~($n=3$)  & 0.233 \\
\bottomrule
\end{tabular}
\end{table}

\subsection{Data sets and statistical analysis}

We confront the IDE scenario with a combination of late-time geometric and structure growth observables. All theoretical predictions used to compute the likelihood are obtained from the modified \texttt{CLASS} solver~\citep{class_ide}.

\noindent
{\it Cosmic Microwave Background (CMB) lensing:}
We use the \emph{Planck} 2018 CMB lensing reconstruction \citep{aghanim2020planck}, which measures the lensing potential power spectrum $C_L^{\phi\phi}$ over multipoles $8 \le L \le 400$. The likelihood is evaluated as
\begin{equation*}
\chi^2_{\mathrm{Planck}} =
(\mathbf{C}^{\mathrm{th}} - \mathbf{C}^{\mathrm{obs}})^{\mathrm{T}}
\mathbf{Cov}^{-1}
(\mathbf{C}^{\mathrm{th}} - \mathbf{C}^{\mathrm{obs}}).
\end{equation*}

\noindent
{\it Redshift-space distortions (RSD):}
Growth constraints are obtained from measurements of $f\sigma_8(z)$ \citep{alam2021completed, avila2022inferring}. The theoretical prediction is given by
\begin{equation}
f\sigma_8(z) = f(z)\,\sigma_8\,\frac{D(z)}{D(0)},
\end{equation}
with likelihood
\begin{equation*}
\chi^2_{\mathrm{RSD}} =
\sum_i
\frac{\left[f\sigma_8^{\mathrm{th}}(z_i)
- f\sigma_8^{\mathrm{obs}}(z_i)\right]^2}{\sigma_i^2}.
\end{equation*}
Here $\sigma_8 \equiv \sigma_8(z=0)$ is the present-day amplitude parameter sampled in the MCMC, and $D(z)/D(0)$ is the linear growth factor ratio computed by the modified \texttt{CLASS} solver for each model realization.

\noindent
{\it Type Ia supernovae:}
Pantheon+SH0ES sample includes the SH0ES Cepheid distance calibration~\citep{brout2022pantheon+}, which constrains the SN~Ia absolute magnitude $M_B$ (and therefore $H_0$) consistently with the local determination. In our analysis, $h$ is treated as a free MCMC parameter with a uniform prior $h\in(0.55,0.85)$, rather than imposing the SH0ES result as an external Gaussian prior. The inferred $h$ posterior therefore corresponds to the full Pantheon+SH0ES likelihood marginalized over the IDE parameters, which broadens the constraint relative to the Cepheid-only uncertainty. The likelihood is evaluated using 
\begin{equation*}
\chi^2_{\mathrm{SN}} =
(\boldsymbol{\mu}^{\mathrm{th}}
- \boldsymbol{\mu}^{\mathrm{obs}})^{\mathrm{T}}
\mathbf{Cov}^{-1}
(\boldsymbol{\mu}^{\mathrm{th}}
- \boldsymbol{\mu}^{\mathrm{obs}}).
\end{equation*}

\medskip

Parameter inference is performed using the affine-invariant ensemble sampler \texttt{emcee} \citep{foreman2013emcee}. The posterior distribution is
\begin{equation*}
\mathcal{P}(\Theta | \mathrm{data}) \propto
\exp\!\left[-\frac{1}{2}\chi^2_{\mathrm{tot}}(\Theta)\right]\Pi(\Theta),
\end{equation*}
where $\Pi(\Theta)$ denotes the uniform priors mentioned in Section~\ref{subsec:model_hierarchy}. This choice of uniform priors applies to the directly sampled parameters; derived quantities such as $S_8 = \sigma_8 \sqrt{\Omega_m / 0.3}$ will therefore have non-uniform induced priors. The KL divergences reported in Tables~\ref{tab:ide_6p_n3_comparison} and~\ref{tab:ide_combined_constraints} are computed relative to these uniform priors in the sampled parameters.

\section{Observational Constraints and Analysis}
\label{sec:observ_constraints}

This section focuses on two related results. First, the functional form of the coupling history is constrained by current data: fixing the activation index to $n=3$ degrades the SN fit by $\Delta\chi^2_{\rm SN}\approx 92$ relative to the flexible model and drives the background parameters toward physically unrealistic values, largely independently of the coupling amplitude. Second, in the heavy-scalar adiabatic regime, IDE-induced modifications to $f(z)$ and $D(z)$ enter $f\sigma_8(z)$ with opposite signs, suppressing the net deviation to $\Delta f\sigma_8/f\sigma_8\lesssim 0.3\%$. We examine these effects using three dataset combinations: RSD only, Planck+RSD, and Planck+RSD+SN.

\subsection{Constraints on the Coupling Activation History}
\label{subsec:activation_shape_tension}

We first examine the sensitivity of current data to the functional form of the coupling history. The flexible six-parameter (6p) IDE model, in which the coupling amplitude $\beta_0$, activation epoch $a_c$, and activation index $n$ are free, is compared with a fixed-index five-parameter realization where $n=3$. The comparison is based on the resulting fit degradation and changes in the posterior structure. Fixing $n=3$ imposes the asymptotic attractor prediction of the $p=1$ class, corresponding to the limit $\phi_{\rm ad}\to v$, rather than the finite-redshift effective index $n_{\rm fit}\approx 1.71$--$2.30$ (see Appendix~\ref{app:logistic_accuracy}). Relative to the 6p model, this removes one continuous shape degree of freedom from the activation history while leaving the priors, datasets, and perturbative consistency conditions unchanged. The comparison, therefore, isolates the observational impact of imposing the asymptotic profile throughout the observational window rather than allowing finite-redshift corrections.

The 6p model shows no statistically significant preference for a nonzero coupling, with $|\beta_0|\lesssim 0.26$ at $95\%$ credibility and posterior support centered within $1\sigma$ of $\beta_0=0$ for all dataset combinations. The minimum $\chi^2$ is identical to that of $\Lambda$CDM, consistent with the maximum likelihood solution occurring at $\beta_0=0$, where the IDE model reduces to $\Lambda$CDM. The resulting $\Delta\mathrm{BIC}=\Delta k\,\ln N$ values are $+7.19$ (RSD only, $N=11$), $+8.99$ (Planck+RSD, $N=20$), and $+22.36$ (Planck+RSD+SN, $N=1721$), with $\Delta k=3$ and $\Delta\chi^2_{\rm min}=0$ throughout, since the maximum-likelihood solution occurs at $\beta_0=0$ where the IDE model reduces to $\Lambda$CDM. These, therefore, arise entirely from the parameter-count penalty as the effective dataset size increases, rather than from any deterioration in fit quality. Posterior constraints (RSD-only and Planck+RSD) for the 6p model are presented in Section~\ref{subsec:flexible_6p_results}.

Table~\ref{tab:ide_6p_n3_comparison} summarizes the marginalized constraints for both realizations. In the 6p case, the background parameters are consistent with standard late-time constraints and exhibit posterior widths comparable to those obtained in $\Lambda$CDM analyses. The corresponding KL divergences remain small for $h$ and $\sigma_8$ ($\mathrm{KL}\lesssim0.01$~nats). By contrast, the marginalized posterior for $\Omega_m$ peaks near $0.361$ (with $\mathrm{KL}=1.006$~nats), which is above the standard $\Lambda$CDM Planck+SN value of $\Omega_m\approx0.31$. Although the maximum likelihood occurs near $\beta_0\approx0$, marginalization over $(a_c,n)$ weakly correlates with the SN distance--redshift relation and shifts the inferred $\Omega_m$ relative to the rigid $\Lambda$CDM limit. However, the fixed-index realization drives the same parameters onto the lower edges of the prior volume, with posterior medians near $\Omega_m\approx0.20$, $h\approx0.55$, and $\sigma_8\approx0.60$, significantly below current constraints. This trend is explicit in the corresponding KL divergences, which increase to $5.822$, $4.336$, and $4.381$~nats for $\Omega_m$, $h$, and $\sigma_8$, respectively due the near-complete boundary collapse quantified in Table~\ref{tab:collapse_diagnostics}. The posterior width of $\beta_0$ also decreases (for $n=3$), from approximately $\pm0.113$ to $\pm0.074$ at $68\%$ credibility. This is due to the restricted parameter volume and the associated model--data tension induced by the fixed activation profile, rather than a substantial increase in direct sensitivity to the interaction amplitude.

\begin{table*}
\centering
\caption{\small Marginalized $68\%$ CI and marginal KL divergences (in nats) for the flexible IDE~(6p) and fixed-index IDE~($n=3$) models under the Planck+RSD+SN dataset combination. The collapse of $\Omega_m$, $h$, and $\sigma_8$ in the fixed-index model toward the lower prior boundaries is evident in the combined data. Large KL values for the background sector in the $n=3$ column are due to the posterior accumulation against the prior boundary rather than genuine information gain from the likelihood; see Table~\ref{tab:collapse_diagnostics} for the boundary diagnostics. Full posterior tables for the RSD-only and Planck+RSD combinations are given in Table~\ref{tab:ide_combined_constraints}.}
\label{tab:ide_6p_n3_comparison}
\setlength{\tabcolsep}{7pt}
\renewcommand{\arraystretch}{1.5}
\begin{tabular}{lcccc}
\toprule
& \multicolumn{2}{c}{IDE~(6p)}
& \multicolumn{2}{c}{IDE~($n=3$)} \\
\cmidrule(lr){2-3}\cmidrule(lr){4-5}
Parameter
  & $68\%$ CI & KL
  & $68\%$ CI & KL \\
\midrule
$\Omega_m$
  & $0.361^{+0.018}_{-0.019}$ & $1.006$
  & $0.2000^{+0.0002}_{-0.0000}$ & $5.822$ \\[3pt]
$h$
  & $0.710^{+0.094}_{-0.108}$ & $0.008$
  & $0.5500^{+0.0017}_{-0.0000}$ & $4.336$ \\[3pt]
$\sigma_8$
  & $0.807^{+0.132}_{-0.138}$ & $0.006$
  & $0.6000^{+0.0022}_{-0.0000}$ & $4.381$ \\[3pt]
$\beta_0$
  & $0.009^{+0.116}_{-0.109}$ & $0.228$
  & $-0.016^{+0.077}_{-0.071}$ & $0.731$ \\[3pt]
$a_c$
  & $0.667^{+0.236}_{-0.285}$ & $0.084$
  & $0.700^{+0.203}_{-0.299}$ & $0.106$ \\[3pt]
$n$
  & $1.91^{+1.91}_{-1.09}$ & $0.262$
  & $3$ (fixed) & --- \\
\midrule
$\chi^2_{\rm SN}$
  & \multicolumn{2}{c}{$1809.0$}
  & \multicolumn{2}{c}{$1901$} \\[2pt]
$\Delta\chi^2_{\rm SN}$
  & \multicolumn{4}{c}{$\approx 92$\quad ($n=3$ excess)} \\
\bottomrule
\end{tabular}
\end{table*}

\begin{figure}
\centering
\includegraphics[width=0.6\linewidth]{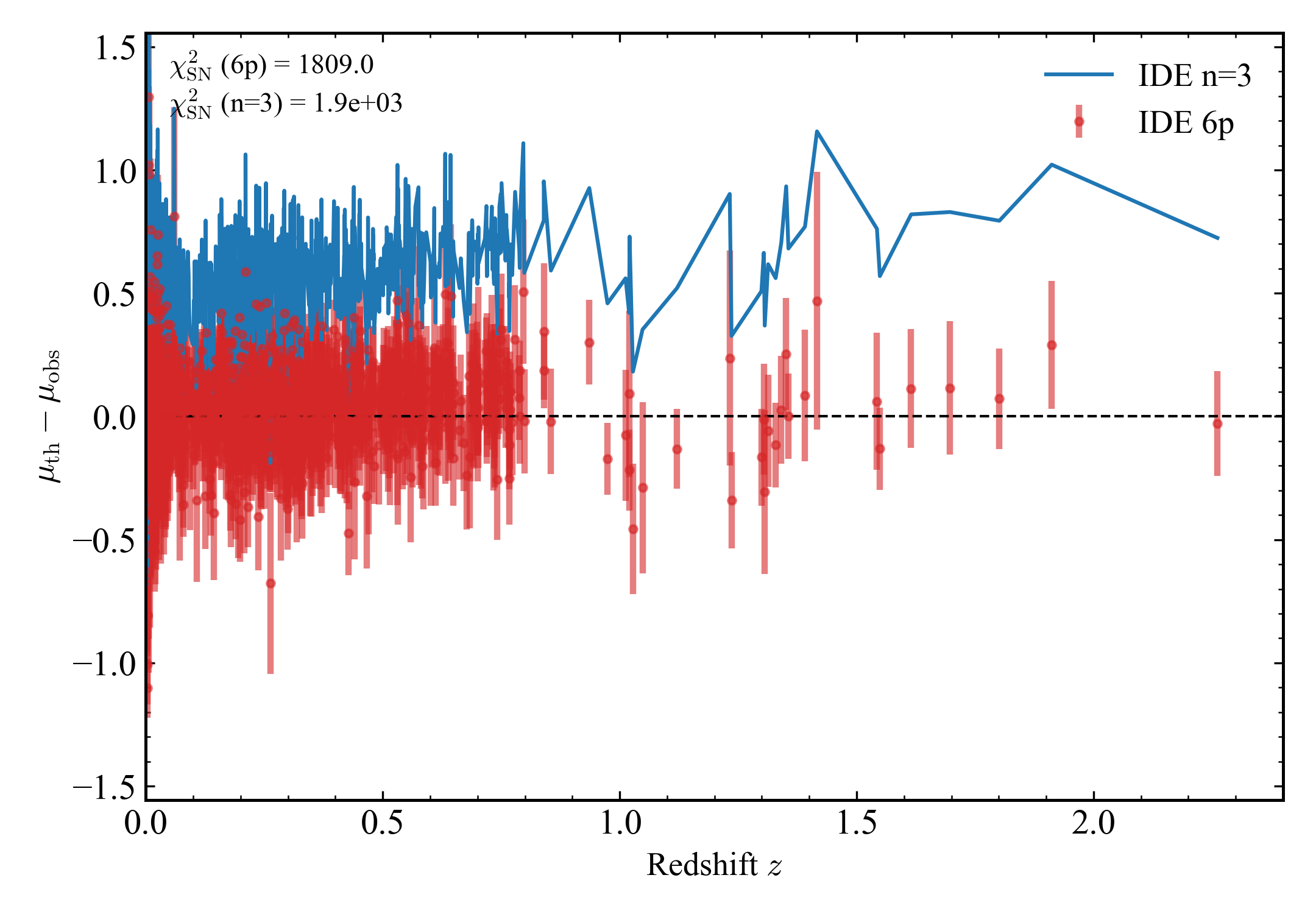}
\caption{\small Supernova Hubble-diagram residuals $\mu_{\rm th}-\mu_{\rm obs}$ for the Planck+RSD+SN best-fit cosmologies. The 6p model yields $\chi^2_{\rm SN}=1809$ for $N_{\rm SN}=1701$ ($\chi^2/\mathrm{dof}\simeq 1.064$), statistically comparable to $\Lambda$CDM. The fixed-index model yields $\chi^2_{\rm SN}\approx 1901$ ($\chi^2/\mathrm{dof}\simeq 1.118$), a degradation of $\Delta\chi^2_{\rm SN}\approx 92$. The absence of any systematic runaway trend in the fixed-index residuals confirms a global geometric incompatibility rather than a localized epoch failure.}
\label{fig:sn_residuals_6p_vs_n3}
\end{figure}

The tension arises from imposing the rigid asymptotic $n=3$ profile across the entire observational window, where finite-redshift corrections remain non-negligible. As shown in Fig.~\ref{fig:sn_residuals_6p_vs_n3}, the fixed-index model yields $\chi^2_{\rm SN}\approx 1901$ at the best-fit point, compared with $\chi^2_{\rm SN}=1809$ for the flexible 6p model, corresponding to a degradation of $\Delta\chi^2_{\rm SN}\approx 92$. The residuals exhibit no localized redshift-dependent runaway behavior, indicating that the discrepancy is distributed across the full SN sample rather than concentrated within a narrow redshift interval. The fixed-index profile is therefore unable to simultaneously reproduce the geometric constraints from Pantheon+SH0ES and the growth constraints from Planck and RSD within the perturbative regime considered here. By contrast, the flexible 6p model achieves $\chi^2_{\rm SN}/\mathrm{dof}\simeq 1.06$, comparable to $\Lambda$CDM, because the activation profile can adjust to finite-redshift departures from the asymptotic attractor form. For the RSD-only and Planck+RSD combinations, the absence of the SN geometric lever arm substantially weakens this tension.

The fixed-index model exhibits an almost complete collapse of the background-parameter posteriors onto their lower prior boundaries. Table~\ref{tab:collapse_diagnostics} quantifies this behavior for the Planck+RSD+SN combination: $100.00\%$ of the $\Omega_m$ posterior, $99.30\%$ of the $h$ posterior, and $99.29\%$ of the $\sigma_8$ posterior lie within $5\%$ of their respective lower prior boundaries, with posterior widths reduced by factors of $27$--$36$ relative to the 6p model. These values fall well below independent late-time constraints; for example, the KiDS-Legacy result $S_8 = 0.815^{+0.016}_{-0.021}$~\citep{wright2025kids} implies values of $\sigma_8$ far above the fixed-index posterior support. 

The minimum $\chi^2$ outside the SN sector remains essentially unchanged, indicating that the boundary accumulation does not correspond to an improved global fit. We have not tested whether this reflects a genuine likelihood preference for values of $(\Omega_m,h,\sigma_8)$ below the current prior boundaries or a feature specific to the adopted prior range; extending these ranges is left for future work. Within the perturbative regime considered here, the fixed-index realization is therefore unable to simultaneously accommodate the combined geometric and growth constraints without migrating almost entirely onto the boundary of parameter space.

\begin{table*}
\centering
\caption{\small Posterior geometry diagnostics for the IDE model with fixed $n=3$ under the Planck+RSD+SN dataset combination. Reported are the marginalized posterior median, standard deviation, distance from the lower prior boundary in units of posterior standard deviation, width ratio relative to the 6p model, and fraction of samples within $5\%$ of each prior boundary.}
\label{tab:collapse_diagnostics}
\setlength{\tabcolsep}{5pt}
\renewcommand{\arraystretch}{1.5}
\begin{tabular}{llccccc}
\toprule
Parameter & Model & Median & Std &
$(\mathrm{Median}-\mathrm{Lower})/\sigma$ &
Width Ratio & \% Near Lower/Upper \\
\midrule
\multirow{2}{*}{$\Omega_m$}
  & IDE~6p    & 0.361 & 0.018 & 8.97 & 1.00
  & $0.00\%$ / $4.86\%$ \\
  & IDE~$n=3$ & 0.200 & 0.0005 & $\approx0$ & 36.0
  & $100.00\%$ / $0.00\%$ \\
\midrule
\multirow{2}{*}{$h$}
  & IDE~6p    & 0.710 & 0.086 & 1.86 & 1.00
  & $4.49\%$ / $5.08\%$ \\
  & IDE~$n=3$ & 0.550 & 0.0032 & $\approx0$ & 26.8
  & $99.30\%$ / $0.00\%$ \\
\midrule
\multirow{2}{*}{$\sigma_8$}
  & IDE~6p    & 0.807 & 0.115 & 1.80 & 1.00
  & $4.70\%$ / $5.21\%$ \\
  & IDE~$n=3$ & 0.600 & 0.0038 & $\approx0$ & 30.3
  & $99.29\%$ / $0.00\%$ \\
\bottomrule
\end{tabular}
\end{table*}

Fixing $n$ modifies the structure of the interaction-sector posterior by removing the $(\beta_0,n)$ degeneracy direction. For the RSD-only dataset, the $68\%$ credible interval for $\beta_0$ decreases from $\pm0.113$ in the flexible 6p model to $\pm0.081$ in the fixed-index realization. Over the same range, the KL divergence for $\beta_0$ increases from $0.234$ to $0.656$~nats. This narrowing arises primarily from the reduced parameter volume associated with the fixed activation profile, rather than from an increase in direct sensitivity to the coupling amplitude. Once the Pantheon+SH0ES dataset is included, the reduced freedom in activation history is accompanied by boundary accumulation, as discussed above, reflecting the incompatibility between the rigid $n=3$ profile and the combined geometric and growth constraints.

\begin{figure*}
\centering
\includegraphics[width=0.8\linewidth]{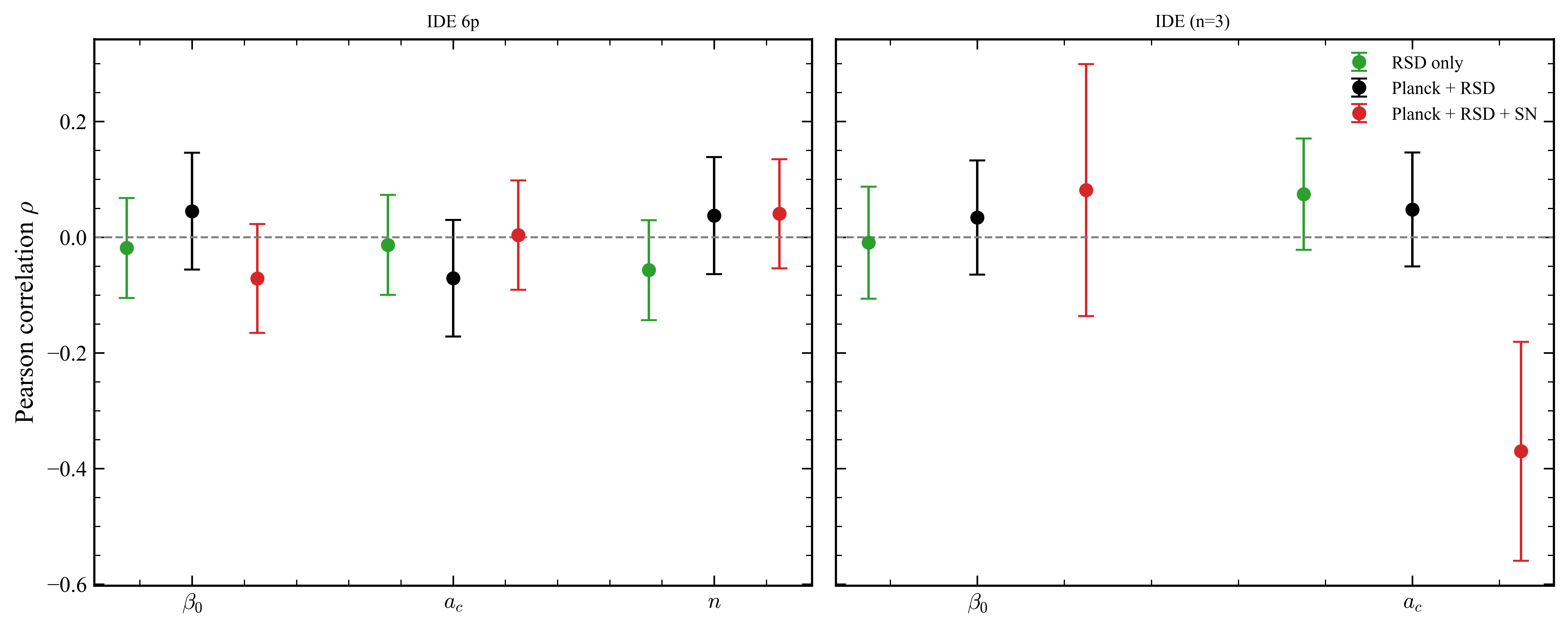}
\caption{\small Pearson correlation coefficient $\rho(S_8,\theta)$ for each interaction parameter $\theta$, with analytic $1\sigma$ uncertainties from the Fisher $z$-transform using the effective sample size. Green, black, and red correspond to RSD only, Planck+RSD, and Planck+RSD+SN, respectively. \textit{Left panel:} in the flexible (6p) model, all correlations are weak and consistent with zero, reflecting an extended degeneracy direction in parameter space. \textit{Right panel:} fixing $n=3$ generates a structured $S_8$--$a_c$ correlation under the Planck+RSD+SN combination. The loss of the shape degree of freedom forces stronger coupling between $a_c$ and $\beta_0$, which is a direct signature of imposing the rigid asymptotic $p=1$ activation profile.}
\label{fig:pearson11}
\end{figure*}

The structural consequence of this choice is further evident in Fig.~\ref{fig:pearson11}, which shows the Pearson correlation between $S_8$ and the interaction parameters for both models. In the flexible 6p model, the correlations remain weak for all dataset combinations, indicating a broad degeneracy structure with limited coupling between the clustering amplitude and the interaction sector. Under the full Planck+RSD+SN combination, however, the fixed-index realization develops a pronounced $S_8$--$a_c$ correlation: shifting the activation epoch to later times suppresses the integrated growth modification, requiring a compensating change in $\beta_0$ to maintain consistency with the RSD constraints and thereby altering the preferred value of $\sigma_8$. This correlation emerges only once the activation profile is restricted to the $p=1$ universality class and is not present in the flexible model.

\subsection{Parameter Constraints and Activation Classes}
\label{subsec:flexible_6p_results}

We now present the marginalized constraints for both model realizations across all three dataset combinations. The resulting posteriors quantify the extent to which current data constrain the interaction sector and assess whether the corresponding universality classes can be observationally distinguished.

Table~\ref{tab:ide_combined_constraints} reports the marginalized constraints for both the flexible and the fixed-index model across the RSD-only and Planck+RSD dataset combinations. In both realizations, the background parameters ($\Omega_m$, $h$) and the clustering amplitude $\sigma_8$ remain consistent with standard late-time cosmological constraints, with posterior widths comparable to those obtained in $\Lambda$CDM analyses. The KL divergences show a clear separation between the background and interaction sectors: for the RSD-only and Planck+RSD combinations, the background parameters remain largely prior-dominated ($\mathrm{KL}\lesssim0.01$~nats), whereas the interaction parameters exhibit modest information gain. In the 6p model, the posteriors of $a_c$ and $n$ retain support across most of the prior range, indicating limited sensitivity to the activation epoch and activation profile. Fixing $n=3$ removes the $(\beta_0,n)$ degeneracy direction and correspondingly tightens the $\beta_0$ posterior, reducing the $68\%$ interval for $\beta_0$ from $\pm0.113$ to $\pm0.081$ for the RSD-only combination and from $\pm0.108$ to $\pm0.083$ for Planck+RSD. The associated KL divergence for $\beta_0$ increases from $0.234$ to $0.656$~nats and from $0.266$ to $0.650$~nats, respectively, due to the reduced parameter volume induced by the fixed activation profile rather than a substantial increase in sensitivity to the coupling.

\begin{table*}
\centering
\caption{\small Marginalized $68\%$ credible intervals and marginal KL divergences (in nats) for the flexible IDE~(6p) and fixed-index IDE~($n=3$) models under the RSD-only and Planck+RSD dataset combinations. The KL divergences are measured relative to the uniform priors of Section~\ref{subsec:model_hierarchy}. The rise in KL for $\beta_0$ from the 6p to the $n=3$ column, from $0.234$ to $0.656$ (RSD-only) and $0.266$ to $0.650$ (Planck+RSD) quantifies the redistribution of posterior information from the shape degree of freedom $n$ onto the coupling amplitude when $n$ is fixed, rather than genuine additional data sensitivity.}
\label{tab:ide_combined_constraints}
\setlength{\tabcolsep}{7pt}
\renewcommand{\arraystretch}{1.45}
\begin{tabular}{lcccc}
\toprule
& \multicolumn{2}{c}{IDE~(6p)}
& \multicolumn{2}{c}{IDE~($n=3$)} \\
\cmidrule(lr){2-3}\cmidrule(lr){4-5}
Parameter
  & $68\%$ CI
  & KL
  & $68\%$ CI
  & KL \\
\midrule
\multicolumn{5}{c}{\textbf{RSD only}} \\
\midrule
$\Omega_m$
  & $0.297^{+0.069}_{-0.067}$ & $0.004$
  & $0.309^{+0.064}_{-0.075}$ & $0.008$ \\[3pt]
$h$
  & $0.705^{+0.102}_{-0.106}$ & $0.004$
  & $0.715^{+0.091}_{-0.107}$ & $0.014$ \\[3pt]
$\sigma_8$
  & $0.811^{+0.129}_{-0.138}$ & $0.009$
  & $0.792^{+0.135}_{-0.131}$ & $0.007$ \\[3pt]
$\beta_0$
  & $-0.001^{+0.113}_{-0.112}$ & $0.234$
  & $0.003^{+0.080}_{-0.082}$  & $0.656$ \\[3pt]
$a_c$
  & $0.667^{+0.235}_{-0.294}$ & $0.073$
  & $0.720^{+0.205}_{-0.309}$ & $0.130$ \\[3pt]
$n$
  & $1.75^{+1.96}_{-0.93}$ & $0.308$
  & $3$ (fixed) & --- \\
\midrule
\multicolumn{5}{c}{\textbf{Planck+RSD}} \\
\midrule
$\Omega_m$
  & $0.303^{+0.065}_{-0.070}$ & $0.006$
  & $0.303^{+0.066}_{-0.068}$ & $0.008$ \\[3pt]
$h$
  & $0.702^{+0.102}_{-0.105}$ & $0.005$
  & $0.714^{+0.092}_{-0.107}$ & $0.013$ \\[3pt]
$\sigma_8$
  & $0.791^{+0.146}_{-0.137}$ & $0.008$
  & $0.795^{+0.139}_{-0.137}$ & $0.004$ \\[3pt]
$\beta_0$
  & $-0.002^{+0.104}_{-0.112}$ & $0.266$
  & $0.002^{+0.082}_{-0.084}$  & $0.650$ \\[3pt]
$a_c$
  & $0.668^{+0.231}_{-0.282}$ & $0.089$
  & $0.719^{+0.203}_{-0.299}$ & $0.129$ \\[3pt]
$n$
  & $1.91^{+1.89}_{-1.07}$ & $0.265$
  & $3$ (fixed) & --- \\
\bottomrule
\end{tabular}
\end{table*}

The marginalized distribution of $n$ peaks near $n\approx1.75$--$1.91$ across all dataset combinations, with $68\%$ credible intervals spanning approximately $[0.8,3.8]$. Although the posterior extends to the upper prior boundary at $n=5$, the robustness analysis of Appendix~\ref{subsec:prior_robustness} shows no evidence for boundary accumulation, indicating that the upper prior limit does not qualitatively affect the inferred constraints on the activation index. At the same time, the KL divergence for $n$ ($0.26$--$0.31$~nats) indicates modest likelihood-driven structure within the interior region.

\begin{figure*}
\centering
\includegraphics[width=0.5\textwidth]{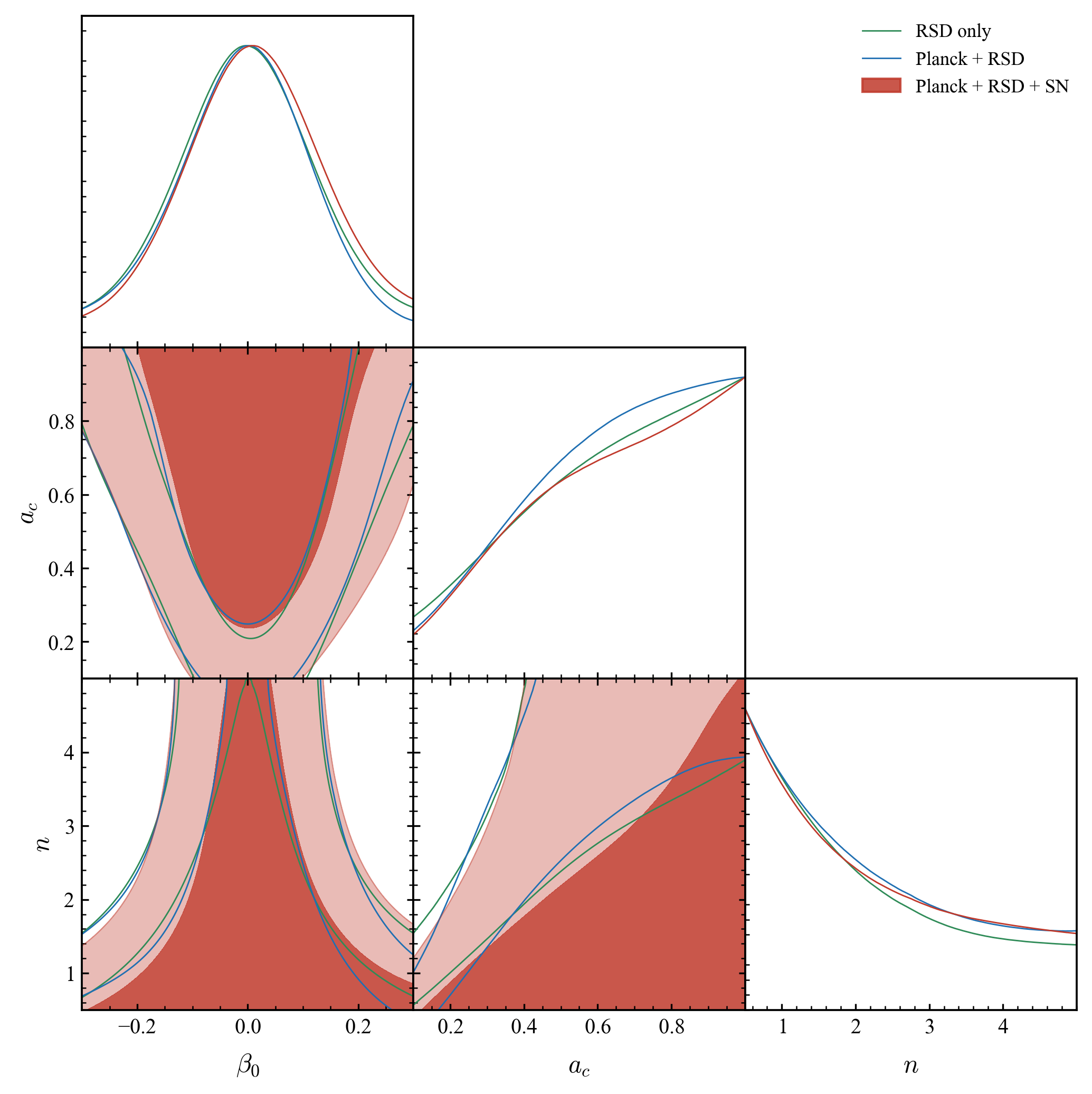}
\caption{\small Marginalized posterior distributions for the interaction-sector parameters $(\beta_0,a_c,n)$ of the flexible IDE (6p) model, using RSD only (green), Planck+RSD (blue), and Planck+RSD+SN (red, filled). Contours denote the $68\%$ and $95\%$ credible regions. The $\beta_0$--$a_c$ and $a_c$--$n$ panels reveal a conditional degeneracy, with $\beta_0$ and $n$ well constrained for small-to-intermediate values of $a_c$ but becoming prior-dominated as $a_c\rightarrow1$.}
\label{fig:ide_triangle}
\end{figure*}

The posterior peak occurs below the asymptotic $p=1$ value. As shown in Appendix~\ref{app:logistic_accuracy} (reported in Table~\ref{tab:logistic_validation}), however, fitting the exact tracking solution of a $q=2$ ($p=1$) potential over the observational range $0\leq z\leq2$ yields effective indices in the range $n_{\rm fit}\approx1.71$--$2.30$, systematically below the asymptotic attractor value $n_{\rm theory}=3$. This shift arises from finite-redshift nonlinear corrections to $V'(\phi)$ before the field fully relaxes to the broken minimum. The observed peak at $n\approx1.75$--$1.91$ is therefore consistent with the pre-asymptotic behavior of the $p=1$ class rather than indicating a statistically significant preference for flatter potentials with $p>1$. Considered on its own, the width of the marginalized posterior, i.e., $68\%$ credible intervals spanning approximately $[0.8,3.8]$, is broad enough to formally encompass all physically motivated classes, $n=3$ ($p=1$), $n=3/2$ ($p=2$), and $n=1$ ($p=3$).

However, the corrected finite-redshift indices for the higher-order classes established in Appendix~\ref{app:logistic_accuracy}, $n_{\rm fit}\in[1.04,1.21]$ for $q=3$ and $n_{\rm fit}\in[0.75,0.82]$ for $q=4$, fall well below the observed peak, whereas the $q=2$ range brackets it closely. Current data, therefore, align specifically with the pre-asymptotic $p=1$ class rather than remaining agnostic among all three. Figure~\ref{fig:ide_triangle} makes the origin of this preference clear. The identifiable region of the $a_c$--$n$ degeneracy occurs at small-to-intermediate values of $a_c$, where the activation index is tightly localized around $n\sim1$--$2$, consistent with the interior $68\%$ credible interval for $a_c$ reported in Table~\ref{tab:ide_6p_n3_comparison}. The broader marginalized envelope of $n$ arises from the additional, weakly informative tail extending toward $a_c\to1$, where $n$ is no longer well constrained, rather than from genuine support for flatter potentials. The preference for the $q=2$ universality class therefore reflects the likelihood structure in the identifiable region of parameter space, rather than an artifact of the prior-dominated tail.

\subsection{Structural Suppression in Growth Observables}
\label{subsec:growth_flexible}

We next examine the behavior of growth observables in the heavy-scalar adiabatic regime. At this limit, the IDE-induced deviations in $f\sigma_8$ are strongly suppressed and remain below the sensitivity of representative Stage-IV growth-rate measurements considered here. In the perturbative adiabatic regime considered here, the homogeneous expansion history remains close to $\Lambda$CDM at leading order. The interaction therefore appears primarily through the linear growth sector, most directly in the observables $f\sigma_8(z)$ and the matter power spectrum $P(k)$. This behavior differs from light-scalar IDE scenarios, where the interaction modifies both the expansion history and the growth sector at the percent level~\citep{pettorino2008coupled, barros2019coupled}.

\begin{figure*}
\centering

\subfloat[]{
\includegraphics[width=0.68\textwidth]{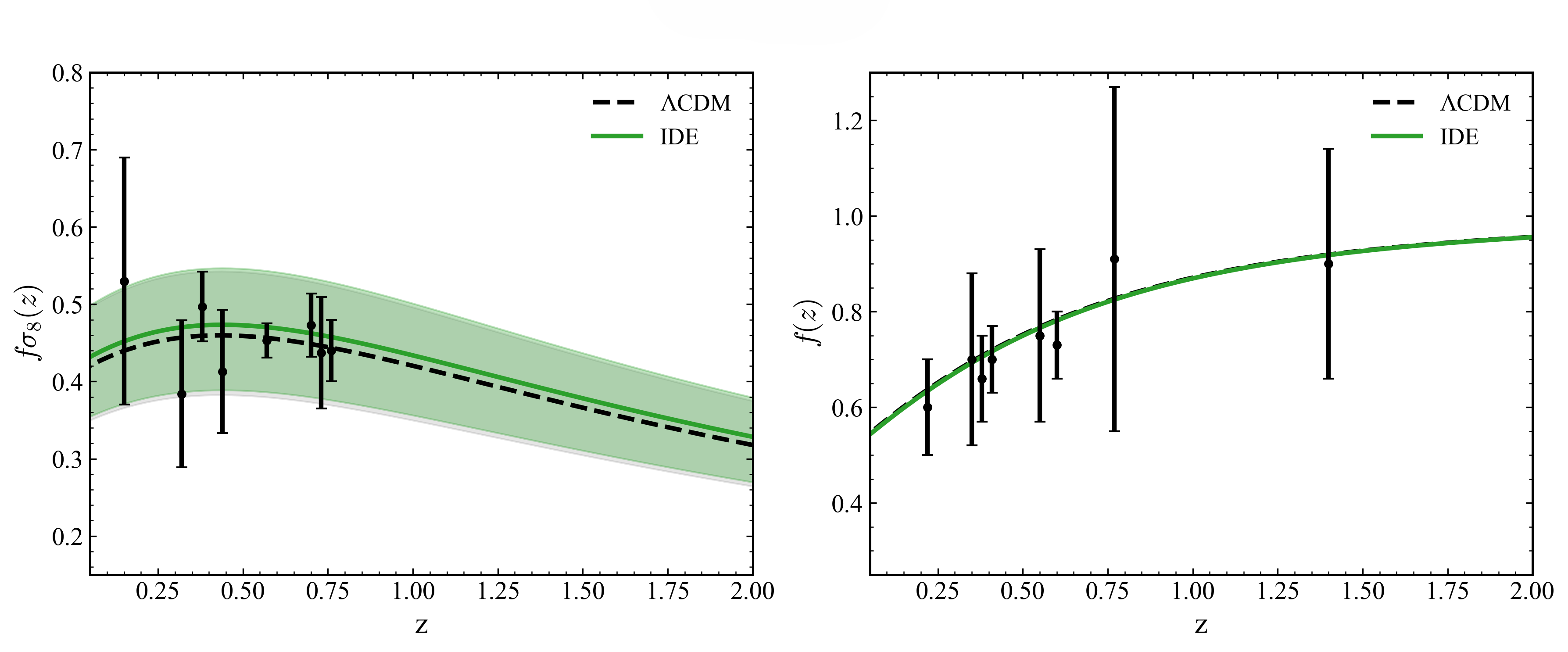}
}

\vspace{0.15cm}

\subfloat[]{
\includegraphics[width=0.68\textwidth]{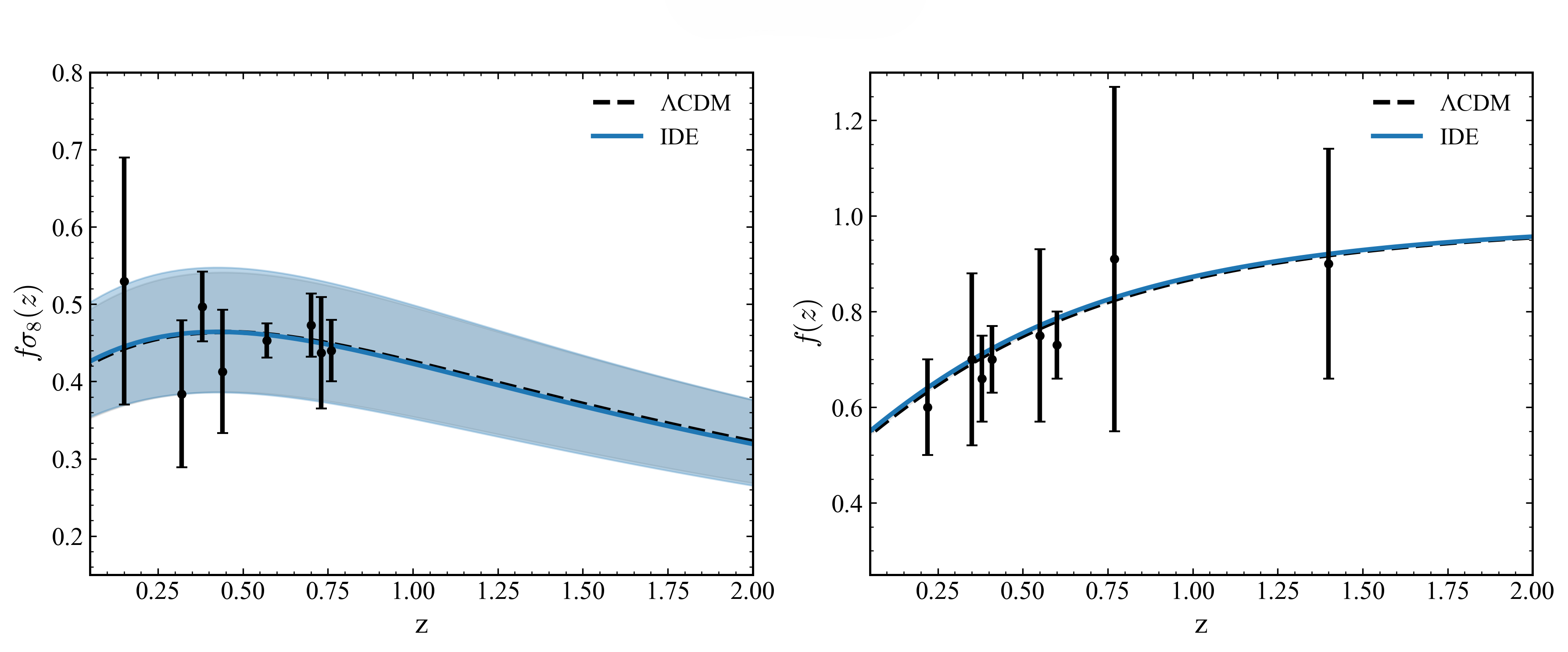}
}

\vspace{0.15cm}

\subfloat[]{
\includegraphics[width=0.68\textwidth]{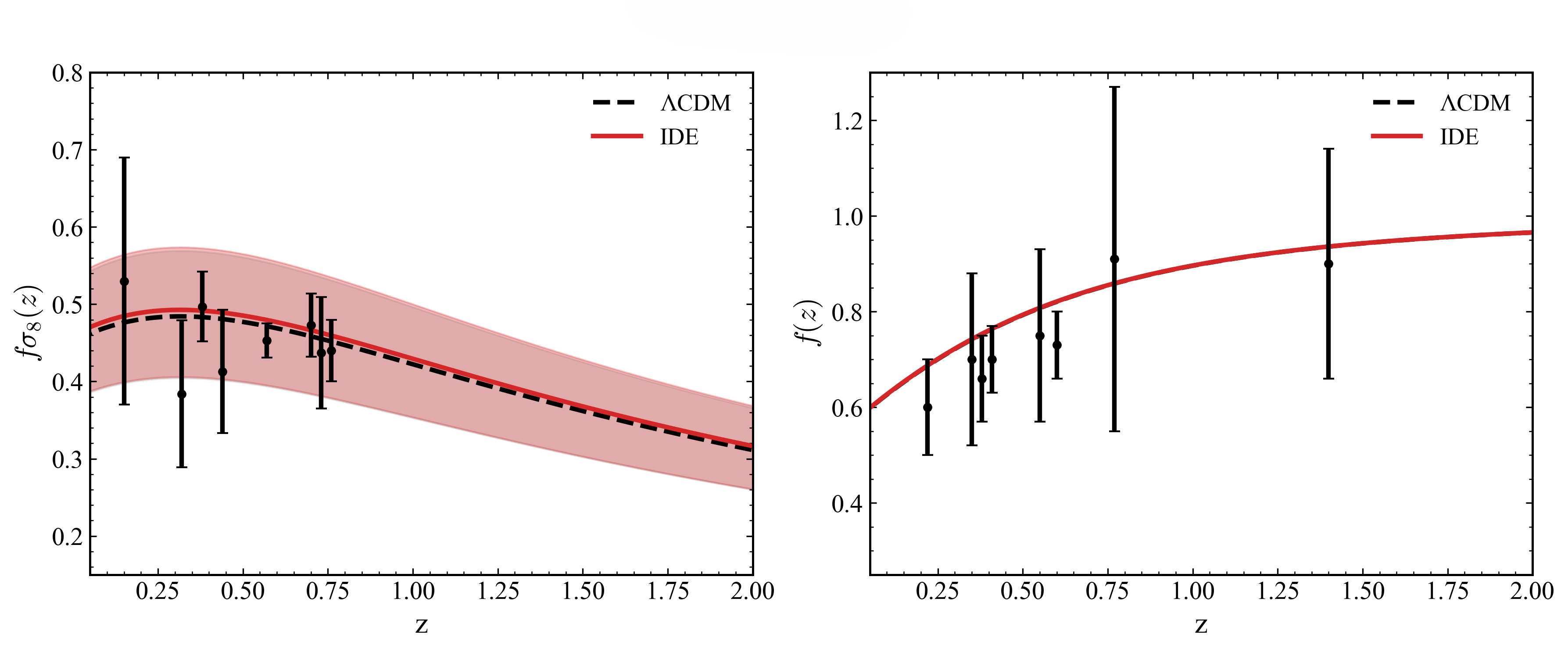}
}

\caption{
Redshift evolution of the linear growth observables $f\sigma_8(z)$ (left) and $f(z)$ (right) in the flexible IDE (6p) model for three dataset combinations: (a) RSD only, (b) Planck+RSD, and (c) Planck+RSD+SN. Solid curves show posterior mean predictions, while shaded regions indicate the $68\%$ credible intervals. The 6p case remains consistent with $\Lambda$CDM within the $68\%$ intervals across the full redshift range. The near-indistinguishability of IDE and $\Lambda$CDM predictions in $f\sigma_8(z)$ reflects the opposite-sign structure of the $f(z)$ and $D(z)$ modifications, rather than negligible individual modifications to either quantity.
}

\label{fig:growth_observables}

\end{figure*}

\begin{figure*}
\centering
\begin{subfigure}[t]{0.6\columnwidth}
\centering
\includegraphics[width=\linewidth]{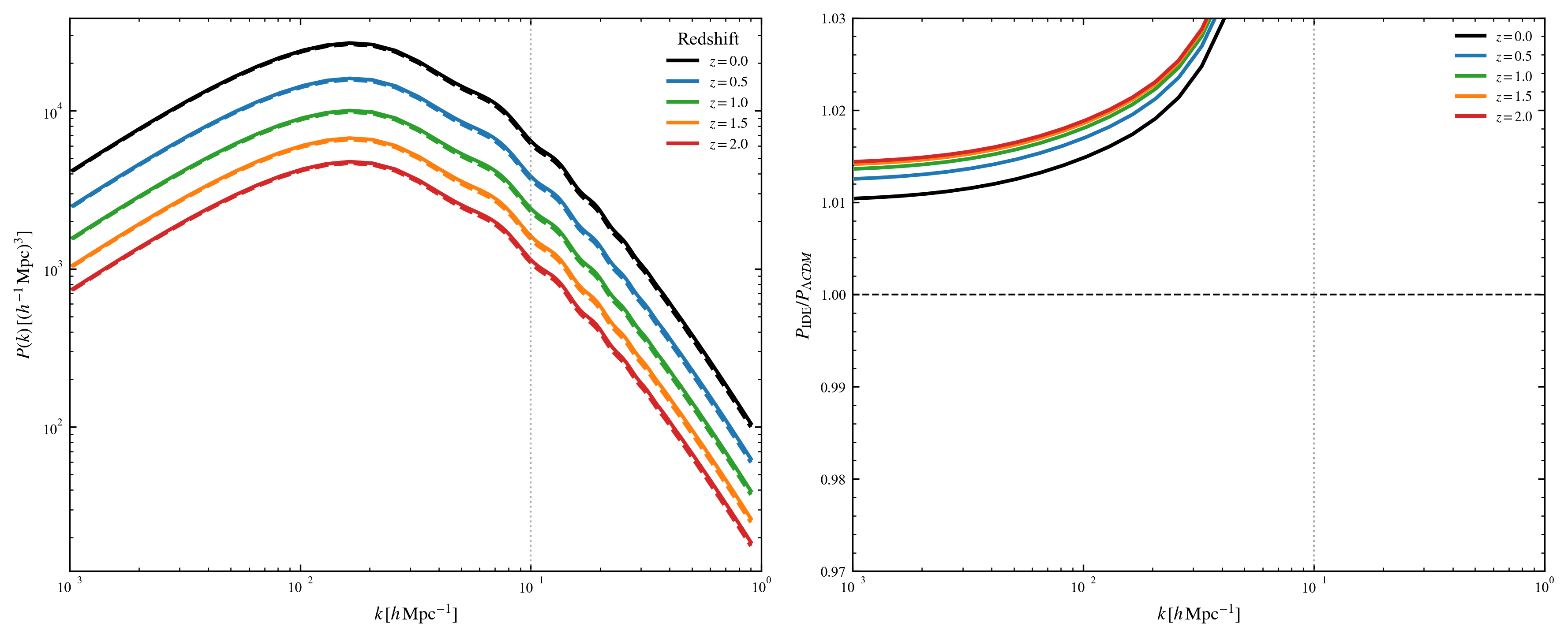}
\caption{}
\end{subfigure}

\vspace{0.1cm}

\begin{subfigure}[t]{0.6\columnwidth}
\centering
\includegraphics[width=\linewidth]{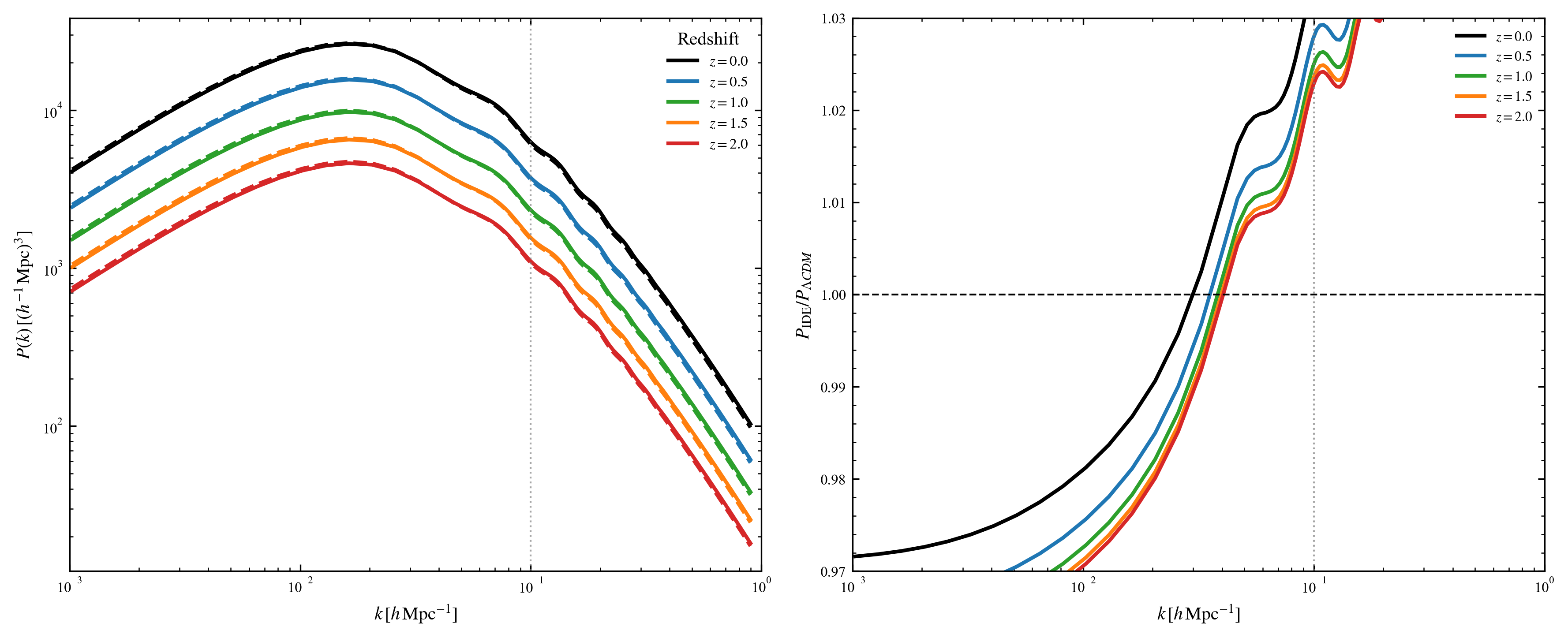}
\caption{}
\end{subfigure}

\vspace{0.1cm}

\begin{subfigure}[t]{0.6\columnwidth}
\centering
\includegraphics[width=\linewidth]{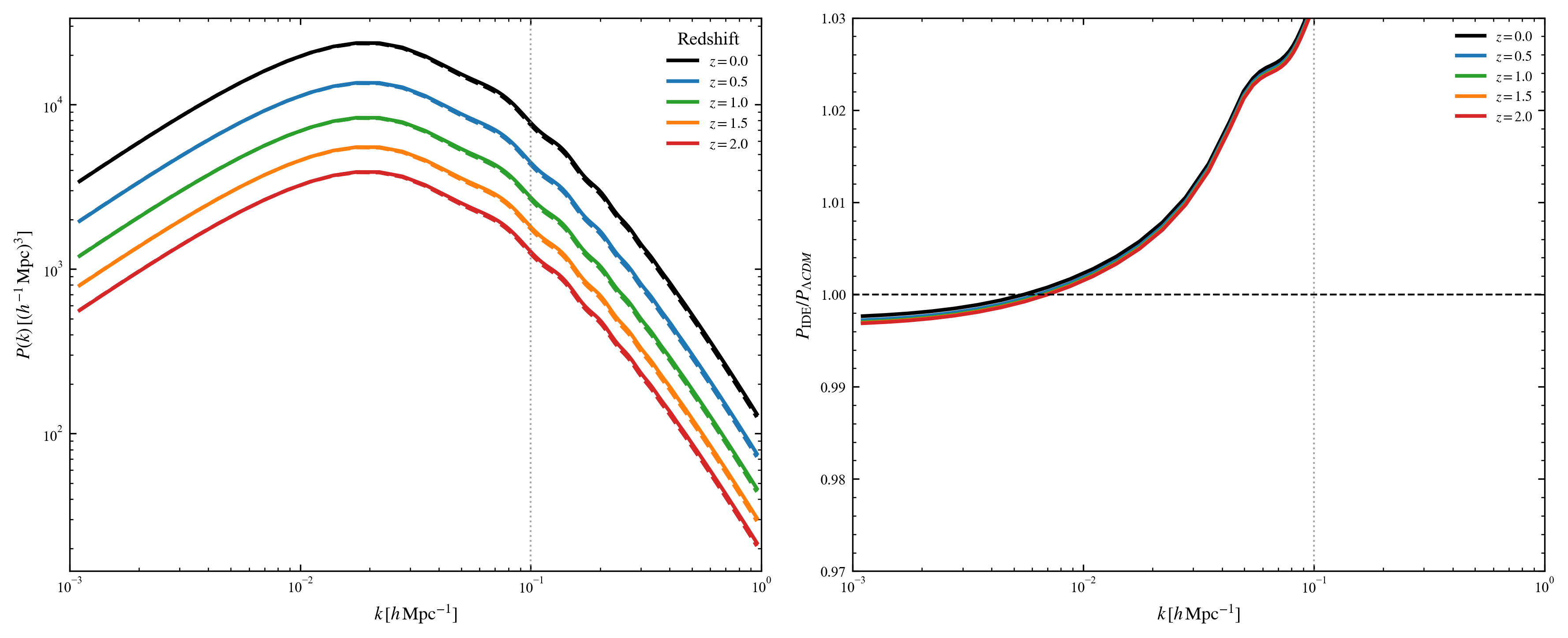}
\caption{}
\end{subfigure}
\caption{\small Evolution of the matter power spectrum $P(k)$ in the IDE~(6p) model compared to $\Lambda$CDM for three dataset combinations: (a)~RSD only, (b)~Planck+RSD, and (c)~Planck+RSD+SN. Each panel shows $P(k)$ (left) and the ratio $P_{\rm IDE}/P_{\Lambda\rm CDM}$ (right). The vertical dotted line marks the linear regime cutoff $k\lesssim 0.1\,h\,\mathrm{Mpc}^{-1}$. The shape of $P(k)$ remains unchanged, and the ratio is nearly constant on linear scales because the interaction is perturbative in the adiabatic regime.}
\label{fig:matter_power}
\end{figure*}

Figure~\ref{fig:growth_observables} shows the redshift evolution of $f\sigma_8(z)$ and $f(z)$ for the flexible 6p realization under the three dataset combinations. The posterior predictions remain consistent with $\Lambda$CDM within the $68\%$ credible intervals over the full observed redshift range. At early times ($z\gtrsim1$), the coupling is strongly suppressed, and the growth history approaches the $\Lambda$CDM limit. Near the activation epoch $a\sim a_c$, the interaction produces percent-level modifications to both $f(z)$ and $D(z)$ and suppress the IDE-induced deviation (at fixed background parameters) to $|\Delta f\sigma_8/f\sigma_8|<0.29\%$ at $95\%$ credibility across the redshift window $0\le z\le 2$. The numerical differences between the IDE and $\Lambda$CDM posterior-mean predictions at $z=0$ ($-0.53\%$ for RSD only, $+1.44\%$ for Planck+RSD, $+0.11\%$ for Planck+RSD+SN) are larger than this bound because they additionally absorb small shifts in the marginalized background parameters $(\Omega_m, \sigma_8)$ between the IDE and $\Lambda$CDM posteriors, rather than reflecting the IDE coupling alone. No dataset combination shows a statistically significant departure from $\Lambda$CDM. Figure~\ref{fig:matter_power} further shows this behavior in the matter power spectrum. The shape of $P(k)$ remains statistically indistinguishable from that of $\Lambda$CDM on linear scales, while the ratio $P_{\rm IDE}/P_{\Lambda\rm CDM}$ stays approximately constant on linear scales ($k\lesssim 0.1\,h\,\mathrm{Mpc}^{-1}$). Since the interaction produces only a small late-time correction to the density growth amplitude, without significantly altering the spectral shape, these results are consistent with the perturbative character of the adiabatic regime.

\begin{figure*}
\centering
\includegraphics[width=0.6\linewidth]{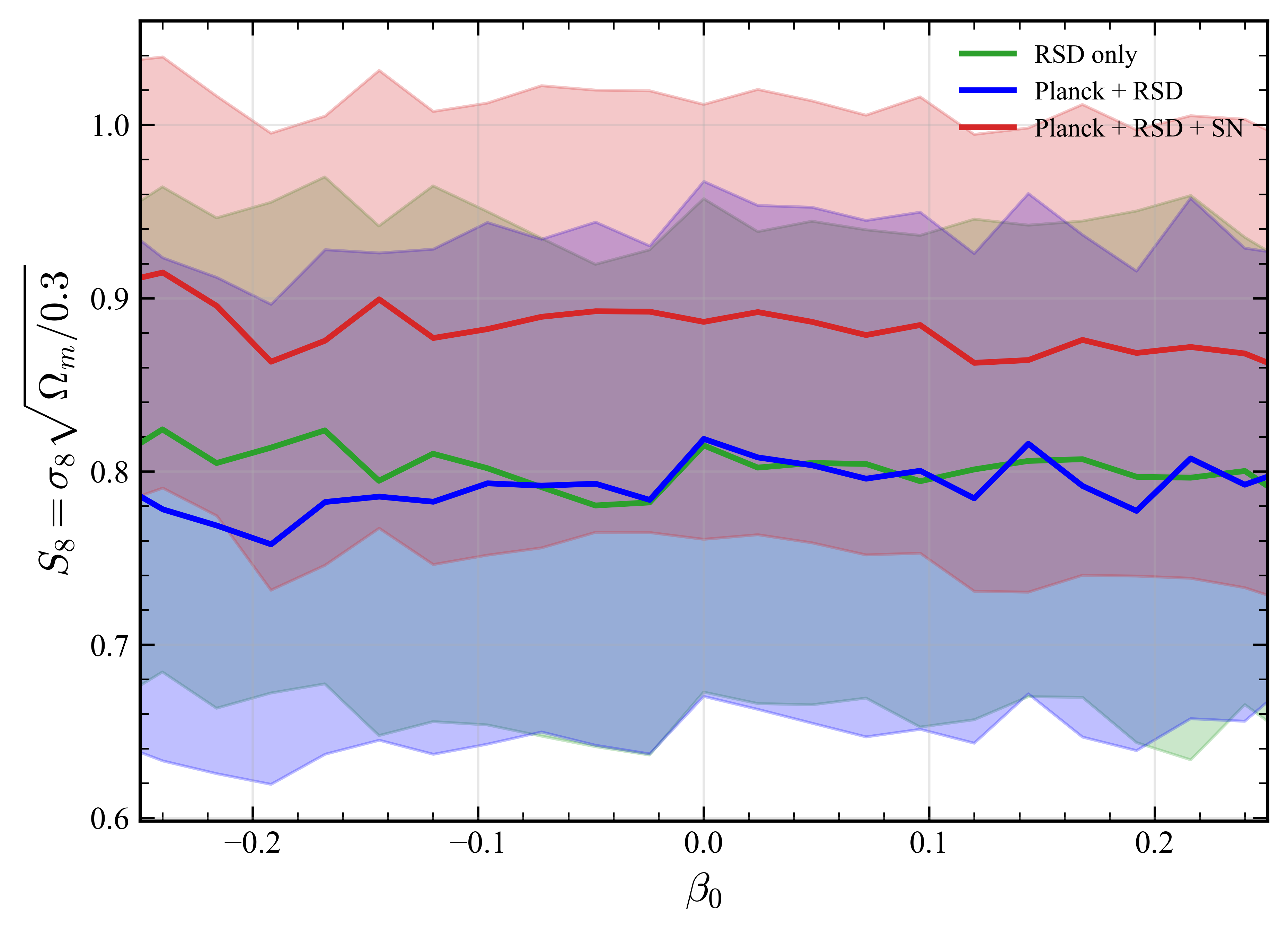}
\caption{\small Posterior-averaged dependence of $S_8 = \sigma_8\sqrt{\Omega_m/0.3}$ on the coupling amplitude $\beta_0$ in the IDE~(6p) model for three dataset combinations. The weak slope and small correlation coefficient ($|\rho|\lesssim 0.07$) confirm the near-orthogonality of the interaction strength and late-time clustering normalization, consistent with the opposite-sign $f(z)$--$D(z)$ structure discussed in Section~\ref{subsec:growth_flexible} rather than parameter tuning.}
\label{fig:s8_beta0}
\end{figure*}

The near-indistinguishability of the 6p realization from $\Lambda$CDM in $f\sigma_8(z)$ arises naturally in the heavy-scalar adiabatic regime and does not require parameter fine-tuning. Over $0\leq z\leq2$, the individual deviations reach $\max|\Delta D/D|\approx1.5\%$ and $\max|\Delta f/f|\approx1.2\%$, remaining of opposite sign throughout this range so that the net deviation in $f\sigma_8$ is suppressed to $\Delta f\sigma_8/f\sigma_8\lesssim0.3\%$ across the posterior. Since $D(a{=}1)\equiv1$ in both models, $\Delta D/D$ vanishes identically at $z=0$ and the low-redshift deviation in $f\sigma_8$ is set by $\Delta f/f$ alone; at higher $z$, $\Delta D/D$ grows as the coupling's cumulative effect accumulates while $\Delta f/f$ recedes, with the two crossing near $z\approx1$. In the present implementation the coupling enters the CDM continuity and Euler equations as a density-proportional correction, not through a modified gravitational term; the opposite-sign structure reflects $D(z)$'s role as a cumulative integral against $f(z)=d\ln D/d\ln a$'s role as a local derivative of that same correction, rather than two independently-sourced channels. This is also visible in the weak posterior correlation between $S_8\equiv\sigma_8\sqrt{\Omega_m/0.3}$ and $\beta_0$, shown in Fig.~\ref{fig:s8_beta0} ($|\rho|\lesssim0.07$ for all dataset combinations). This differs from momentum-exchange models~\citep{valiviita2008large, cruickshank2025forecasts}, where $f\sigma_8$ remains directly sensitive to the interaction; a comparison with light-scalar IDE scenarios~\citep{barros2019coupled, pettorino2008coupled} is left for future work.

\subsection{Exploratory Forecasts for Future Surveys}
\label{subsec:detectability}

The suppression of the IDE signal in standard growth observables highlights the need for a quantitative assessment of its detectability in future Stage-IV surveys. We therefore present an exploratory, order-of-magnitude estimate of the signal-to-noise ratio (SNR) associated with the interaction signal in the growth-rate observable $(f\sigma_8(z))$ and in the weak-lensing convergence power spectrum for representative Euclid- and LSST-like survey configurations. The calculation employs simplified single-source-plane and diagonal-covariance approximations and is intended to compare the relative sensitivity of different observables rather than to provide a definitive survey forecast. A full tomographic analysis incorporating realistic survey systematics and covariance modelling is deferred to future work. Details of the forecasting methodology are provided in Appendix~\ref{app:forecast}.

The IDE model induces percent-level modifications in the individual growth quantities, with $\max|\Delta D/D|\approx 1.5\%$ and $\max|\Delta f/f|\approx 1.2\%$ over $0\leq z\leq 2$. The opposite-sign structure of $\Delta D/D$ and $\Delta f/f$ nevertheless suppresses the corresponding deviation in $f\sigma_8(z)$ to $0.034\%$ for Euclid and $0.058\%$ for LSST at the Planck+RSD+SN best-fit point, with a posterior mean of $0.082\%$ and a $95\%$ upper bound of $0.29\%$. Assuming representative fractional per-bin uncertainties of $1\%$ for Euclid and $2\%$ for LSST, the resulting SNR for the $f\sigma_8$ signal remains well below unity for the fiducial configuration (Table~\ref{tab:snr_estimates}) and does not exceed $\sim2$ across the posterior range explored here. The predicted signal therefore remains below the projected sensitivity of Stage-IV growth-rate measurements based on $f\sigma_8(z)$ alone~\citep{teixeira2023forecasts, joseph2023forecast, cruickshank2025forecasts}.

Weak-lensing observables retain greater sensitivity to this IDE class because the convergence power spectrum, $C_\ell^{\kappa\kappa}\propto\int D^2(z)\, W^2(z)\,d\chi$, depends on the integrated growth factor along the line of sight and does not directly involve $f(z)$. Within the simplified forecast adopted here, the resulting deviation reaches $\Delta C_\ell/C_\ell \sim 2$--$3\%$, consistent with the analytic estimate $\Delta C_\ell/C_\ell \approx 2\,\Delta D/D$, and is substantially larger than the corresponding $f\sigma_8$ signal. Using a simplified single source plane LSST-like configuration ($A_{\rm sky}=14{,}000\,\mathrm{deg}^2$, $n_{\rm gal}=20\,\mathrm{arcmin}^{-2}$, $z_{\rm max}=2.5$), we obtain an indicative weak-lensing sensitivity of $\mathrm{SNR}_{\rm WL}\sim1$--$2$. Although still below conventional detection thresholds, this represents an improvement of roughly a factor of $20$--$30$ relative to the projected $f\sigma_8$ sensitivity. A full tomographic analysis with multiple source redshift bins would likely further strengthen these constraints~\citep{mazoun2024probing, mazoun2025interacting}, and is left for future work.

\begin{table*}
\centering
\caption{\small Indicative SNR for the IDE signal in the growth-rate ($f\sigma_8$) and weak-lensing ($C^{\kappa\kappa}_\ell$) observables. The Planck+RSD+SN case is the fiducial scenario; the RSD-only case provides a growth-optimized upper bound. The weak-lensing SNR uses a single source plane at $z_{\rm max}=2.5$ with an LSST-like configuration and represents a conservative lower bound; tomographic analyses are likely to improve it by a factor of a few. All values are order-of-magnitude indicators given the approximations described in Appendix~\ref{app:forecast}.}
\label{tab:snr_estimates}
\setlength{\tabcolsep}{6pt}
\renewcommand{\arraystretch}{1.5}
\begin{tabular}{lccc}
\toprule
Scenario &
  \makecell{Euclid SNR\\($f\sigma_8$)} &
  \makecell{LSST SNR\\($f\sigma_8$)} &
  \makecell{WL SNR} \\
\midrule
Planck+RSD+SN (best-fit)   & $\sim 0.05$ & $\sim 0.04$ &
  $\sim 1$--$2\,^{a}$ \\
Planck+RSD+SN (post. mean) & $\sim 0.1$  & $\sim 0.08$ &
  $\sim 1$--$2\,^{a}$ \\
RSD-only (best-fit)        & $\sim 0.3$  & $\sim 0.2$  & --- \\
RSD-only (post. mean)      & $\sim 0.5$  & $\sim 0.4$  & --- \\
\bottomrule
\end{tabular}
\begin{minipage}{0.95\linewidth}
\footnotesize
$^{a}$Computed at the Planck+RSD+SN best-fit parameters using a single effective source plane; this is a conservative lower bound on a tomographic analysis.
\end{minipage}
\end{table*}

\section{Microphysical Interpretation of the Observational Constraints}
\label{sec:theoretical_implications}

The preceding sections established that the data do not favor a nonzero coupling and that the posterior is primarily bounded by perturbative consistency rather than by the data likelihood. The perturbative consistency condition restricts the posterior to configurations satisfying $|\epsilon(a)|\ll 1$ across the observational window. This bound directly translates into a constraint on the effective scalar mass at the activation epoch. 

Recalling from Eq.~\eqref{eq:epsilon_def_method} that
\begin{equation}
\epsilon(a)\sim 3\,\beta^2(\phi)\,\frac{H^2(a)}{m_{\rm eff}^2(a)},
\label{eq:epsilon_micro_position}
\end{equation}
and using $\beta(a_c)=\beta_0/2$ from the logistic profile Eq.~\eqref{eq:beta_logistic}, the perturbative bound
$|\epsilon(a_c)|\lesssim\epsilon_*=10^{-2}$ becomes
\begin{equation}
m_{\rm eff}(a_c)\gtrsim \mathcal{O}(10)\,\beta_0\,H(a_c).
\label{eq:mass_hierarchy_position}
\end{equation}
For the posterior preferred $|\beta_0|\lesssim 0.15$, this requires $m_{\rm eff}(a_c)\gtrsim \mathcal{O}(1)\,H(a_c)$, and for the $95\%$ credibility upper bound $|\beta_0|\lesssim 0.26$, the bound tightens to $m_{\rm eff}(a_c)\gtrsim \mathcal{O}(3)\,H(a_c)$. In both regimes, the scalar remains heavier than the Hubble scale at the activation epoch, and tracks the instantaneous minimum of the effective potential rather than undergoing slow-roll evolution.

\begin{figure*}
    \centering
    \includegraphics[width=0.6\linewidth]{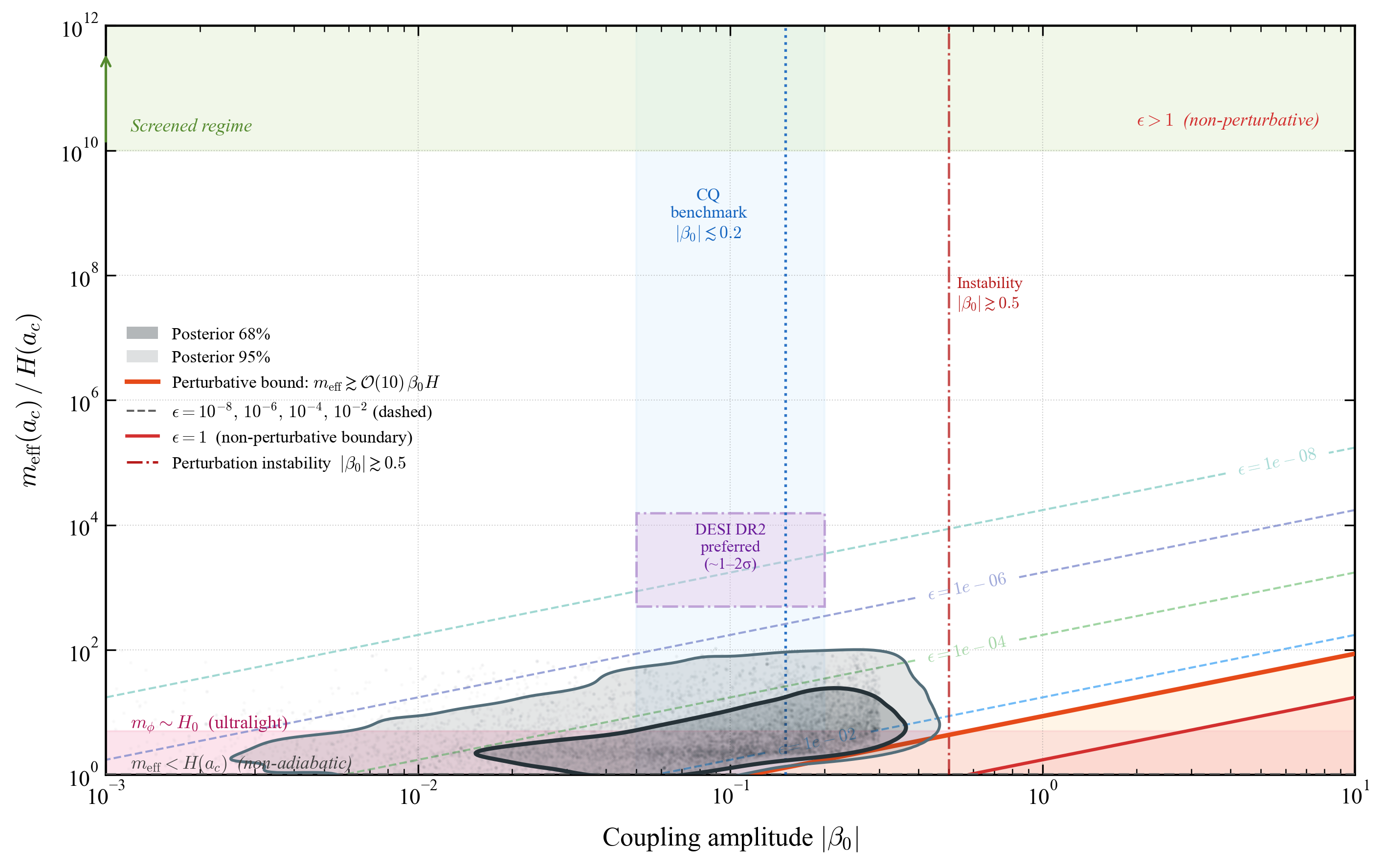}
\caption{\small Constraint plane in the theory space. Dashed contours show constant values of $\epsilon \propto \beta_0^2/(m_{\rm eff}/H)^2$, while the orange line denotes the perturbative hierarchy bound from Eq.~\eqref{eq:mass_hierarchy_position}. Filled contours show the $68\%$ and $95\%$ credible regions of the Planck+RSD+SN posterior after applying the perturbative filter. The allowed region lies well above the ultralight quintessence scale $m_\phi \sim H_0$, below the screened regime, and away from the perturbative-instability boundary $|\beta_0| \gtrsim 0.5$. The posterior therefore occupies a heavy-scalar, adiabatic, and perturbative sector of parameter space, demonstrating that the observationally allowed IDE solutions correspond to a controlled EFT regime rather than to screened or ultralight realizations.}
    \label{fig:constrain_space}
\end{figure*}

Figure~\ref{fig:constrain_space} maps the Planck+RSD+SN posterior onto the constraint plane $(|\beta_0|,\,m_{\rm eff}(a_c)/H(a_c))$. We caution that the proximity of the posterior to the $\epsilon\sim 10^{-2}$ contour is partly a consequence of the perturbative filter itself, which by construction excludes samples above this threshold; the concentration should therefore not be interpreted as a preference for near-critical configurations. Within this region, the allowed parameter space occupies a comparable coupling range to coupled quintessence models~\citep{patil2024coupled, potting2022coupled}, lies well above the ultralight quintessence scale $m_\phi\sim H_0$~\citep{hui2017ultralight, hlovzek2017future}, and is clearly separated from any screened scalar regime in which the fifth force is suppressed by environmental density rather than by the adiabatic hierarchy. The present model therefore occupies a structurally distinct region of scalar-tensor theory space and activation dynamics, driven by late-time CDM dilution.

\section{Discussion}
\label{sec:discussion}

\subsection{Theoretical positioning and EFT naturalness}

The heavy scalar attractor class studied here occupies a distinct region of scalar-tensor theory space. In screened scalar theories and environmentally dependent dilaton models, the local structure of $V(\phi)$ and $A(\phi)$ near a density-dependent minimum governs the dynamics in the high-density regime. The present construction applies the same organizing principle in the low-density phase, where the local restoring order $p$ controls the late-time evolution. In both cases, the dynamics follow from the adiabatic balance condition evaluated in different density regimes. The most notable similarity appears in $\alpha$-attractor models~\citep{galante2015unity, akrami2018dark}. However, the analogy is limited; $\alpha$-attractor universality arises from the geometry of the kinetic term during inflationary evolution, whereas the present universality is governed by the local restoring order of the scalar potential during matter-dominated adiabatic tracking in an entirely different physical regime.

Another important consideration is the naturalness of the different universality classes under EFT. For $p > 1$ (equivalently $n < 3$, or $q > 2$ in the polynomial family), the minimum is degenerate at tree level with $V''(v) = 0$. The one-loop correction to the potential near the degenerate minimum is of order $\delta V \sim \lambda^2 v^4/(16\pi^2)$, which generates an effective curvature $\delta V''(v) \sim \lambda^2 v^2/(8\pi^2)$. For typical coupling values $\lambda \sim 10^{-2}$-$10^{-1}$, this correction exceeds the tree-level curvature $V''(v) = 0$ by construction, shifting the minimum to a non-degenerate configuration and placing the corrected potential in the $p = 1$ class. The higher-order classes are therefore radiatively unstable unless protected by an additional symmetry. By contrast, the $p = 1$ class has a non-degenerate quadratic minimum with $V''(v) \neq 0$. The Coleman-Weinberg and axion-like potentials are concrete realizations of this class and remain stable under radiative corrections within their respective frameworks. 

Within this regime, the perturbative bound $\max|\epsilon(a)| < 10^{-2}$ implies a constraint on the effective scalar mass. For the posterior-preferred range $|\beta_0| \lesssim 0.15$, Eq.~\eqref{eq:mass_hierarchy_position} gives $m_{\rm eff}(a_c) \gtrsim \mathcal{O}(1)\,H(a_c)$, and for the $95\%$ credibility upper bound $|\beta_0| \lesssim 0.26$, the bound tightens to $m_{\rm eff}(a_c) \gtrsim \mathcal{O}(3)\,H(a_c)$. This bound is somewhat weaker than constraints reported for constant-coupling coupled quintessence models, which typically find $\beta\lesssim0.05$--$0.10$ from Planck combined with late-time probes~\citep{xia2009constraint, xia2013new}, with a comparable Planck-only central value $\beta=0.084^{+0.052}_{-0.044}$~\citep{da2022simple}. We attribute this difference to the additional shape parameters $(a_c,n)$, since marginalization over their conditional degeneracy broadens the posterior of $\beta_0$ relative to constant-coupling models. In this context, the overlap between the allowed parameter space and the coupling range of coupled quintessence models is further restricted by the adiabatic hierarchy, which traps the scalar near its effective potential minimum. This constraint is structurally different from fifth-force and equivalence-principle bounds on baryonic couplings since the late-time activation of the coupling is not directly probed by those tests.

\subsection{Observational consequences of the \texorpdfstring{$f$-$D$}{f-D} sign structure}

In the heavy scalar adiabatic regime, the IDE-induced modifications to the growth rate $f(z)$ and growth factor $D(z)$ enter $f\sigma_8(z)=f(z)\,D(z)\,\sigma_8$ with opposite sign throughout $0\le z\le2$, suppressing the net deviation to $\Delta f\sigma_8/f\sigma_8 \lesssim 0.3\%$ across the posterior despite percent-level changes in the individual quantities, $\max|\Delta D/D|\approx 1$--$2\%$ and $\max|\Delta f/f|\approx 1.2\%$. As discussed in Section~\ref{subsec:growth_flexible}, this sign structure follows from the distinction between $D(z)$, sourced cumulatively by the density-proportional correction to the CDM equations, and $f(z)=d\ln D/d\ln a$, its local derivative. It is therefore a structural feature of the perturbative adiabatic regime rather than a result of parameter tuning. This behavior distinguishes the present scenario from momentum-exchange models~\citep{valiviita2008large,cruickshank2025forecasts}, which preserve the $\Lambda$CDM background but generate a net suppression of $f\sigma_8(z)$ without the opposite-sign structure identified here, since the growth rate and growth factor are modified asymmetrically; $f\sigma_8(z)$ therefore remains an informative probe in that class. In the present heavy-scalar attractor class, by contrast, the opposite-sign structure appears intrinsic to the implemented correction, and improvements in survey precision alone are unlikely to recover the signal from $f\sigma_8$ without complementary probes.

These considerations have direct implications for survey strategy. Isolating the interaction signal requires probes that constrain $f(z)$ and $D(z)$ independently~\citep{mazoun2024probing, mazoun2025interacting}, rather than in the combination $f\sigma_8$ where the opposite-sign structure suppresses the net signal. Weak-lensing tomography is a natural candidate, since the convergence power spectrum $C_\ell^{\kappa\kappa} \propto \int D^2(z)\,W^2(z)\,\mathrm{d}\chi$ is sensitive to $D(z)$ integrated along the line of sight and is not subject to this suppression. A simplified single source plane estimate yields $\mathrm{SNR}_{\rm WL} \approx 1.5$ for an LSST-like configuration, with a maximum fractional deviation in $C_\ell^{\kappa\kappa}$ of $\approx 3.3\%$, consistent with the analytic expectation $\Delta C_\ell/C_\ell \approx 2\,\Delta D/D$. While sub-threshold at current sensitivity, this represents a factor of $\sim 20$-$30$ improvement over the $f\sigma_8$ signal.

Complementary probes include cluster number counts from eROSITA~\citep{ghirardini2024srg, artis2025srg} and Euclid~\citep{scaramella2022euclid, mellier2025euclid}, which constrain $\sigma_8(z)$ through the halo mass function and are not subject to the $f(z)$--$D(z)$ suppression discussed above. The ISW cross-correlation with galaxy density~\citep{seraille2024constraining} probes the time evolution of the metric potentials near the activation epoch $a_c$ while being sensitive to the transition redshift that neither lensing nor cluster counts directly access. DESI DR2 BAO measurements~\citep{petri2026dark, qu2025atacama} will sharpen the geometric constraints on $\Omega_m$ and $H(a_c)$, which in turn tighten Eq.~\eqref{eq:mass_hierarchy_position} independently of the growth sector. Together, these probes, each sensitive to a different projection of the IDE signal define a minimal observational strategy required to test this class of models. A rigorous multi-probe Fisher forecast is left to future work.

\subsection{Current constraints, model limitations, and cosmological tensions}

While the flexible model remains statistically consistent with $\Lambda$CDM, the fixed-index benchmark reveals a substantial tension between the rigid asymptotic $n=3$ realization and the combined geometric and growth data. This is evident in the degradation of the SN fit and the migration of cosmological parameters toward the boundaries of the allowed parameter space. This indicates that the observable interaction history cannot be directly identified with the asymptotic attractor solution, and that finite-redshift evolution plays an essential role in connecting the underlying scalar dynamics to cosmological observations.

The adiabatic tracking condition constrains background deviations to  $|\Delta H/H| \lesssim \epsilon_\star \sim 10^{-2}$. Since alleviating the Hubble tension at the level of $\Delta H_0/H_0 \approx 9\%$ would require modifications to the expansion history far exceeding the perturbative bound, the heavy scalar attractor regime studied here cannot address the Hubble tension. We stress that this conclusion applies specifically to the adiabatic limit and does not preclude light scalar IDE from contributing to tension resolution. On the $S_8$ side, because of the opposite-sign $f(z)$--$D(z)$ structure discussed above, the clustering amplitude remains close to $\Lambda$CDM with the posterior correlation $|\rho(S_8, \beta_0)| \lesssim 0.07$ across all dataset combinations. This is consistent with the recent KiDS-Legacy result, $S_8 = 0.815^{+0.016}_{-0.021}$~\citep{wright2025kids}, which reduces the previously reported $\sim 3\sigma$ discrepancy~\citep{heymans2021kids} to $<1\sigma$; see also the independent review by~\citep{pantos2026status}. If the true clustering amplitude is indeed close to this value, IDE models designed primarily to suppress $S_8$ are increasingly disfavored, and the structural suppression mechanism studied here is more consistent with this emerging picture.

We note three further limitations of the present analysis. First, the perturbation treatment assumes adiabatic tracking, and is applied over $z \lesssim 2$ where this condition holds for most posterior samples. At higher redshifts, the power-law suppression of $\beta_{\rm eff}(a) = \beta_0\,y(a)$ ensures that the IDE source terms in the perturbation equations are negligible and standard adiabatic initial conditions remain valid. Second, the coupled CDM sector in the present implementation receives a density-proportional correction to its continuity and Euler equations, sourced by the same rate entering the background exchange term; a self-consistently sourced scalar-field perturbation, and the resulting fifth-force contribution familiar from coupled quintessence, are left for future implementation. Third, the present work is restricted to linear scales $k \lesssim 0.1\,h\,\mathrm{Mpc}^{-1}$ and does not model nonlinear structure formation. In coupled scalar-tensor models, nonlinear effects can modify the halo mass function and the matter power spectrum on small scales~\citep{baldi2010hydrodynamical, baldi2012dark}. A full assessment of the model's predictions in the nonlinear regime would require dedicated $N$-body simulations or extensions to the halo model, which we leave for future work.

\section{Conclusion}
\label{sec:conclusion}

In this work, we have shown that the late-time activation profile of the conformally coupled scalar--tensor IDE is not an arbitrary phenomenological ansatz but, in the adiabatic tracking regime, follows from the local restoring structure of the symmetry-breaking potential. The asymptotic normal form is characterized by the relation \(n=3/p\), where \(p\) is the local restoring order near the broken minimum, thereby organizing distinct scalar potentials into identical asymptotic attractor behavior. In particular, structurally different realizations such as quartic, Coleman--Weinberg, and axion-like potentials belong to the radiatively stable \(p=1\) class and share the canonical asymptotic index \(n=3\). The logistic coupling profile employed in the observational analysis should therefore be understood as a controlled global extension of this asymptotic behavior, valid over the redshift range where the tracking approximation holds and consistent with the exact solution to good accuracy over the observational window.

Testing this framework with Planck CMB lensing, RSD, and Pantheon+SH0ES SN data, the flexible six-parameter realization yields $|\beta_0|\lesssim 0.26$ at the $95\%$ credibility level. By contrast, imposing the rigid $n=3$ asymptotic profile produces a substantial degradation of the SN fit, $\Delta\chi^2_{\rm SN}\approx 92$, and drives $\Omega_m$ and $\sigma_8$ toward the lower boundaries of the allowed parameter space. Current observations do not provide statistically significant evidence for a nonzero interaction and only weakly constrain the activation history. Nevertheless, they show that the observable coupling history cannot be identified directly with its asymptotic attractor form, and that a realistic implementation of the $p=1$ class requires finite-redshift corrections beyond the rigid asymptotic limit. The principal result of the present work is therefore not the detection of IDE, but the identification of a dynamical classification linking late-time coupling phenomenology to the local restoring structure of the scalar potential.

A key implication of the heavy-scalar attractor regime is that the absence of a detectable signal in $f\sigma_8$ cannot be interpreted as the absence of an interaction. Although the coupling modifies both the growth rate $f(z)$ and the growth factor $D(z)$ at the percent level, these corrections enter $f\sigma_8$ with opposite signs, suppressing the net observable deviation to $\lesssim 0.3\%$. The resulting signal lies below the projected sensitivity of Stage-IV growth-rate measurements and reflects a structural property of the attractor dynamics rather than a limitation of current data. In this class of models, the relevant question is therefore not whether an interaction is present, but whether the observables employed can resolve the underlying growth modifications. Probes that constrain $D(z)$ independently of $f(z)$, particularly weak-lensing tomography, provide a more direct test of the interaction and provide substantially greater sensitivity than growth-rate measurements alone.

The present work, therefore, should be regarded as a first step toward a dynamical classification of IDE models. Its conclusions are restricted to the heavy-scalar adiabatic regime and to linear cosmological observables, where the interaction produces only small departures from $\Lambda$CDM and does not address the Hubble or $(S_8)$ tensions. Several important directions remain open, including a treatment of departures from adiabatic tracking, nonlinear structure formation, tomographic weak-lensing forecasts, and extensions to broader classes of scalar-field interactions. Establishing whether the universality structure identified here persists beyond the perturbative regime and determining how it manifests in future high-precision observations will be essential for assessing the viability of dynamically regulated IDE scenarios.

\section*{Acknowledgements}
PKMV, KNS, and KA acknowledge that part of the numerical computations presented in this work were carried out using the Pegasus computing cluster at IUCAA, Pune, India. PKMV acknowledges financial support from the Seed Money Scheme of CHRIST (Deemed to be University) under Sanction No.~CU-ORS-SM-24/29. The authors thank the anonymous referee for a careful reading of the manuscript and for constructive comments that significantly improved the clarity and robustness of the analysis.

\section*{Data Availability Statement}

All observational datasets analyzed in this work are publicly available and can be accessed through the original references cited in the manuscript. The MCMC chains generated for this study are available from the authors upon reasonable request. The modified \texttt{CLASS} solver used in this work has been made publicly available and can be accessed via Ref.~\citep{class_ide}.

\appendix
\section{Autonomous Flow for Polynomial Potentials}
\label{app:polynomial}
This appendix provides the detailed specialization of the general formalism of Section~\ref{sec:covariant_framework} to the $\mathbb{Z}_2$-symmetric polynomial potentials that realize density‑triggered SSB. The derivation of autonomous flow and balance equation used in Section~\ref{sec:attractor_structure} is presented here to keep the main text concise while preserving a self‑contained account of the model-specific dynamics.

Following \citep{ghilencea2019spontaneous, nakagawa2023early, katz2014higgs, zhang2020obtaining}, we consider the general polynomial class, which is the minimal two-term realization of $\mathbb{Z}_2$-symmetric SSB
\begin{equation}
V(\phi) = -\frac12 \mu^2 \phi^2 + \frac{\lambda}{2q}\phi^{2q},
\qquad q\ge2,
\label{eq:general_ssb_potential}
\end{equation}
with $\mu^2>0$ and $\lambda>0$. For $q=2$, it is algebraically equivalent to the familiar Mexican-hat potential $\frac{\lambda}{4}(\phi^2-v^2)^2$ after dropping the field-independent constant $\frac{\lambda}{4}v^4$, which does not affect the dynamics. The degenerate vacuum minima are
\begin{equation}
\phi=\pm v, \qquad v \equiv \left(\frac{\mu^2}{\lambda}\right)^{\frac{1}{2q-2}},\label{eq:vev_definition}
\end{equation}
with curvature
\begin{equation}
m_\phi^2\equiv V''(v)= 2(q-1)\mu^2 \ \neq 0 \quad \text{for every } q\ge2.
\label{eq:mass_definition}
\end{equation}
Equation~\eqref{eq:mass_definition} shows that the two-term potential has a non-degenerate quadratic minimum for \emph{every} $q$: it therefore belongs to the $p=1$ universality class for all $q$, and does not by itself realize the higher-order classes $p=q-1$ shown in Fig.~\ref{fig:universal_attractor} and Table~\ref{tab:universality_mapping}. It is, in this sense, the generic $p=1$ representative of $\mathbb{Z}_2$-symmetric SSB, consistent with the radiative-stability argument that one-loop corrections generate exactly the mass term of Eq.~\eqref{eq:general_ssb_potential}, driving any fine-tuned higher-order minimum back toward this form. To realize the $p=q-1$ classes explicitly, we instead work directly with the compact potential used in the main text, Eq.~\eqref{eq:poly_potential},
\[
V(\phi)=\frac{\lambda}{2q}\bigl(\phi^2-v^2\bigr)^q, \qquad q\ge2,
\]
whose degenerate tree-level minimum, $V''(v)=0$ for $q>2$, is the fine-tuned configuration required to reach $p>1$; the two potentials coincide only at $q=2$.

To bridge this potential with screening-type attractor dynamics, we specialize to the case of constant coupling $\beta(\phi)=\beta_c$. The effective potential can be written as
\begin{equation}
V_{\rm eff}(\phi;\rho_{\rm DM})= \frac{\lambda}{2q}(\phi^2-v^2)^q- \frac{\beta_c}{M_{\rm Pl}}\rho_{\rm DM}\phi,
\label{eq:Veff_specific}
\end{equation}
with the tracking condition,
\begin{equation}
\lambda\,\phi\,(\phi^2-v^2)^{q-1}- \frac{\beta_c}{M_{\rm Pl}}\rho_{\rm DM}=0.
\label{eq:ssb_balance}
\end{equation}
It follows that for $\beta_c>0$, the physical solution lies on the $\phi>0$ branch and is continuously connected to the positive vacuum minimum \citep{zhang2020obtaining, hinterbichler2010screening}. This is a consequence of the $\mathbb{Z}_2$ symmetry being broken by the DM source i.e., $(\beta_c/M_{\rm Pl})\rho_{\rm DM}\phi$ selects $\phi>0$ for $\beta_c>0$. Introducing
\begin{equation}
x \equiv \frac{\phi_{\rm ad}}{v}, \qquad
\xi(a)\equiv \frac{\beta_c \rho_{\rm DM}(a)} {\lambda v^{2q-1} M_{\rm Pl}},
\label{eq:dimensionless_defs}
\end{equation}
and substituting $\phi = v x$, the balance condition takes the dimensionless form
\begin{equation}
x\,(x^2-1)^{q-1} = \xi(a).
\label{eq:dimensionless_balance}
\end{equation}
which defines a one-dimensional density-controlled trajectory.

In the perturbatively small interaction regime, the DM density scales as $\rho_{\rm DM}\propto a^{-3}$ to leading order, giving $\xi(a)=\xi_0 a^{-3}$ with $\xi_0 \equiv \xi(a=1)$. The solution evolves monotonically from $x \gg 1$ at early times toward $x \to 1$ at late times, marking the epoch at which the scalar field approaches the symmetry-breaking minimum.

Differentiating Eq.~\eqref{eq:dimensionless_balance} with respect to $\ln a$ yields
\begin{equation}
(x^2-1)^{q-2}\left[(2q-1)x^{2}-1\right]\frac{dx}{d\ln a} = -3\xi(a),
\label{eq:dx_derivation}
\end{equation}
and eliminating $\xi(a)$ using Eq.~\eqref{eq:dimensionless_balance} gives the autonomous flow
\begin{equation}
\frac{dx}{d\ln a} = \frac{-3x\left(x^{2}-1\right)} {(2q-1)x^{2}-1}.
\label{eq:autonomous_flow_general}
\end{equation}
For $q=2$, this reduces to
\begin{equation}
\frac{dx}{d\ln a} = \frac{-3x(x^2-1)} {3x^2-1},
\label{eq:flow_quartic}
\end{equation}
with fixed points corresponding to the extrema of the effective potential.

On the physical $\phi>0$ branch, which is selected by $\beta_c>0$, the point $x=1$ is a stable attractor corresponding to the positive symmetry-breaking minimum. Linearizing Eq.~\eqref{eq:autonomous_flow_general} about $x=1+\varepsilon$ gives $d\varepsilon/d\ln a = -[3/(q-1)]\,\varepsilon+\mathcal{O}(\varepsilon^2)$, reproducing the eigenvalue $\gamma_p=-3/(q-1)$ of Eq.~\eqref{eq:n_poly}. The roots $\phi=0$ and $\phi=-v$ of the bare potential, $V'(\phi)=\lambda\phi(\phi^2-v^2)^{q-1}=0$, correspond to $x=0$ and $x=-1$; both lie outside the physical density-tracking branch $x>1$ and are not accessed by the trajectory for $\beta_c>0$. This attractor structure indeed gives a logistic-type evolution, characterized by a rapid transition near $x \sim \mathcal{O}(1)$ and asymptotic saturation at late times \citep{strogatz2024nonlinear}.

\section{Early-Time Asymptotics and Adiabatic Consistency}
\label{app:adiabatic_asymptotics}

In this Appendix, we derive the analytic high-redshift behavior of the adiabatic mass hierarchy introduced in Section~\ref{sec:covariant_framework} and verify that the posterior solutions obtained in Section~\ref{sec:observ_constraints} remain within the adiabatic tracking regime over the observationally relevant redshift range.

\subsection{Effective mass along the density-controlled trajectory}

The effective mass governing small fluctuations about the density-controlled minimum is defined as \(m_{\rm eff}^2 =\left.{\partial^2 V_{\rm eff}}/{\partial \phi^2}\right|_{\phi=\phi_{\rm ad}},\) where $V_{\rm eff}$ is given in Eq.~\eqref{eq:Veff_general}, specialized here to the potential of Eq.~\eqref{eq:poly_potential}, $V(\phi)=\frac{\lambda}{2q}(\phi^2-v^2)^q$. Differentiating $V'(\phi)=\lambda\phi(\phi^2-v^2)^{q-1}$ once more gives
\begin{equation}
m_{\rm eff}^2 = \lambda\,(\phi_{\rm ad}^2-v^2)^{q-2}\left[(2q-1)\phi_{\rm ad}^2-v^2\right].
\label{eq:meff_raw_app}
\end{equation}

Using the dimensionless ratio $x\equiv\phi_{\rm ad}/v$, Eq.~\eqref{eq:meff_raw_app} becomes
\begin{equation}
m_{\rm eff}^2 = \lambda v^{2q-2}\,(x^2-1)^{q-2}\left[(2q-1)x^2-1\right].
\label{eq:meff_dimensionless_app}
\end{equation}
Note that the DM coupling term does not contribute to $m_{\rm eff}^2$ since it is linear in $\phi$; the effective mass is therefore set entirely by the potential curvature at $\phi_{\rm ad}$. Along the adiabatic tracking trajectory, the algebraic balance condition~\eqref{eq:dimensionless_balance} reads
\begin{equation}
x\,(x^2-1)^{q-1}=\xi(a), \qquad \xi(a)=\xi_0 a^{-3},
\label{eq:tracking_app}
\end{equation}
as defined in Eq.~\eqref{eq:dimensionless_defs}.

\subsection{High-redshift asymptotics}

At sufficiently high redshift ($a\ll1$), the Universe is matter-dominated and the Hubble rate is given by \(H^2(a)\simeq H_0^2 \Omega_m a^{-3}.\) In this regime $\xi(a)\gg1$, so $x\gg1$ and the balance condition~\eqref{eq:tracking_app} is dominated by its leading power, $(x^2-1)^{q-1}\simeq x^{2q-2}$, giving \(x\cdot x^{2q-2}=x^{2q-1}\simeq \xi_0 a^{-3}.\) Solving for the field displacement yields
\begin{equation}
x(a) \simeq \xi_0^{1/(2q-1)} a^{-3/(2q-1)}.
\label{eq:x_asymptotic_app}
\end{equation}
Substituting this into Eq.~\eqref{eq:meff_dimensionless_app} and retaining only the dominant terms at large $x$, $(x^2-1)^{q-2}\simeq x^{2q-4}$ and $(2q-1)x^2-1\simeq(2q-1)x^2$, we obtain
\begin{equation}
m_{\rm eff}^2 \simeq \lambda v^{2q-2}(2q-1)\,x^{2q-2}
\simeq \lambda v^{2q-2}(2q-1)\, \xi_0^{\frac{2q-2}{2q-1}}
a^{-\frac{3(2q-2)}{2q-1}},
\label{eq:meff_asymptotic_app}
\end{equation}
which describes the early-time ($x\gg 1$) regime where the field is far from the broken minimum. This coincides with the high-redshift limit of Eq.~\eqref{eq:meff_dimensionless_app} precisely because both the compact potential and its two-term counterpart of Appendix~\ref{app:polynomial} approach the same leading behavior, $V(\phi)\to(\lambda/2q)\phi^{2q}$, for $\phi\gg v$; the two potentials differ only near the minimum, where the local restoring order $p=q-1$ is set. The late-time scaling near $x\to 1$ is given by Eq.~\eqref{eq:poly_mass_scaling} of the main text, which applies in the regime $|\delta|/v\ll 1$.

If we divide this result by the Hubble rate, the hierarchy ratio becomes
\begin{equation}
\frac{m_{\rm eff}^2}{H^2} \simeq \frac{\lambda v^{2q-2}}{H_0^2} \frac{(2q-1)}{\Omega_m} \xi_0^{\frac{2q-2}{2q-1}} a^{\frac{3}{2q-1}}.
\label{eq:ratio_asymptotic_app}
\end{equation}
Notice that this yields a universal early-time scaling,
\begin{equation}
\frac{m_{\rm eff}^2}{H^2} \propto a^{\frac{3}{2q-1}} = (1+z)^{-\frac{3}{2q-1}}.
\label{eq:final_scaling_app}
\end{equation}
Since $2q-1>0$, the exponent is positive and therefore
\begin{equation}
\lim_{z\to\infty} \frac{m_{\rm eff}^2}{H^2} =0.
\end{equation}
In other words, the strict adiabatic hierarchy $m_{\rm eff}^2 \gg H^2$ cannot hold for arbitrarily long times in the past. The Hubble rate $H^2 \propto a^{-3}$ simply grows more steeply than $m_{\rm eff}^2 \propto a^{-3 + 3/(2q-1)}$ as $a \to 0$, so the ratio inevitably approaches zero at high redshift.

We emphasize that the logistic normal form derived in Section~\ref{sec:attractor_structure} is intended to capture the late-time attractor behavior, not to serve as an exact solution at early times. This breakdown at high redshift is expected and does not affect our analysis, since the logistic parametrization is applied only over the redshift range $z \lesssim 2$ where the adiabatic condition remains satisfied. The consistency of this choice is verified numerically in Appendix~\ref{app:verification_adiabatic}.

\subsection{Consistency with BBN and recombination}
\label{app:bbn_consistency}

The early-time suppression derived in Appendix~\ref{app:adiabatic_asymptotics} has implications for compatibility with BBN and recombination constraints~\citep{cima2026probing,goh2024observational}. From Eq.~\eqref{eq:x_asymptotic_app}, the field displacement follows the exact early-time scaling $x(a)\propto a^{-3/(2q-1)}$ at high redshift, so that $y=1/x\propto a^{3/(2q-1)}$. This scaling is distinct from the near-attractor exponent $n=3/p$, which governs the logistic normal form only over the observationally relevant regime $0\le z\le2$ and should not be extrapolated to the early Universe. For the radiatively stable $q=2$ universality class, the early-time solution reduces to $y\propto a$.

Using the tracking solution defined by Eq.~\eqref{eq:tracking_app}, we evaluate the interaction strength at recombination ($a_{\rm rec}=9.08\times10^{-4}$, $z_{\rm rec}=1100$) and during BBN ($a_{\rm BBN}=2.35\times10^{-10}$, corresponding to $T\sim1\,\mathrm{MeV}$). The calculation is performed for $\xi_0$ values corresponding to the $68\%$ posterior interval of the activation epoch (Table~\ref{tab:ide_6p_n3_comparison}, $a_c\in[0.38,0.90]$), using the relation $y(a_c)=1/2\Leftrightarrow\xi_0=2\cdot3^{q-1}a_c^3$.
This yields
\begin{equation}
\frac{\beta_{\rm eff}(a_{\rm rec})}{\beta_0}
\sim6\times10^{-4}\text{--}2\times10^{-3},
\qquad
\frac{\beta_{\rm eff}(a_{\rm BBN})}{\beta_0}
\sim1\times10^{-10}\text{--}4\times10^{-10},
\label{eq:bbn_suppression}
\end{equation}
across the posterior range. The corresponding cumulative background deformation $\int \epsilon\, d\ln a,$ remains many orders of magnitude below current observational sensitivity at both epochs.

These results should be interpreted as an order-of-magnitude consistency check rather than a dedicated confrontation with BBN or recombination data. A quantitative analysis would require specifying the global form of the conformal coupling function $\beta(\phi)$, rather than only its local behavior near the symmetry-broken minimum, and is therefore left for future work.
\subsection{Consistency over the observational window}
\label{app:verification_adiabatic}

\begin{figure*}[t]
\centering

\subfloat[]{
\includegraphics[width=0.45\textwidth]{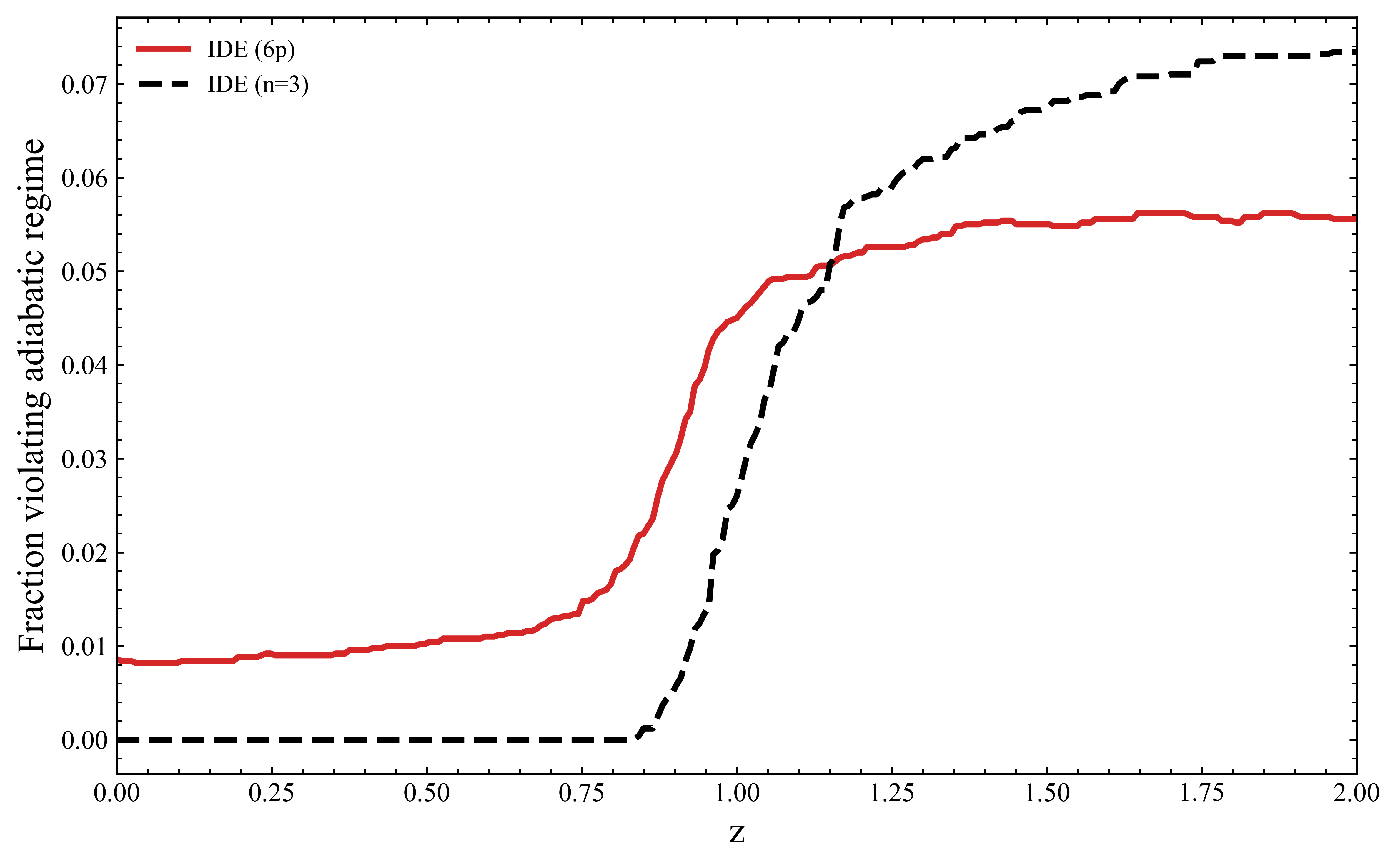}
\label{fig:adiabatic_violation}
}
\hfill
\subfloat[]{
\includegraphics[width=0.45\textwidth]{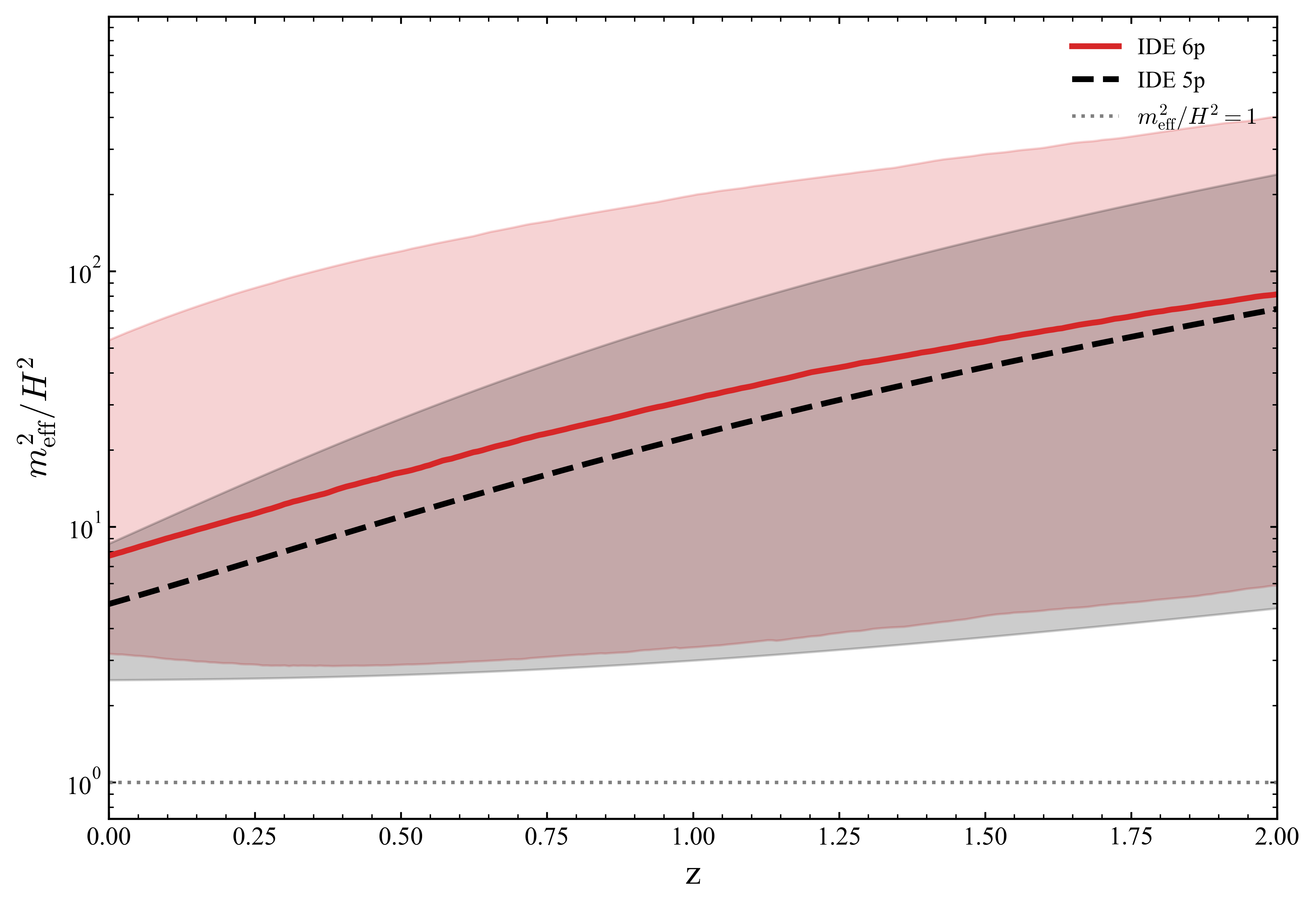}
\label{fig:adiabatic_ratio}
}

\caption{\small 
Adiabatic tracking diagnostics from Planck+RSD+SN posterior samples. Left panel: fraction of samples violating $m_{\rm eff}^2/H^2>1$. Right panel: median and $68\%$ credible interval of $m_{\rm eff}^2/H^2$. The ratio decreases toward high redshift, consistent with Eq.~\eqref{eq:final_scaling_app}. Across the range $0 \leq z \leq 2$, the solutions remain predominantly adiabatic and follow the tracking approximation.
}
\label{fig:adiabatic_diagnostics}

\end{figure*}

Here, we numerically show that the posterior solutions analyzed in Sec .~\ref {sec:observ_constraints} remain within the adiabatic tracking regime. We reconstruct the ratio $m_{\rm eff}^2/H^2$ using Eq.~\eqref{eq:meff_dimensionless_app}, evaluated along the tracking solution of Eq.~\eqref{eq:tracking_app}, for each accepted parameter vector in the full MCMC chains for the Planck+RSD+SN dataset combination in Figure~\ref{fig:adiabatic_diagnostics}: (i) the redshift evolution of the fraction of posterior samples violating the adiabatic condition $m_{\rm eff}^2/H^2>1$, and (ii) the median and $68\%$ credible region of the hierarchy ratio. Moreover, the Planck+RSD+SN combination provides the most conservative test of the adiabatic hierarchy, while the RSD-only and Planck+RSD cases likewise satisfy the hierarchy, with only a small fraction of samples approaching the boundary.

At $z=0$, all posterior samples satisfying the perturbative filter $|\epsilon(a)|<10^{-2}$ also satisfy $m_{\rm eff}^2/H^2>1$ by construction, since the filter requires $m_{\rm eff}^2\gtrsim \mathcal{O}(10^2)\beta_0^2 H^2$ from Eq.~\eqref{eq:mass_hierarchy_position}. The non-trivial test is at $z\sim 2$, where the ratio $m_{\rm eff}^2/H^2$ is smallest. Figure~\ref{fig:adiabatic_diagnostics} shows that the median ratio remains above unity over the full range $0\leq z\leq 2$, indicating the self-consistency of the adiabatic tracking approximation across the observational window. A minority of samples approach the boundary $m_{\rm eff}^2/H^2\simeq1$ at the high-redshift end, but no systematic breakdown of the hierarchy is observed. We stress that this adiabatic condition ($m_{\rm eff}^2/H^2 \gg 1$) governs the validity of the algebraic tracking reduction, whereas the filter $|\epsilon(a)| < \epsilon_\star$ separately ensures that the background deformation remains small.

\section{Convergence Diagnostics and Robustness Tests}
\label{sec:diagnostics}
\subsection{MCMC Convergence Diagnostics}
\label{sec:mcmc_diagnostics}

Given the boundary-hugging character of the fixed-index posterior (Section~\ref{subsec:activation_shape_tension}), a regime known to challenge ensemble-sampler mixing, we report explicit convergence diagnostics for all six emcee runs (two model realizations $\times$ three dataset combinations). For each dataset--model combination, two independently initialized chains (32 walkers each) were run. Post burn-in ($N_{\rm burn}=300$), we compute the integrated autocorrelation time $\tau$, the split-$\hat R$ statistic, and the classic Gelman--Rubin $\hat R$ for every sampled parameter, together with a two-sample Kolmogorov--Smirnov comparison of the marginalized posterior between the two independent replicates.

Table~\ref{tab:mcmc_diagnostics} summarizes the results. For five of the six combinations, both replicates pass a split-$\hat R < 1.3$ gate and agree closely: the maximum KS statistic across all sampled parameters is $D \le 0.10$ in every case, with $68\%$ credible intervals overlapping for every parameter, indicating that independent chains converge to statistically indistinguishable posteriors despite $N_{\rm steps}/\tau$ remaining below the conservative $\gtrsim 50$ threshold recommended for autocorrelation-time reliability.

For the fixed-index model under Planck+RSD+SN --- the combination with the most extreme boundary pileup (Table~\ref{tab:collapse_diagnostics}) --- one of the two replicates failed the convergence gate (split-$\hat R$ up to $5.5$, driven by persistent disagreement between walkers rather than a stable but shifted mode) and was excluded from the reported posterior. The surviving replicate's marginal posteriors were verified to show a monotonic increase in density up to the prior boundary with no discontinuity or secondary structure, consistent with a genuine likelihood-driven boundary pileup rather than a sampler artifact at the wall.

\begin{table*}[ht]
\centering
\caption{\small MCMC convergence diagnostics for all six IDE emcee runs. $N_{\rm w}$: walkers; $N_{\rm burn}$: discarded burn-in steps; $\bar\tau$: mean integrated autocorrelation time across sampled parameters and kept replicates; $N_{\rm steps}/\tau$: post-burn-in steps per autocorrelation time (range across kept replicates); $\hat R_{\rm split}$, $\hat R_{\rm classic}$: maximum values across parameters and kept replicates; KS $D$: maximum two-sample Kolmogorov--Smirnov statistic across parameters between the two independent replicates (where only one replicate survived the convergence gate).}
\label{tab:mcmc_diagnostics}
\setlength{\tabcolsep}{5pt}
\renewcommand{\arraystretch}{1.3}
\begin{tabular}{llcccccc}
\toprule
Dataset & Model & $N_{\rm w}$ & $N_{\rm burn}$ & $\bar\tau$ & $N_{\rm steps}/\tau$ & $\hat R_{\rm split}$ & $\hat R_{\rm classic}$ \\
\midrule
RSD only        & IDE~(6p)   & 32 & 300 & 370 & 19.7--19.8 & 1.116 & 1.064 \\
RSD only        & IDE~($n{=}3$)  & 32 & 300 & 140 & 16.7--22.4 & 1.185 & 1.087 \\
Planck+RSD      & IDE~(6p)   & 32 & 300 & 178 & 15.1--15.6 & 1.203 & 1.110 \\
Planck+RSD      & IDE~($n{=}3$)  & 32 & 300 & 147 & 17.6--19.3 & 1.132 & 1.064 \\
Planck+RSD+SN   & IDE~(6p)   & 32 & 300 & 126 & 17.2--18.0 & 1.156 & 1.082 \\
Planck+RSD+SN   & IDE~($n{=}3$)  & 32$^{a}$ & 300 & 151 & 14.7 & 1.125 & 1.013 \\
\bottomrule
\end{tabular}
\begin{minipage}{0.95\linewidth}
\footnotesize
$^{a}$One of two replicates for this combination failed the split-$\hat R<1.3$ convergence gate (max $\hat R_{\rm split}=5.5$) and was excluded; reported diagnostics are for the single surviving, converged replicate.
\end{minipage}
\end{table*}

\subsection{Robustness to Prior Choices and Perturbative Filtering}
\label{subsec:prior_robustness}

In Section~\ref{subsec:model_hierarchy}, we noted that the posterior support for the activation index $n$ extends to the upper prior boundary at $n=5$, reflecting the limited sensitivity of the current data to steep activation histories rather than a genuine preference for arbitrarily large $n$. To test this interpretation, we examine the shape of the marginalized $n$ posterior in the top decile of the prior range, $n\in[4.55,5.0]$. If the boundary at $n=5$ were truncating a genuinely rising, likelihood-preferred density --- that is, if the unconstrained posterior favored $n>5$ --- the marginal density would increase monotonically into the wall, and the fraction of samples in this decile would exceed the $10\%$ expected under a locally flat, prior-dominated distribution.

\begin{table}[ht]
\centering
\caption{\small Posterior behavior near the upper prior boundary of the activation index $n$ for the flexible six-parameter (6p) model. Shown are the posterior densities in the last four histogram bins ($n\in[4.1,5.0]$, bin width $0.225$) together with the fraction of posterior mass in the top decile of the prior range, $n\in[4.55,5.0]$. A uniform (fully prior-dominated) marginal would yield a density of $1/(5.0-0.5)\approx0.222$ in every bin and exactly $10\%$ of the posterior mass in the top decile.}
\label{tab:prior_robustness}
\setlength{\tabcolsep}{6pt}
\renewcommand{\arraystretch}{1.3}
\begin{tabular}{lccccc}
\toprule
Dataset & \multicolumn{4}{c}{Density, last 4 bins ($n\to5.0$)} & \% in top decile \\
\midrule
RSD only        & 0.112 & 0.132 & 0.112 & 0.121 & $5.9\%$ \\
Planck+RSD      & 0.128 & 0.131 & 0.139 & 0.134 & $6.9\%$ \\
Planck+RSD+SN   & 0.134 & 0.131 & 0.130 & 0.139 & $6.8\%$ \\
\bottomrule
\end{tabular}
\end{table}
The posterior exhibits no evidence of accumulating against the upper prior boundary on the activation index. As summarized in Table~\ref{tab:prior_robustness}, the marginal density over the last four histogram bins ($n\in[4.1,5.0]$, bin width $0.225$) remains flat to mildly non-monotonic, at roughly half the density ($0.11$--$0.14$) expected for a strictly uniform marginal over $n\in[0.5,5.0]$ ($1/4.5\simeq0.222$). The fraction of posterior mass contained in the top decile is only $5.9\%$--$6.9\%$, well below the $10\%$ benchmark, indicating that extending the prior beyond $n=5$ would not qualitatively alter the inferred constraints on the activation index.

As reported in Table~\ref{tab:ablation}, relaxing the threshold from $\epsilon_\star=5\times10^{-3}$ to $2\times10^{-2}$ increases the acceptance fraction from $0.33$ to $0.56$, while leaving the marginalized constraints on the background cosmological parameters $(\Omega_m,h,\sigma_8)$ statistically unchanged. The corresponding comparison of the interaction-sector parameters in Table~\ref{tab:ablation_interaction} demonstrates that the perturbative filter primarily affects the interaction history rather than the background cosmology. Across all three dataset combinations, the $68\%$ credible interval for $\beta_0$ is reduced by approximately $44$--$45\%$, while the preferred activation history shifts toward larger $a_c$ and smaller $n$.

This behavior follows directly from the mass-hierarchy relation, according to which the perturbative parameter $\epsilon(a)$ depends explicitly on $\beta_0$ and on the ratio $m_{\rm eff}(a_c,n)/H(a_c)$. Models with an earlier activation epoch (small $a_c$), where the adiabatic mass hierarchy is intrinsically weaker, and models with a steeper activation profile (large $n$), which generate a larger instantaneous $\dot\phi$ near $a_c$, are both more likely to violate $|\epsilon(a)|<\epsilon_\star$. Imposing $\epsilon(a)<\epsilon_\star$ therefore preferentially removes models with stronger coupling and earlier, faster symmetry restoration, while leaving the inferred background cosmology essentially unchanged since $\epsilon(a)$ has no direct dependence on $(\Omega_m,h,\sigma_8)$.

\begin{table*}[t]
\centering
\caption{ \small Robustness of the inferred cosmological parameters to the perturbative consistency filter for the flexible six-parameter (6p) IDE model. For the Planck+RSD+SN dataset we compare the posterior obtained after convergence selection only (Raw) with the filtered posterior for three values of the perturbative threshold $\epsilon_\star$. The inferred background parameters remain statistically stable, indicating that the perturbative filter primarily restricts the interaction sector while leaving the inferred background cosmology robust.}
\label{tab:ablation}
\renewcommand{\arraystretch}{1.25}
\setlength{\tabcolsep}{8pt}
\begin{tabular}{lcccc}
\toprule
Stage &
Acceptance &
$\Omega_m$ &
$h$ &
$\sigma_8$ \\
\midrule
Raw (no $\epsilon$ filter)
&
1.000
&
$0.3613^{+0.0182}_{-0.0193}$
&
$0.7053^{+0.0957}_{-0.1047}$
&
$0.8073^{+0.1327}_{-0.1414}$
\\

$\epsilon_\star=5\times10^{-3}$
&
0.334
&
$0.3617^{+0.0178}_{-0.0194}$
&
$0.7098^{+0.0936}_{-0.1076}$
&
$0.8079^{+0.1317}_{-0.1396}$
\\

$\epsilon_\star=10^{-2}$
&
0.437
&
$0.3614^{+0.0180}_{-0.0194}$
&
$0.7095^{+0.0937}_{-0.1076}$
&
$0.8071^{+0.1324}_{-0.1379}$
\\

$\epsilon_\star=2\times10^{-2}$
&
0.559
&
$0.3611^{+0.0181}_{-0.0191}$
&
$0.7091^{+0.0937}_{-0.1059}$
&
$0.8076^{+0.1327}_{-0.1405}$
\\
\bottomrule
\end{tabular}
\end{table*}
\begin{table*}[t]
\centering
\caption{\small Comparison of the raw (R\^R-gated) and perturbatively filtered ($\epsilon_\star=10^{-2}$) posterior constraints on the IDE parameters $(\beta_0,a_c,n)$ for the flexible (6p) model, shown for all three dataset combinations.}
\label{tab:ablation_interaction}
\renewcommand{\arraystretch}{1.25}
\setlength{\tabcolsep}{6pt}
\begin{tabular}{llccc}
\toprule
Dataset & Stage & $\beta_0$ & $a_c$ & $n$ \\
\midrule
\multirow{2}{*}{RSD only}
 & Raw      & $-0.0052^{+0.2008}_{-0.2019}$ & $0.5264^{+0.3145}_{-0.2912}$ & $2.649^{+1.593}_{-1.482}$ \\
 & Filtered & $-0.0014^{+0.1131}_{-0.1119}$ & $0.6671^{+0.2352}_{-0.2937}$ & $1.751^{+1.956}_{-0.928}$ \\
\midrule
\multirow{2}{*}{Planck+RSD}
 & Raw      & $\phantom{-}0.0086^{+0.1942}_{-0.2000}$ & $0.5216^{+0.3126}_{-0.2819}$ & $2.689^{+1.615}_{-1.486}$ \\
 & Filtered & $-0.0023^{+0.1037}_{-0.1121}$ & $0.6684^{+0.2308}_{-0.2821}$ & $1.911^{+1.894}_{-1.073}$ \\
\midrule
\multirow{2}{*}{Planck+RSD+SN}
 & Raw      & $\phantom{-}0.0144^{+0.1945}_{-0.2086}$ & $0.5297^{+0.3198}_{-0.2835}$ & $2.687^{+1.571}_{-1.518}$ \\
 & Filtered & $\phantom{-}0.0086^{+0.1154}_{-0.1094}$ & $0.6674^{+0.2361}_{-0.2848}$ & $1.914^{+1.911}_{-1.085}$ \\
\bottomrule
\end{tabular}
\end{table*}

\section{ Accuracy of the Global Logistic Parametrization}
\label{app:logistic_accuracy}

The logistic coupling profile $\beta_{\rm eff}(a) = \beta_0(a/a_c)^n/[1+(a/a_c)^n]$ is derived in Section~\ref{sec:logistic_activation} as a local normal form valid asymptotically near the broken minimum. In the likelihood analysis of Section~\ref{sec:observ_constraints}, this profile is applied across $0 \leq z \leq 2$ with $(a_c, n)$ as free parameters. Here we validate this by comparing the logistic family against the exact numerical solution of the balance condition,
\begin{equation}
    x\,(x^2-1)^{q-1} = \xi_0\, a^{-3},
    \label{eq:balance_C}
\end{equation}
where $x(a) \equiv \phi_{\rm ad}(a)/v$ and
\begin{equation}
    \xi_0 \equiv \frac{\beta_c\,\rho_{{\rm DM},0}}
    {\lambda\, v^{2q-1}\,M_{\rm Pl}}
    \label{eq:xi0_C}
\end{equation}
parametrizes the interaction strength. The order parameter is 
$y_{\rm exact}(a) = 1/x(a)$.

We solve Eq.~\eqref{eq:balance_C} numerically for $q = 2, 3, 4$ and $\xi_0 \in \{0.1, 0.3, 0.5, 1.0\}$, covering the posterior-relevant range. The logistic form,
\begin{equation}
    y_{\rm logistic}(a;\,a_c,n) = \frac{(a/a_c)^n}{1 + (a/a_c)^n},
    \label{eq:logistic_fit_C}
\end{equation}
is fitted to $y_{\rm exact}(a)$ by minimizing the least-squares residual
\begin{equation}
    \mathcal{R}(a_c, n) = \sum_{i}
    \left[y_{\rm logistic}(a_i;\,a_c,n) 
    - y_{\rm exact}(a_i)\right]^2
    \label{eq:residual_C}
\end{equation}
over $500$ uniformly spaced points in $a \in [1/3,\,1]$. This allows both $a_c$ and $n$ to vary freely, testing the accuracy of the logistic \emph{functional form} rather than of the asymptotic index $n_{\rm theory} = 3/p$ at finite redshift. The maximum fractional deviation is defined as
\begin{equation}
    \Delta_{\rm max} \equiv \max_{a\,\in\,[1/3,\,1]}
    \left|\frac{y_{\rm logistic}(a) - y_{\rm exact}(a)}
    {y_{\rm exact}(a)}\right|.
    \label{eq:delta_max_C}
\end{equation}
The results are collected in Table~\ref{tab:logistic_validation}.

\begin{table}[htbp]
\centering
\renewcommand{\arraystretch}{1.4}
\caption{\small Least-squares comparison between the logistic parametrization and the exact numerical solution of Eq.~\eqref{eq:balance_C} for $q = 2,3,4$  and representative values of $\xi_0$. The best-fit parameters $(a_c^{\rm fit},  n_{\rm fit})$ minimize the residual~\eqref{eq:residual_C} over $0 \leq z \leq 2$. The maximum absolute coupling error is evaluated at $\beta_0 = 0.15$.}
\label{tab:logistic_validation}
\begin{tabular}{ccccccc}
\hline\hline
$q$ & $n_{\rm theory}$ & $\xi_0$ & $a_c^{\rm fit}$ & 
$n_{\rm fit}$ & $\Delta_{\rm max}$ (\%) & 
$|\Delta\beta_{\rm eff}|_{\rm max}$ \\
\hline
\multirow{4}{*}{2} & \multirow{4}{*}{3.00}
  & 0.10 & 0.28 & 2.30 & 2.60 & $1.11\times10^{-2}$ \\
  & & 0.30 & 0.38 & 2.01 & 4.54 & $1.93\times10^{-2}$ \\
  & & 0.50 & 0.44 & 1.88 & 5.17 & $2.20\times10^{-2}$ \\
  & & 1.00 & 0.54 & 1.71 & 5.56 & $2.36\times10^{-2}$ \\
\hline
\multirow{4}{*}{3} & \multirow{4}{*}{1.50}
  & 0.10 & 0.20 & 1.21 & 0.76 & $5.68\times10^{-3}$ \\
  & & 0.30 & 0.27 & 1.13 & 1.08 & $8.32\times10^{-3}$ \\
  & & 0.50 & 0.31 & 1.09 & 1.21 & $9.50\times10^{-3}$ \\
  & & 1.00 & 0.39 & 1.04 & 1.35 & $1.09\times10^{-2}$ \\
\hline
\multirow{4}{*}{4} & \multirow{4}{*}{1.00}
  & 0.10 & 0.14 & 0.82 & 0.34 & $3.66\times10^{-3}$ \\
  & & 0.30 & 0.19 & 0.79 & 0.44 & $4.85\times10^{-3}$ \\
  & & 0.50 & 0.22 & 0.77 & 0.49 & $5.41\times10^{-3}$ \\
  & & 1.00 & 0.27 & 0.75 & 0.54 & $6.13\times10^{-3}$ \\
\hline\hline
\end{tabular}
\end{table}

The logistic functional form reproduces the exact tracking solution to within $5.6\%$ for $q=2$, $1.4\%$ for $q=3$, and $0.5\%$ for $q=4$ across all tested values of $\xi_0$. This shows that the logistic family is an accurate representation of the exact dynamics when $(a_c,n)$ are free and increasingly so for flatter potentials, whose slower approach to the attractor produces a more gradual crossover that a single logistic function captures more precisely.

Two further results follow from Table~\ref{tab:logistic_validation}. First, the best-fit index $n_{\rm fit}$ is systematically below $n_{\rm theory} = 3/p$, with the deficit decreasing as $\xi_0 \to 0$. This is a pre-asymptotic correction: $n_{\rm theory}$ is the exact attractor eigenvalue only in the limit $\phi_{\rm ad} \to v$, while at finite redshift nonlinear corrections to $V'(\phi)$ shift the effective steepness below its asymptotic value. Since the flexible 6p model treats $n$ as a free parameter, the MCMC naturally absorbs this correction without bias. Second, the maximum absolute coupling error,
\begin{equation}
    |\Delta\beta_{\rm eff}|_{\rm max} 
    = \beta_0\,\Delta_{\rm max}\,
    \max_a |y_{\rm exact}(a)|,
    \label{eq:abs_err_C}
\end{equation}
is bounded by $2.36\times10^{-2}$ for $q=2$ and below $1.1\times10^{-2}$ for $q=3,4$ at $\beta_0 = 0.15$. Since the IDE signal in $f\sigma_8$ is suppressed to $\Delta f\sigma_8/f\sigma_8 \lesssim 0.3\%$ at this amplitude, the coupling error propagates to at most
\begin{equation}
    \delta\!\left(\frac{\Delta f\sigma_8}{f\sigma_8}\right)
    \approx \frac{2.36\times10^{-2}}{0.15}\times 0.3\%
    \approx 0.05\%,
    \label{eq:fsig8_err_C}
\end{equation}
which is two orders of magnitude below current observational precision. The observational conclusions of Section~\ref{sec:observ_constraints} are therefore robust to the extrapolation of the logistic parametrization.

\section{Signal-to-Noise Estimates for Euclid, LSST, and Weak Lensing}
\label{app:forecast}

\subsection{Methodology}
\label{app:wl_forecast}
The growth observable is constructed from the \texttt{CLASS} output as \(f\sigma_8(z)=f(z)\,D(z)\,\sigma_8,\) where \(f(z)\) is the growth rate and \(D(z)\) is the linear growth factor, normalized to unity at \(z=0\). The present-day clustering amplitude, \(\sigma_8\), is held fixed at the same fiducial value in both \(\Lambda\)CDM and IDE. The signal is evaluated at representative redshift points spanning \(z \approx 0.5\)-\(1.5\) for Euclid~\citep{scaramella2022euclid} and \(z \approx 0.3\)-\(1.1\) for LSST~\citep{ivezic2019lsst}, corresponding to the redshift range where growth measurements are most sensitive. These points are used as a simplified proxy for survey binning rather than a full survey binning scheme.

We consider two fiducial IDE scenarios: the full six-parameter (6p) model constrained by Planck+RSD+SN, and the RSD-only variant that maximizes sensitivity to growth by relaxing geometric constraints. For each model, the predicted \(f\sigma_8(z)\) is interpolated and evaluated at the representative survey bins. The conservative fractional uncertainties on \(f\sigma_8(z)\) of \(1\%\) for Euclid and \(2\%\) for LSST per redshift bin, assuming diagonal covariance.  The  SNR is defined as~\citep{tegmark1997karhunen}
\begin{equation}
\mathrm{SNR} = \sqrt{\sum_i \frac{\left[\Delta f\sigma_8(z_i)\right]^2}{\sigma_i^2}},
\end{equation}
where \(\Delta f\sigma_8 \equiv f\sigma_8^{\mathrm{IDE}} - f\sigma_8^{\Lambda\mathrm{CDM}}\) and \(\sigma_i\) is the forecasted uncertainty in the \(i\)-th redshift bin. This simplification ignores bin-to-bin correlations and cosmic variance, and the resulting SNR should be viewed as an upper bound on detectability.

In addition to the growth-rate observable, we estimate the weak-lensing signal using the convergence power spectrum $C_\ell^{\kappa\kappa}$ computed under the Limber approximation,
\begin{equation}
C_\ell^{\kappa\kappa} = \int_0^{\chi_{\rm max}} \frac{W^2(\chi)}{\chi^2}
P_{\rm m}\!\left(\frac{\ell+0.5}{\chi}, z(\chi)\right)\, d\chi,
\end{equation}
where $P_{\rm m}(k,z)$ is extracted directly from the \texttt{CLASS} matter power spectrum outputs at discrete redshift slices, and $W(\chi)$ is the lensing efficiency kernel; taken here in the single-source-plane approximation. For this forecast, we adopt a simplified single effective source plane at $z_{\rm max}=2.5$, corresponding to a conservative approximation of an LSST-like survey ($A_{\rm sky}=14000\,{\rm deg}^2$, $n_{\rm gal}=20\,{\rm arcmin}^{-2}$).

The covariance of the lensing power spectrum is approximated as Gaussian,
\begin{equation}
{\rm Var}(C_\ell^{\kappa\kappa}) = 
\frac{2\left(C_\ell^{\kappa\kappa}+N_\ell\right)^2}
{(2\ell+1)\,f_{\rm sky}},
\end{equation}
where $N_\ell=\sigma_\epsilon^2/n_{\rm gal}$ is the shape-noise power spectrum. The corresponding SNR is defined as~\citep{hoekstra2008weak}
\begin{equation}
\mathrm{SNR}_{\rm WL} = \left[\sum_\ell \frac{(\Delta C_\ell)^2}{\mathrm{Var}(C_\ell)}\right]^{1/2}.
\end{equation}
This estimate neglects tomographic binning and cross-correlations, and therefore provides a conservative lower bound on the sensitivity of weak lensing to the IDE signal.

\subsection{Intrinsic deviations in the growth sector}

Before examining the observable combinations, we quantify the intrinsic modifications to the growth dynamics. We define $\Delta X \equiv X_{\mathrm{IDE}} - X_{\Lambda\mathrm{CDM}}$ and compute relative deviations using the \texttt{CLASS} background outputs over the redshift range relevant for large-scale structure ($0 \lesssim z \lesssim 2$). The linear growth factor $D(z)$ is normalized to unity at $z=0$. The maximum relative deviations between the IDE model and $\Lambda$CDM are found to be
\begin{align}
\max\left|\frac{\Delta D}{D}\right| &\approx 1.5\%,\\
\max\left|\frac{\Delta f}{f}\right| &\approx 1.2\%.
\end{align}

These percent-level differences show that the interaction induces genuine modifications to the growth dynamics. At the same time, the evolution remains strongly correlated with $\Lambda$CDM, so the two models remain highly degenerate when considering the individual growth observables $D(z)$ and $f(z)$ at current data precision. Moreover, consistency with the growth relation \(f(z)=d\ln D/d\ln a\) is satisfied to high precision with the average absolute difference between the growth rate obtained by numerical differentiation of $D(z)$ and that reported directly by \texttt{CLASS} being \(\langle |f_{\rm num}-f_{\rm CLASS}|\rangle \sim 10^{-7}\). This confirms that the deviations are physical and not numerical artifacts.

\subsection{Suppression in the observable \texorpdfstring{$f\sigma_8(z)$}{fσ8(z)}}

Despite the percent-level modifications in $D(z)$ and $f(z)$ identified in the previous subsection, the combined observable is strongly suppressed. Over the redshift $0 \lesssim z \lesssim 2$, the fiducial Planck+RSD+SN case yields a best-fit deviation of
\begin{equation}
\frac{\Delta f\sigma_8}{f\sigma_8} \approx 0.034\% \text{ (Euclid)} \quad \text{and} \quad 0.058\% \text{ (LSST)},
\end{equation}
with a posterior mean of $0.082\%$ and a $95\%$ upper bound of $0.29\%$. For the RSD-only scenario, the best-fit deviation rises to $0.20\%$ (Euclid) and $0.31\%$ (LSST), with a posterior mean of $0.40\%$. This suppression follows from the opposite-sign structure of $\Delta D/D$ and $\Delta f/f$, whose fractional modifications enter with opposite sign and largely offset in the product $f\sigma_8(z)$. The effect is therefore strongest in the observable combination, even though the underlying growth equations are genuinely modified. This understanding is reinforced by the small coupling amplitude favored by current constraints and by the fact that \(\sigma_8\) is held fixed between the IDE and \(\Lambda\)CDM predictions. The resulting near-degeneracy explains why the model remains difficult to distinguish from \(\Lambda\)CDM in large-scale structure observables.

\section{Declaration of Generative AI and AI-assisted technologies in the writing process}
During the preparation of this work, the author, Pradosh Keshav, used an AI/LLM to correct grammatical mistakes and improve the structure of specific paragraphs. After using this tool/service, the author reviewed and edited the content as needed and takes full responsibility for the content of the publication.

\bibliography{main}{}

@article{avila2022inferring,
  title={Inferring S 8 (z) and $\gamma$ (z) with cosmic growth rate measurements using machine learning},
  author={Avila, Felipe and Bernui, Armando and Bonilla, Alexander and Nunes, Rafael C},
  journal={The European Physical Journal C},
  volume={82},
  number={7},
  pages={594},
  year={2022},
  publisher={Springer}
}

@article{d2016quantum,
  title={Quantum field theory of interacting dark matter and dark energy: Dark monodromies},
  author={D’Amico, Guido and Hamill, Teresa and Kaloper, Nemanja},
  journal={Physical Review D},
  volume={94},
  number={10},
  pages={103526},
  year={2016},
  publisher={APS}
}

@article{di2017can,
  title={Can interacting dark energy solve the H 0 tension?},
  author={Di Valentino, Eleonora and Melchiorri, Alessandro and Mena, Olga},
  journal={Physical Review D},
  volume={96},
  number={4},
  pages={043503},
  year={2017},
  publisher={APS}
}

@article{di2021interacting,
  title={Interacting Dark Energy in a closed universe},
  author={Di Valentino, Eleonora and Melchiorri, Alessandro and Mena, Olga and Pan, Supriya and Yang, Weiqiang},
  journal={Monthly Notices of the Royal Astronomical Society: Letters},
  volume={502},
  number={1},
  pages={L23--L28},
  year={2021},
  publisher={Oxford University Press}
}

@article{di2020nonminimal,
  title={Nonminimal dark sector physics and cosmological tensions},
  author={Di Valentino, Eleonora and Melchiorri, Alessandro and Mena, Olga and Vagnozzi, Sunny},
  journal={Physical Review D},
  volume={101},
  number={6},
  pages={063502},
  year={2020},
  publisher={APS}
}

@article{yang2018interacting,
  title={Interacting dark energy with time varying equation of state and the H 0 tension},
  author={Yang, Weiqiang and Mukherjee, Ankan and Di Valentino, Eleonora and Pan, Supriya},
  journal={Physical Review D},
  volume={98},
  number={12},
  pages={123527},
  year={2018},
  publisher={APS}
}

@article{amendola2000coupled,
  title={Coupled quintessence},
  author={Amendola, Luca},
  journal={Physical Review D},
  volume={62},
  number={4},
  pages={043511},
  year={2000},
  publisher={APS}
}

@article{hui2017ultralight,
  title={Ultralight scalars as cosmological dark matter},
  author={Hui, Lam and Ostriker, Jeremiah P and Tremaine, Scott and Witten, Edward},
  journal={Physical Review D},
  volume={95},
  number={4},
  pages={043541},
  year={2017},
  publisher={APS}
}

@article{farrar2004interacting,
  title={Interacting dark matter and dark energy},
  author={Farrar, Glennys R and Peebles, P James E},
  journal={The Astrophysical Journal},
  volume={604},
  number={1},
  pages={1},
  year={2004},
  publisher={IOP Publishing}
}

@article{wetterich1988cosmology,
  title={Cosmology and the fate of dilatation symmetry},
  author={Wetterich, Christof},
  journal={Nuclear Physics B},
  volume={302},
  number={4},
  pages={668--696},
  year={1988},
  publisher={Elsevier}
}

@article{valiviita2008large,
  title={Large-scale instability in interacting dark energy and dark matter fluids},
  author={V{\"a}liviita, Jussi and Majerotto, Elisabetta and Maartens, Roy},
  journal={Journal of Cosmology and Astroparticle Physics},
  volume={2008},
  number={07},
  pages={020},
  year={2008},
  publisher={IOP Publishing}
}

@article{bean2008_prd,
  author = "R. Bean and E. E. Flanagan and M. Trodden",
  title = "{Adiabatic instability in coupled dark energy-dark matter models: constraints and implications}",
  journal = "Phys. Rev. D",
  volume = "78",
  pages = "023009",
  year = "2008",
  eprint = "arXiv:0709.1128",
  archivePrefix = "arXiv",
  primaryClass = "astro-ph",
}

@article{keshav2024interacting,
  title = {Interacting Dark Energy and Its Implications for Unified Dark Sector},
  author = {Keshav MV, P. and Arun, K.},
  journal = {International Journal of Theoretical Physics},
  volume = {63},
  pages = {265},
  year = {2024},
  doi = {10.1007/s10773-024-05794-6}
}

@article{tan2019dark,
  title={Dark energy and spontaneous mirror symmetry breaking},
  author={Tan, Wanpeng},
  journal={arXiv preprint arXiv:1908.11838},
  year={2019}
}

@article{aghanim2020planck,
  title={Planck 2018 results-VI. Cosmological parameters},
  author={Aghanim, Nabila and Akrami, Yashar and Ashdown, Mark and Aumont, Jonathan and Baccigalupi, Carlo and Ballardini, Mario and Banday, Anthony J and Barreiro, RB and Bartolo, Nicola and Basak, S and others},
  journal={Astronomy \& Astrophysics},
  volume={641},
  pages={A6},
  year={2020},
  publisher={EDP sciences}
}

@article{katz2014higgs,
  title={Higgs couplings and electroweak phase transition},
  author={Katz, Andrey and Perelstein, Maxim},
  journal={Journal of High Energy Physics},
  volume={2014},
  number={7},
  pages={1--24},
  year={2014},
  publisher={Springer}
}

@article{adame2024desi,
  title={DESI 2024 VII: cosmological constraints from the full-shape modeling of clustering measurements},
  author={Adame, AG and Aguilar, J and Ahlen, S and Alam, S and Alexander, DM and Allende Prieto, C and Alvarez, M and Alves, O and Anand, A and Andrade, U and others},
  journal={Journal of Cosmology and Astroparticle Physics},
  volume={2025},
  number={07},
  pages={028},
  year={2025},
  publisher={IOP Publishing}
}

@article{hlovzek2017future,
  title={Future CMB tests of dark matter: Ultralight axions and massive neutrinos},
  author={Hlo{\v{z}}ek, Ren{\'e}e and Marsh, David JE and Grin, Daniel and Allison, Rupert and Dunkley, Jo and Calabrese, Erminia},
  journal={Physical Review D},
  volume={95},
  number={12},
  pages={123511},
  year={2017},
  publisher={APS}
}

@article{wang2024further,
  title={Further understanding the interaction between dark energy and dark matter: current status and future directions},
  author={Wang, Bin and Abdalla, Elcio and Atrio-Barandela, Fernando and Pavon, Diego},
  journal={Reports on Progress in Physics},
  year={2024},
  publisher={IOP Publishing}
}

@article{riess2022comprehensive,
  title={A comprehensive measurement of the local value of the Hubble constant with 1 km s- 1 Mpc- 1 uncertainty from the Hubble Space Telescope and the SH0ES team},
  author={Riess, Adam G and Yuan, Wenlong and Macri, Lucas M and Scolnic, Dan and Brout, Dillon and Casertano, Stefano and Jones, David O and Murakami, Yukei and Anand, Gagandeep S and Breuval, Louise and others},
  journal={The Astrophysical journal letters},
  volume={934},
  number={1},
  pages={L7},
  year={2022},
  publisher={IOP Publishing}
}

@article{heymans2021kids,
  title={KiDS-1000 Cosmology: Multi-probe weak gravitational lensing and spectroscopic galaxy clustering constraints},
  author={Heymans, Catherine and Tr{\"o}ster, Tilman and Asgari, Marika and Blake, Chris and Hildebrandt, Hendrik and Joachimi, Benjamin and Kuijken, Konrad and Lin, Chieh-An and S{\'a}nchez, Ariel G and Van Den Busch, Jan Luca and others},
  journal={Astronomy \& Astrophysics},
  volume={646},
  pages={A140},
  year={2021},
  publisher={EDP Sciences}
}

@article{damour1994string,
  title={The string dilation and a least coupling principle},
  author={Damour, Thibault and Polyakov, Alexander M},
  journal={Nuclear Physics B},
  volume={423},
  number={2-3},
  pages={532--558},
  year={1994},
  publisher={Elsevier}
}

@article{gleyzes2015effective,
  title={Effective theory of interacting dark energy},
  author={Gleyzes, J{\'e}r{\^o}me and Langlois, David and Mancarella, Michele and Vernizzi, Filippo},
  journal={Journal of Cosmology and Astroparticle Physics},
  volume={2015},
  number={08},
  pages={054},
  year={2015},
  publisher={IOP Publishing}
}

@article{gleyzes2016effective,
  title={Effective theory of dark energy at redshift survey scales},
  author={Gleyzes, J{\'e}r{\^o}me and Langlois, David and Mancarella, Michele and Vernizzi, Filippo},
  journal={Journal of Cosmology and Astroparticle Physics},
  volume={2016},
  number={02},
  pages={056},
  year={2016},
  publisher={IOP Publishing}
}

@article{coleman1973radiative,
  title={Radiative corrections as the origin of spontaneous symmetry breaking},
  author={Coleman, Sidney and Weinberg, Erick},
  journal={Physical Review D},
  volume={7},
  number={6},
  pages={1888},
  year={1973},
  publisher={APS}
}

@article{amendola2008quintessence,
  title={Quintessence cosmologies with a growing matter component},
  author={Amendola, Luca and Baldi, Marco and Wetterich, Christof},
  journal={Physical Review D—Particles, Fields, Gravitation, and Cosmology},
  volume={78},
  number={2},
  pages={023015},
  year={2008},
  publisher={APS}
}

@article{baldi2012dark,
  title={Dark energy simulations},
  author={Baldi, Marco},
  journal={Physics of the Dark Universe},
  volume={1},
  number={1-2},
  pages={162--193},
  year={2012},
  publisher={Elsevier}
}

@article{zhang2020obtaining,
  title={Obtaining a scalar fifth force via a symmetry-breaking coupling between the scalar field and matter},
  author={Zhang, Hai-Chao},
  journal={Physical Review D},
  volume={101},
  number={4},
  pages={044020},
  year={2020},
  publisher={APS}
}

@article{bamba2013spontaneous,
  title={Spontaneous symmetry breaking in cosmos: The hybrid symmetron as a dark energy switching device},
  author={Bamba, K and Gannouji, R and Kamijo, M and Nojiri, S and Sami, M},
  journal={Journal of Cosmology and Astroparticle Physics},
  volume={2013},
  number={07},
  pages={017},
  year={2013},
  publisher={IOP Publishing}
}

@article{nakagawa2023early,
  title={Early dark energy by a dark Higgs field and axion-induced nonthermal trapping},
  author={Nakagawa, Shota and Takahashi, Fuminobu and Yin, Wen},
  journal={Physical Review D},
  volume={107},
  number={6},
  pages={063016},
  year={2023},
  publisher={APS}
}

@article{clemson2012interacting,
  title={Interacting dark energy: Constraints and degeneracies},
  author={Clemson, Timothy and Koyama, Kazuya and Zhao, Gong-Bo and Maartens, Roy and V{\"a}liviita, Jussi},
  journal={Physical Review D—Particles, Fields, Gravitation, and Cosmology},
  volume={85},
  number={4},
  pages={043007},
  year={2012},
  publisher={APS}
}

@article{pettorino2008coupled,
  title={Coupled and Extended Quintessence: theoretical differences and structure formation},
  author={Pettorino, Valeria and Baccigalupi, Carlo},
  journal={Physical Review D—Particles, Fields, Gravitation, and Cosmology},
  volume={77},
  number={10},
  pages={103003},
  year={2008},
  publisher={APS}
}

@article{patt2006higgs,
  title={Higgs-field portal into hidden sectors},
  author={Patt, Brian and Wilczek, Frank},
  journal={arXiv preprint hep-ph/0605188},
  year={2006}
}

@article{hinterbichler2010screening,
  title={Screening long-range forces through local symmetry restoration},
  author={Hinterbichler, Kurt and Khoury, Justin},
  journal={Physical review letters},
  volume={104},
  number={23},
  pages={231301},
  year={2010},
  publisher={APS}
}

@article{hinterbichler2011symmetron,
  title={Symmetron cosmology},
  author={Hinterbichler, Kurt and Khoury, Justin and Levy, Aaron and Matas, Andrew},
  journal={Physical Review D—Particles, Fields, Gravitation, and Cosmology},
  volume={84},
  number={10},
  pages={103521},
  year={2011},
  publisher={APS}
}

@article{khoury2004chameleon,
  title={Chameleon fields: Awaiting surprises for tests of gravity in space},
  author={Khoury, Justin and Weltman, Amanda},
  journal={Physical review letters},
  volume={93},
  number={17},
  pages={171104},
  year={2004},
  publisher={APS}
}

@article{amendola2014multifield,
  title={Multifield coupled quintessence},
  author={Amendola, Luca and Barreiro, Tiago and Nunes, Nelson J},
  journal={Physical Review D},
  volume={90},
  number={8},
  pages={083508},
  year={2014},
  publisher={APS}
}

@article{upadhye2013symmetron,
  title={Symmetron dark energy in laboratory experiments},
  author={Upadhye, Amol},
  journal={Physical review letters},
  volume={110},
  number={3},
  pages={031301},
  year={2013},
  publisher={APS}
}

@article{potting2022coupled,
  title={Coupled quintessence with a generalized interaction term},
  author={Potting, Robertus and S{\'a}, Paulo M},
  journal={International Journal of Modern Physics D},
  volume={31},
  number={15},
  pages={2250111},
  year={2022},
  publisher={World Scientific}
}

@article{curtin2021twin,
  title={Twin Higgs portal dark matter},
  author={Curtin, David and Gryba, Shayne},
  journal={Journal of High Energy Physics},
  volume={2021},
  number={8},
  pages={1--45},
  year={2021},
  publisher={Springer}
}

@article{baldi2010hydrodynamical,
  title={Hydrodynamical N-body simulations of coupled dark energy cosmologies},
  author={Baldi, Marco and Pettorino, Valeria and Robbers, Georg and Springel, Volker},
  journal={Monthly Notices of the Royal Astronomical Society},
  volume={403},
  number={4},
  pages={1684--1702},
  year={2010},
  publisher={The Royal Astronomical Society}
}

@article{pradosh2025loop,
  title={Loop-corrected scalar potentials and late-time acceleration in f (R) gravity},
  author={Pradosh Keshav, MV and Kenath, Arun},
  journal={The European Physical Journal C},
  volume={85},
  number={9},
  pages={990},
  year={2025},
  publisher={Springer}
}

@article{cruickshank2025forecasts,
  title={Forecasts for interacting dark energy with time-dependent momentum exchange},
  author={Cruickshank, Nathan and Crittenden, Robert and Koyama, Kazuya and Bruni, Marco},
  journal={arXiv preprint arXiv:2504.03555},
  year={2025}
}

@article{brout2022pantheon+,
  title={The Pantheon+ analysis: SuperCal-fragilistic cross calibration, retrained SALT2 light-curve model, and calibration systematic uncertainty},
  author={Brout, Dillon and Taylor, Georgie and Scolnic, Dan and Wood, Charlotte M and Rose, Benjamin M and Vincenzi, Maria and Dwomoh, Arianna and Lidman, Christopher and Riess, Adam and Ali, Noor and others},
  journal={The Astrophysical Journal},
  volume={938},
  number={2},
  pages={111},
  year={2022},
  publisher={IOP Publishing}
}

@article{V:2025oex,
    author = "V., Pradosh Keshav M. and Kenath, Arun",
    title = "{Quintessence and false vacuum: Two sides of the same coin?}",
    eprint = "2504.18611",
    archivePrefix = "arXiv",
    primaryClass = "gr-qc",
    doi = "10.1007/s12043-025-02910-x",
    journal = "Pramana",
    volume = "99",
    number = "2",
    pages = "58",
    year = "2025"
}

@article{blas2011cosmic,
  title={The cosmic linear anisotropy solving system (CLASS). Part II: Approximation schemes},
  author={Blas, Diego and Lesgourgues, Julien and Tram, Thomas},
  journal={Journal of Cosmology and Astroparticle Physics},
  volume={2011},
  number={07},
  pages={034--034},
  year={2011}
}

@article{sabogal2025sign,
  title={Sign switching in dark sector coupling interactions as a candidate for resolving cosmological tensions},
  author={Sabogal, Miguel A and Silva, Emanuelly and Nunes, Rafael C and Kumar, Suresh and Di Valentino, Eleonora},
  journal={Physical Review D},
  volume={111},
  number={4},
  pages={043531},
  year={2025},
  publisher={APS}
}

@article{sabogal2024quantifying,
  title={Quantifying the S 8 tension and evidence for interacting dark energy from redshift-space distortion measurements},
  author={Sabogal, Miguel A and Silva, Emanuelly and Nunes, Rafael C and Kumar, Suresh and Di Valentino, Eleonora and Giar{\`e}, William},
  journal={Physical Review D},
  volume={110},
  number={12},
  pages={123508},
  year={2024},
  publisher={APS}
}

@article{pourtsidou2013models,
  title={Models of dark matter coupled to dark energy},
  author={Pourtsidou, Alkistis and Skordis, C and Copeland, EJ},
  journal={Physical Review D—Particles, Fields, Gravitation, and Cosmology},
  volume={88},
  number={8},
  pages={083505},
  year={2013},
  publisher={APS}
}

@article{gasperini1994dilaton,
  title={Dilaton production in string cosmology},
  author={Gasperini, Maurizio and Veneziano, Gabriele},
  journal={Physical Review D},
  volume={50},
  number={4},
  pages={2519},
  year={1994},
  publisher={APS}
}

@article{joyce2015beyond,
  title={Beyond the cosmological standard model},
  author={Joyce, Austin and Jain, Bhuvnesh and Khoury, Justin and Trodden, Mark},
  journal={Physics Reports},
  volume={568},
  pages={1--98},
  year={2015},
  publisher={Elsevier}
}

@article{mv2025spontaneous,
doi = {10.1088/1475-7516/2026/05/019},
year = {2026},
month = {may},
publisher = {IOP Publishing},
volume = {2026},
number = {05},
pages = {019},
author = {MV, Pradosh Keshav and Kavya, NS and Arun, Kenath},
title = {Spontaneous symmetry breaking as a late-time trigger for interacting dark energy},
journal = {Journal of Cosmology and Astroparticle Physics}
}

@article{damour1992tensor,
  title={Tensor-multi-scalar theories of gravitation},
  author={Damour, Thibault and Esposito-Farese, Gilles},
  journal={Classical and Quantum Gravity},
  volume={9},
  number={9},
  pages={2093--2176},
  year={1992}
}

@incollection{faraoni2004scalar,
  title={Scalar-tensor gravity},
  author={Faraoni, Valerio},
  booktitle={Cosmology in Scalar-Tensor Gravity},
  pages={1--53},
  year={2004},
  publisher={Springer}
}

@article{brax2012unified,
  title={Unified description of screened modified gravity},
  author={Brax, Philippe and Davis, Anne-Christine and Li, Baojiu and Winther, Hans A},
  journal={Physical Review D—Particles, Fields, Gravitation, and Cosmology},
  volume={86},
  number={4},
  pages={044015},
  year={2012},
  publisher={APS}
}

@article{alam2021completed,
  title={The completed SDSS-IV extended Baryon Oscillation Spectroscopic Survey: N-body mock challenge for the eBOSS emission line galaxy sample},
  author={Alam, Shadab and De Mattia, Arnaud and Tamone, Am{\'e}lie and Avila, S and Peacock, John A and Gonzalez-Perez, V and Smith, Alex and Raichoor, Anand and Ross, Ashley J and Bautista, Julian E and others},
  journal={Monthly Notices of the Royal Astronomical Society},
  volume={504},
  number={4},
  pages={4667--4686},
  year={2021},
  publisher={Oxford University Press}
}

@book{strogatz2024nonlinear,
  title={Nonlinear dynamics and chaos: with applications to physics, biology, chemistry, and engineering},
  author={Strogatz, Steven H},
  year={2024},
  publisher={Chapman and Hall/CRC}
}

@article{linder2005many,
  title={How many dark energy parameters?},
  author={Linder, Eric V and Huterer, Dragan},
  journal={Physical Review D—Particles, Fields, Gravitation, and Cosmology},
  volume={72},
  number={4},
  pages={043509},
  year={2005},
  publisher={APS}
}

@article{linder2004reconstructing,
  title={Reconstructing and deconstructing dark energy},
  author={Linder, Eric V},
  journal={Physical Review D—Particles, Fields, Gravitation, and Cosmology},
  volume={70},
  number={6},
  pages={061302},
  year={2004},
  publisher={APS}
}

@article{di2017constraining,
  title={Constraining dark energy dynamics in extended parameter space},
  author={Di Valentino, Eleonora and Melchiorri, Alessandro and Linder, Eric V and Silk, Joseph},
  journal={Physical Review D},
  volume={96},
  number={2},
  pages={023523},
  year={2017},
  publisher={APS}
}

@article{ghilencea2019spontaneous,
  title={Spontaneous breaking of Weyl quadratic gravity to Einstein action and Higgs potential},
  author={Ghilencea, Dumitru M},
  journal={Journal of High Energy Physics},
  volume={2019},
  number={3},
  pages={1--15},
  year={2019},
  publisher={Springer}
}

@article{dine1981simple,
  title={A simple solution to the strong CP problem with a harmless axion},
  author={Dine, Michael and Fischler, Willy and Srednicki, Mark},
  journal={Physics letters B},
  volume={104},
  number={3},
  pages={199--202},
  year={1981},
  publisher={Elsevier}
}

@article{turner1990windows,
  title={Windows on the axion},
  author={Turner, Michael S},
  journal={Physics Reports},
  volume={197},
  number={2},
  pages={67--97},
  year={1990},
  publisher={Elsevier}
}

@article{zaregonbadi2023cosmological,
  title={Cosmological solutions of chameleon scalar field model},
  author={Zaregonbadi, Raziyeh and Saba, Nasim and Farhoudi, Mehrdad},
  journal={The European Physical Journal C},
  volume={83},
  number={10},
  pages={975},
  year={2023},
  publisher={Springer}
}

@article{foreman2013emcee,
  title={emcee: the MCMC hammer},
  author={Foreman-Mackey, Daniel and Hogg, David W and Lang, Dustin and Goodman, Jonathan},
  journal={Publications of the Astronomical Society of the Pacific},
  volume={125},
  number={925},
  pages={306--312},
  year={2013},
  publisher={University of Chicago Press}
}

@article{scaramella2022euclid,
  title={Euclid preparation-I. The Euclid wide survey},
  author={Scaramella, Roberto and Amiaux, J and Mellier, Y and Burigana, Carlo and Carvalho, CS and Cuillandre, J-C and Da Silva, A and Derosa, A and Dinis, J and Maiorano, E and others},
  journal={Astronomy \& Astrophysics},
  volume={662},
  pages={A112},
  year={2022},
  publisher={EDP sciences}
}

@article{galante2015unity,
  title={Unity of cosmological inflation attractors},
  author={Galante, Mario and Kallosh, Renata and Linde, Andrei and Roest, Diederik},
  journal={Physical review letters},
  volume={114},
  number={14},
  pages={141302},
  year={2015},
  publisher={APS}
}

@article{akrami2018dark,
  title={Dark energy, $\alpha$-attractors, and large-scale structure surveys},
  author={Akrami, Yashar and Kallosh, Renata and Linde, Andrei and Vardanyan, Valeri},
  journal={Journal of Cosmology and Astroparticle Physics},
  volume={2018},
  number={06},
  pages={041--041},
  year={2018}
}

@article{abdul2025desi,
  title={DESI DR2 results. II. Measurements of baryon acoustic oscillations and cosmological constraints},
  author={Abdul Karim, M and Aguilar, J and Ahlen, Steven and Alam, Shadab and Allen, Lori and Prieto, C Allende and Alves, Ot{\'a}vio and Anand, Abhijeet and Andrade, Uendert and Armengaud, E and others},
  journal={Physical Review D},
  volume={112},
  number={8},
  pages={083515},
  year={2025},
  publisher={APS}
}

@article{jackson2009large,
  title={On the large-scale instability in interacting dark energy and dark matter fluids},
  author={Jackson, Brendan M and Taylor, Andy and Berera, Arjun},
  journal={Physical Review D—Particles, Fields, Gravitation, and Cosmology},
  volume={79},
  number={4},
  pages={043526},
  year={2009},
  publisher={APS}
}

@article{qu2025atacama,
  title={Atacama Cosmology Telescope DR6 and DESI: Structure growth measurements from the cross-correlation of DESI legacy imaging galaxies and CMB lensing from ACT DR6 and P lanck PR4},
  author={Qu, Frank J and Hang, Qianjun and Farren, Gerrit and Bolliet, Boris and Aguilar, Jessica Nicole and Ahlen, Steven and Alam, Shadab and Brooks, David and Cai, Yan-Chuan and Calabrese, Erminia and others},
  journal={Physical Review D},
  volume={111},
  number={10},
  pages={103503},
  year={2025},
  publisher={APS}
}

@article{you2025dynamical,
  title={Dynamical dark energy implies a coupled dark sector: Insights from DESI DR2 via a data-driven approach},
  author={You, Changyu and Wang, Dan and Yang, Tao},
  journal={Physical Review D},
  volume={112},
  number={4},
  pages={043503},
  year={2025},
  publisher={APS}
}

@article{giare2024interacting,
  title={Interacting dark energy after DESI baryon acoustic oscillation measurements},
  author={Giar{\`e}, William and Sabogal, Miguel A and Nunes, Rafael C and Di Valentino, Eleonora},
  journal={Physical Review Letters},
  volume={133},
  number={25},
  pages={251003},
  year={2024},
  publisher={APS}
}

@article{burrage2018tests,
  title={Tests of chameleon gravity},
  author={Burrage, Clare and Sakstein, Jeremy},
  journal={Living reviews in relativity},
  volume={21},
  number={1},
  pages={1},
  year={2018},
  publisher={Springer}
}

@article{nojiri2011unified,
  title={Unified cosmic history in modified gravity: from F (R) theory to Lorentz non-invariant models},
  author={Nojiri, Shin’ichi and Odintsov, Sergei D},
  journal={Physics Reports},
  volume={505},
  number={2-4},
  pages={59--144},
  year={2011},
  publisher={Elsevier}
}

@article{elizalde2004late,
  title={Late-time cosmology in a (phantom) scalar-tensor theory: dark energy and the cosmic speed-up},
  author={Elizalde, Emilio and Nojiri, Shin’ichi and Odintsov, Sergei D},
  journal={Physical Review D},
  volume={70},
  number={4},
  pages={043539},
  year={2004},
  publisher={APS}
}

@article{xia2009constraint,
  title={Constraint on coupled dark energy models from observations},
  author={Xia, Jun-Qing},
  journal={Physical Review D—Particles, Fields, Gravitation, and Cosmology},
  volume={80},
  number={10},
  pages={103514},
  year={2009},
  publisher={APS}
}

@article{xia2013new,
  title={New limits on coupled dark energy from Planck},
  author={Xia, Jun-Qing},
  journal={Journal of Cosmology and Astroparticle Physics},
  volume={2013},
  number={11},
  pages={022--022},
  year={2013}
}

@article{da2022simple,
  title={A simple parametrisation for coupled dark energy},
  author={da Fonseca, Vitor and Barreiro, Tiago and Nunes, Nelson J},
  journal={Physics of the Dark Universe},
  volume={35},
  pages={100940},
  year={2022},
  publisher={Elsevier}
}

@article{lebedev2023dark,
  title={Dark matter production via a non-minimal coupling to gravity},
  author={Lebedev, Oleg and Solomko, Timofey and Yoon, Jong-Hyun},
  journal={Journal of Cosmology and Astroparticle Physics},
  volume={2023},
  number={02},
  pages={035},
  year={2023},
  publisher={IOP Publishing}
}

@article{lerner2009gauge,
  title={Gauge singlet scalar as inflaton and thermal relic dark matter},
  author={Lerner, Rose N and McDonald, John},
  journal={Physical Review D—Particles, Fields, Gravitation, and Cosmology},
  volume={80},
  number={12},
  pages={123507},
  year={2009},
  publisher={APS}
}

@misc{class_ide,
  author       = {Pradosh Keshav MV},
  title        = {class\_ide: Interacting Dark Energy Analysis Code},
  year         = {2026},
  howpublished = {\url{https://github.com/pradosh93keshav-code/class_ide}},
  note         = {GitHub repository, accessed: 2026-04-06}
}

@article{brax2021testing,
  title={Testing screened modified gravity},
  author={Brax, Philippe and Casas, Santiago and Desmond, Harry and Elder, Benjamin},
  journal={Universe},
  volume={8},
  number={1},
  pages={11},
  year={2021},
  publisher={MDPI}
}

@article{alvarez2019einstein,
  title={Einstein Yang--Mills Higgs dark energy revisited},
  author={{\'A}lvarez, Miguel and Bayron Orjuela-Quintana, J and Rodriguez, Yeinzon and Valenzuela-Toledo, Cesar A},
  journal={Classical and Quantum Gravity},
  volume={36},
  number={19},
  pages={195004},
  year={2019},
  publisher={IOP Publishing}
}

@article{touboul2022result,
  title={Result of the MICROSCOPE weak equivalence principle test},
  author={Touboul, Pierre and Metris, Gilles and Rodrigues, Manuel and Berge, Joel and Robert, Alain and Baghi, Quentin and Andre, Yves and Bedouet, Judicael and Boulanger, Damien and Bremer, Stefanie and others},
  journal={Classical and Quantum Gravity},
  volume={39},
  number={20},
  pages={204009},
  year={2022},
  publisher={IOP Publishing}
}

@article{barros2019coupled,
  title={Coupled quintessence with a $\Lambda$CDM background: removing the $\sigma$8 tension},
  author={Barros, Bruno J and Amendola, Luca and Barreiro, Tiago and Nunes, Nelson J},
  journal={Journal of Cosmology and Astroparticle Physics},
  volume={2019},
  number={01},
  pages={007--007},
  year={2019}
}

@article{barros2020spherical,
  title={Spherical collapse in coupled quintessence with a $\Lambda$ CDM background},
  author={Barros, Bruno J and Barreiro, Tiago and Nunes, Nelson J},
  journal={Physical Review D},
  volume={101},
  number={2},
  pages={023502},
  year={2020},
  publisher={APS}
}

@article{joseph2023forecast,
  title={Forecast analysis on interacting dark energy models from future generation PICO and DESI missions},
  author={Joseph, Albin and Saha, Rajib},
  journal={Monthly Notices of the Royal Astronomical Society},
  volume={519},
  number={2},
  pages={1809--1822},
  year={2023},
  publisher={Oxford University Press}
}

@article{teixeira2023forecasts,
  title={Forecasts on interacting dark energy with standard sirens},
  author={Teixeira, Elsa M and Daniel, Richard and Frusciante, Noemi and van de Bruck, Carsten},
  journal={Physical Review D},
  volume={108},
  number={8},
  pages={084070},
  year={2023},
  publisher={APS}
}

@article{wright2025kids,
  title={KiDS-Legacy: Cosmological constraints from cosmic shear with the complete Kilo-Degree Survey},
  author={Wright, Angus H and St{\"o}lzner, Benjamin and Asgari, Marika and Bilicki, Maciej and Giblin, Benjamin and Heymans, Catherine and Hildebrandt, Hendrik and Hoekstra, Henk and Joachimi, Benjamin and Kuijken, Konrad and others},
  journal={Astronomy \& Astrophysics},
  volume={703},
  pages={A158},
  year={2025},
  publisher={EDP Sciences}
}

@article{silva2025new,
  title={New constraints on interacting dark energy from DESI DR2 BAO observations},
  author={Silva, Emanuelly and Sabogal, Miguel A and Scherer, Mateus and Nunes, Rafael C and Di Valentino, Eleonora and Kumar, Suresh},
  journal={Physical Review D},
  volume={111},
  number={12},
  pages={123511},
  year={2025},
  publisher={APS}
}

@article{petri2026dark,
  title={Dark degeneracy in DESI DR2 data: Interacting or evolving dark energy?},
  author={Petri, Vitor and Marra, Valerio and Von Marttens, Rodrigo},
  journal={Physical Review D},
  volume={113},
  number={2},
  pages={023504},
  year={2026},
  publisher={APS}
}

@article{cima2026probing,
  title={Probing non-minimal dark sectors via the 21 cm line at cosmic dawn},
  author={Cima, Federico and D'Eramo, Francesco},
  journal={Journal of Cosmology and Astroparticle Physics},
  volume={2026},
  number={02},
  pages={020},
  year={2026},
  publisher={IOP Publishing}
}

@article{goh2024observational,
  title={Observational constraints on early coupled quintessence},
  author={Goh, Lisa WK and Bachs-Esteban, Joan and G{\'o}mez-Valent, Adri{\`a} and Pettorino, Valeria and Rubio, Javier},
  journal={Physical Review D},
  volume={109},
  number={2},
  pages={023530},
  year={2024},
  publisher={APS}
}

@article{ivezic2019lsst,
  title={LSST: from science drivers to reference design and anticipated data products},
  author={Ivezi{\'c}, {\v{Z}}eljko and Kahn, Steven M and Tyson, J Anthony and Abel, Bob and Acosta, Emily and Allsman, Robyn and Alonso, David and AlSayyad, Yusra and Anderson, Scott F and Andrew, John and others},
  journal={The Astrophysical Journal},
  volume={873},
  number={2},
  pages={111},
  year={2019},
  publisher={The American Astronomical Society}
}

@article{mazoun2025interacting,
  title={Interacting dark sector within ETHOS: Cosmological constraints from SPT cluster abundance with DES and HST weak lensing data},
  author={Mazoun, Asmaa and Bocquet, Sebastian and Mohr, Joseph J and Garny, Mathias and Rubira, Henrique and Klein, Matthias and Bleem, LE and Grandis, Sebastian and Schrabback, Tim and Aguena, M and others},
  journal={Physical Review D},
  volume={111},
  number={8},
  pages={083543},
  year={2025},
  publisher={APS}
}

@article{mazoun2024probing,
  title={Probing interacting dark sector models with future weak lensing-informed galaxy cluster abundance constraints from SPT-3G and CMB-S4},
  author={Mazoun, Asmaa and Bocquet, Sebastian and Garny, Mathias and Mohr, Joseph J and Rubira, Henrique and Vogt, Sophie ML},
  journal={Physical Review D},
  volume={109},
  number={6},
  pages={063536},
  year={2024},
  publisher={APS}
}

@article{seraille2024constraining,
  title={Constraining dark energy with the integrated Sachs-Wolfe effect},
  author={Seraille, Emeric and Noller, Johannes and Sherwin, Blake D},
  journal={Physical Review D},
  volume={110},
  number={12},
  pages={123525},
  year={2024},
  publisher={APS}
}

@article{vardanyan2023modeling,
  title={Modeling and testing screening mechanisms in the laboratory and in space},
  author={Vardanyan, Valeri and Bartlett, Deaglan J},
  journal={Universe},
  volume={9},
  number={7},
  pages={340},
  year={2023},
  publisher={MDPI}
}

@article{ghirardini2024srg,
  title={The SRG/eROSITA all-sky survey-Cosmology constraints from cluster abundances in the western Galactic hemisphere},
  author={Ghirardini, V and Bulbul, E and Artis, E and Clerc, N and Garrel, C and Grandis, S and Kluge, M and Liu, A and Bahar, YE and Balzer, F and others},
  journal={Astronomy \& Astrophysics},
  volume={689},
  pages={A298},
  year={2024},
  publisher={EDP Sciences}
}

@article{silva2026one,
  title={One-loop power spectrum corrections in interacting dark energy cosmologies},
  author={Silva, Emanuelly and Hartmann, Gabriel and Nunes, Rafael C},
  journal={Physical Review D},
  volume={113},
  number={6},
  pages={063548},
  year={2026},
  publisher={APS}
}

@article{van2024interacting,
  title={Interacting dark energy: clarifying the cosmological implications and viability conditions},
  author={van der Westhuizen, Marcel A and Abebe, Amare},
  journal={Journal of Cosmology and Astroparticle Physics},
  volume={2024},
  number={01},
  pages={048},
  year={2024},
  publisher={IOP Publishing}
}

@article{van2025linear,
  title={I. Linear interacting dark energy: Analytical solutions and theoretical pathologies},
  author={Van Der Westhuizen, Marcel and Abebe, Amare and Di Valentino, Eleonora},
  journal={Physics of the Dark Universe},
  pages={102119},
  year={2025},
  publisher={Elsevier}
}

@article{mellier2025euclid,
  title={Euclid. I. Overview of the Euclid mission},
  author={Mellier, Yannick and Abdurro{\'u}f, Abdurro{\'u}f and Barroso, JA Acevedo and Ach{\'u}carro, A and Adamek, J and Adam, Ren{\'e} and Addison, Graeme E and Aghanim, N and Aguena, M and Ajani, V and others},
  journal={Astronomy \& Astrophysics},
  year={2025},
  publisher={EDP sciences}
}

@article{artis2025srg,
  title={The SRG/eROSITA All-Sky Survey-Constraints on the structure growth from cluster number counts},
  author={Artis, E and Bulbul, E and Grandis, S and Ghirardini, V and Clerc, N and Seppi, R and Comparat, J and Cataneo, M and Von Der Linden, A and Bahar, YE and others},
  journal={Astronomy \& Astrophysics},
  volume={696},
  pages={A5},
  year={2025},
  publisher={EDP Sciences}
}

@article{aoki2026effective,
  title={Effective field theory of coupled dark energy and dark matter},
  author={Aoki, Katsuki and Beltr{\'a}n Jim{\'e}nez, Jose and Pookkillath, Masroor C and Tsujikawa, Shinji},
  journal={Physical Review D},
  volume={113},
  number={4},
  pages={044053},
  year={2026},
  publisher={APS}
}

@article{pan2026interacting,
  title={Interacting dark energy after DESI DR2: A challenge for the $\Lambda$ CDM paradigm?},
  author={Pan, Supriya and Paul, Sivasish and Saridakis, Emmanuel N and Yang, Weiqiang},
  journal={Physical Review D},
  volume={113},
  number={2},
  pages={023515},
  year={2026},
  publisher={APS}
}

@article{patil2024coupled,
  title={Coupled quintessence scalar field model in light of observational datasets},
  author={Patil, Trupti and Ruchika and Panda, Sukanta},
  journal={Journal of Cosmology and Astroparticle Physics},
  volume={2024},
  number={05},
  pages={033},
  year={2024},
  publisher={IOP Publishing}
}

@article{hoekstra2008weak,
  title={Weak gravitational lensing and its cosmological applications},
  author={Hoekstra, Henk and Jain, Bhuvnesh},
  journal={Annual Review of Nuclear and Particle Science},
  volume={58},
  number={1},
  pages={99--123},
  year={2008},
  publisher={Annual Reviews}
}

@article{tegmark1997karhunen,
  title={Karhunen-Loeve eigenvalue problems in cosmology: How should we tackle large data sets?},
  author={Tegmark, Max and Taylor, Andy N and Heavens, Alan F},
  journal={The Astrophysical Journal},
  volume={480},
  number={1},
  pages={22--35},
  year={1997}
}

@article{pantos2026status,
  title={Status of the S8 Tension: A 2026 Review of Probe Discrepancies},
  author={Pantos, Ioannis and Perivolaropoulos, Leandros},
  journal={Physics of the Dark Universe},
  pages={102286},
  year={2026},
  publisher={Elsevier}
}
\bibliographystyle{unsrt}

\end{document}